\documentclass[10pt,A4]{article}
\usepackage{amsfonts}
\usepackage{epsfig}
\usepackage{graphicx}
\usepackage{latexsym}
\usepackage{amsmath}
\usepackage{amssymb}
\usepackage{hyperref}
\DeclareGraphicsExtensions{.ps,.eps}
\newcommand{\qed}{\hfill \rule {1ex}{1ex} \smallskip}
\newtheorem{defin}{Definition}
\newtheorem{lemma}{Lemma}
\newtheorem{theorem}{Theorem}

\newtheorem{proposition}{Proposition}
\newcommand{\resetsect}{\setcounter{section}{1}}
\newcommand{\resetequ}{\setcounter{equation}{0}}

\begin{document}
\title{Quasi-diffusion  in a 3D Supersymmetric\\
Hyperbolic Sigma Model}

\author{M. Disertori$^a$%
\footnote{e-mail: Margherita.Disertori@univ-rouen.fr},
T. Spencer$^b$, M.R. Zirnbauer$^c$ \\
$a$) Laboratoire de Math\'ematiques Rapha\"el Salem, UMR CNRS 6085\\
Universit\'e de Rouen, 76801, France\\
$b$) Institute for Advanced Study, Einstein Drive, \\
Princeton, NJ 08540,  USA\\
$c$) Institut f\"ur Theoretische Physik, Universit\"at zu K\"oln,\\
Z\"ulpicher Stra{\ss}e 77, 50937 K\"oln, Germany \\}
\maketitle
\begin{abstract}
Abstract: We study a lattice field model which qualitatively reflects
the phenomenon of Anderson localization and delocalization for real
symmetric band matrices. In this statistical mechanics model, the
field takes values in a supermanifold based on the hyperbolic plane.
Correlations in this model may be described in terms of a random walk
in a highly correlated random environment. We prove that in three or
more dimensions the model has a `diffusive' phase at low
temperatures. Localization is expected at high temperatures. Our
analysis uses estimates on non-uniformly elliptic Green's functions
and a family of Ward identities coming from internal supersymmetry.
\end{abstract}

\section{Introduction}

\subsection{Some history and motivation}

It has been known since the pioneering work of Wegner
\cite{wegner,sw} that information about the spectral and transport
properties of random band matrices and random Schr{\"o}dinger
operators can be inferred from the correlation functions of
statistical mechanical models of a certain kind. These models have a
hyperbolic symmetry, typically  a noncompact group such as
$\mathrm{O}(p,q)$ or $\mathrm{U}(p,q)$, and were originally studied
in the limit of $p = q = 0$ replicas.

The connection between random Schr\"odinger operators and statistical
mechanics models was made more precise by Efetov \cite{efetov-Adv},
who introduced the so-called supersymmetry method to avoid the use of
replicas. In Efetov's formulation one employs both commuting (or
bosonic) and anticommuting (or Grassmann) integration variables, and
these are related by a natural symmetry that makes the emerging
statistical mechanics system supersymmetric (SUSY). The simplest
class of these models has a $\mathrm{U}(1,1|2)$ symmetry. This means
that for the bosonic variables there exists a hyperbolic
symmetry $\mathrm{U}(1,1)$ preserving an indefinite Hermitian form on
$\mathbb{C}^2$, and the  Grassmann variables are governed by
a compact $\mathrm{U}(2)$ symmetry. Moreover, there exist odd
symmetries mixing Grassmann and bosonic variables.

The fields $\mathbb{Z}^d \ni j \mapsto Q_j$ of the supersymmetric
models introduced by Efetov are 4 by 4 supermatrices built from
bosonic as well as Grassmann entries. In the physics literature one
usually assumes the sigma model approximation, which is believed to
capture the essential features of the energy correlations and
transport properties of the underlying quantum system. The sigma
model approximation constrains the matrix field $Q$ by $Q_j^2 =
\mathrm{Id}$ for all $j$. This constraint is similar to the
constraints appearing in the Ising or Heisenberg models, where $S_j
\cdot S_j = 1\,$. We refer the reader to
\cite{efetov-book,mirlin,fyodorov-NM,disertori-GUE,LSZ} for an
introduction to these ideas.

The models described above are difficult to analyse with mathematical
rigor in more than one dimension. In this paper we study  a simpler
SUSY model. Our exposition will be essentially self-contained and the
full supersymmetric formalism alluded to here will serve primarily as
a source of motivation.

\subsection{Probabilistic representation of our model}\label{sect:1.2}

In this paper we analyze a lattice field model which may be thought
of as a simplified version of one of Efetov's nonlinear sigma models.
More precisely, it is related to the model that derives from real
symmetric matrices, see Section \ref{sect:origin}. In this
statistical mechanics model the field at site $j$ has four degrees of
freedom. Two of these, $t_j$ and $s_j\,$, parametrize a hyperboloid
and the other two, $\bar\psi_j$ and $\psi_j\,$, are Grassmann (i.e.,
anticommuting) variables. Technically speaking, the field takes
values in a target space denoted by $\mathrm{H}^{2|2}$, which is a
supermanifold extension of the hyperbolic plane $\mathrm{H}^2$; see
Section \ref{sect:defs}. This model was introduced by one of us in
\cite{toymodel,migdal-kad}, and localization was established in one
dimension (1D) in the sense that the conductance was proven to decay
exponentially in the system size \cite{toymodel}. The model is
expected to reflect the qualitative behavior of random band matrices
-- namely localization and diffusion -- in any dimension. However,
most of our discussion will be restricted to three dimensions.

Our supersymmetric hyperbolic nonlinear sigma model, called the
$\mathrm{H}^{2|2}$ model for short, will be formulated on a lattice
cube $\Lambda \subset \mathbb{Z}^d$ of side $L\,$. We shall see (in
Section \ref{sect:horo}) that the action of the field variables is
quadratic in $\psi$, $\bar\psi$, and $s$. This feature is special to
the horospherical coordinate system that we use. It enables us to
reduce the $\mathrm{H}^{2|2}$ model to the statistical mechanics of a
\emph{single} field $t :\, \Lambda \to \mathbb{R}\,$, $j \mapsto
t_j\,$. Its free energy or effective action, $F(t)$, is real, so the
resulting statistical mechanical model has a probabilistic
interpretation.

In order to specify $F(t)$, first consider the finite-difference
elliptic operator $D_{\beta,\varepsilon}(t)$ defined by the quadratic
form
\begin{equation}\label{eq:1.1}
    [v\,;D_{\beta,\varepsilon}(t)\,v]_\Lambda = \beta
    \sum\nolimits_{(ij)} \mathrm{e}^{t_i+t_j} (v_i-v_j)^2 +
    \varepsilon \sum\nolimits_{k\in\Lambda} \mathrm{e}^{t_k} v_k^2\;.
\end{equation}
This operator plays a central role in our analysis. The first sum is
over nearest neighbor pairs in $\Lambda$, and $[\,;\, ]_\Lambda$
denotes the usual 
scalar product in $\ell^2 (\Lambda)$. We see that $D_{1,0}(\mathbf{0}
)$ is the finite-difference Laplacian. The regularization parameter
$\varepsilon > 0$ will serve to make the theory well-defined. One may
interpret $D_{\beta,\varepsilon} (t)$ as the generator of a random
walk in an environment given by the fluctuating field $t$, with a
death rate of $\varepsilon\, \mathrm{e}^{t_j}$ at site $j$. Note that
the operator $D$ is elliptic but not uniformly so, as $t_j \in
\mathbb{R}$ has unbounded range.

The free energy or effective action $F_{\beta,\varepsilon}(t)$ is now
expressed by
\begin{align}
    F_{\beta,\varepsilon} (t) &= \beta \sum\nolimits_{(ij)}
    (\cosh(t_i - t_j) - 1) \cr &- \ln \mathrm{Det}^{1/2}
    D_{\beta,\varepsilon} (t) + \sum\nolimits_{k\in\Lambda}
    (t_k -\varepsilon +\varepsilon\,\cosh t_k )\label{eq:1.2}\;.
\end{align}
If $d\mu_\Lambda$ denotes the product measure
\begin{equation}\label{eq:1.3}
    d\mu_\Lambda = \prod_{k\in\Lambda}\frac{dt_k}{\sqrt{2\pi}}
\end{equation}
on $\mathbb{R}^{|\Lambda|}$, then the partition function is
\begin{equation}\label{eq:1.4}
    Z_{\Lambda}(\beta,\varepsilon) = \int_{\mathbb{R}^{|\Lambda|}}
    \mathrm{e}^{- F_{\beta,\varepsilon}}\, d\mu_\Lambda = 1\;.
\end{equation}
The partition function is identically equal to unity independent of
$\beta$, $\varepsilon$ even when $\beta$ depends on the edge $(ij)$
and $\varepsilon$ depends on the lattice point $k\,$; see
\eqref{eq:var-par}. This is a reflection of an internal supersymmetry
which will be explained in later sections. There exist many variants
of this identity. One of them gives us easy control of nearest
neighbor fluctuations of the field $t$ (cf.\ Section \ref{sect:4}).

The expectation of an observable function $t \mapsto f(t)$ is defined
by
\begin{equation}\label{eq:1.5}
    \langle f \rangle_{\Lambda,\beta,\varepsilon} = \int f\,
    \mathrm{e}^{- F_{\beta,\varepsilon}} d\mu_{\Lambda} \;.
\end{equation}

Let us make a few comments on these expository definitions.
\begin{enumerate}
\item The action or free energy $F_{\beta,\varepsilon}(t)$ is
    nonlocal due to the presence of the term $- \ln
    \mathrm{Det}^{1/2} D(t)$. This nonlocality arises from
    integrating out three massless free fields, one ($s$) of
    bosonic and two ($\bar\psi, \psi$) of Grassmann type.
\item $F_{\beta, \varepsilon}(t)$ is not convex as a function of
    $t$ and therefore the Brascamp-Lieb estimates used in earlier
    work on a related model \cite{SZ} do not apply. The lack of
    convexity is an important feature and opens the possibility
    for a localization-delocalization transition to occur.
\item When $\varepsilon=0$, $F_{\beta,0}(t)$ is invariant under
    shifts $t_j \to t_j + c$ by any constant $c \in \mathbb{R}
    \,$. To see this, note that for $\varepsilon = 0$ we have
    $D_{\beta,0}(t + c) = \mathrm{e}^{2 c} D_{\beta,0}(t)$ by
    \eqref{eq:1.1}. The resulting additional term $- |\Lambda| c$
    from $- \ln \mathrm{Det}^{1/2} D_{\beta,0} (t)$ in
    \eqref{eq:1.2} is canceled by another such term, which arises
    from shifting $\sum_{k \in \Lambda} t_k\,$. This symmetry
    (which is a formal one, since the integral is ill-defined for
    $\varepsilon = 0$) is associated with the presence of a
    massless mode. The importance of the regularization
    $\varepsilon$, which was omitted from the present argument,
    becomes evident from the saddle point discussed below.
\item The model at hand describes a disordered quantum system at
    \emph{zero temperature}. Nevertheless, adopting the familiar
    language of statistical mechanics and thermodynamics, we
    refer to the field stiffness $\beta$ as the inverse
    `temperature'. ($\beta$ is actually the dimensionless
    conductance for an Ohmic system of size $L = 1$ as measured
    in lattice units.)
\end{enumerate}

\subsection{Main result}

The main goal of this paper is to estimate the fluctuations of the
field $t$ for large values of the parameter $\beta$ and dimension $d
= 3$.  This will enable us to prove that the random walk in the
random environment drawn from $F(t)$ is transient. More precisely, we
will prove the following. (Similar estimates hold for all dimensions
$d \ge 3$.)
\begin{theorem}\label{thm:1}
For $d = 3$, there is a $\bar\beta \geq 1$ such that if $\beta \geq
\bar\beta$, the fluctuations of the field $t$ are uniformly bounded
in $x$, $y$, and $\Lambda:$
\begin{equation}\label{eq:1.6-TS}
    \langle \cosh^m (t_x - t_y ) \rangle_{\Lambda,\beta,\,\varepsilon}
    \leq 2 \;,
\end{equation}
provided that $m \leq \beta^{1/8}$.
\end{theorem}
This theorem implies that for any $x$ and $y$, $|t_{x}-t_{y}|$ is
very unlikely to be large. A stronger version of \eqref{eq:1.6-TS} is
given in \eqref{Ih1}. We will use this result to prove
\begin{theorem}\label{thm:2}
Under the hypothesis of Theorem \ref{thm:1} the average field is
bounded:
\begin{equation}\label{eq:1.7-TS}
    \langle \cosh^p (t_{x}) \rangle_{\Lambda,\beta,\,\varepsilon}
    \leq   \frac{5}{2}  \;,
\end{equation}
provided $p\leq 10$ and $|\Lambda|^{1-\alpha/3 }\,\varepsilon \geq 1$
with $\alpha \geq 1 / \ln \beta$.  Thus in the thermodynamic limit
$|\Lambda| \to \infty$ we may send $\varepsilon \to 0$ while
maintaining the bound on $\langle \cosh^{p}t_{x} \rangle$.
\end{theorem}

To investigate the localized or extended nature of the energy
eigenstates of a disordered quantum system with Hamiltonian $H$, one
looks at the average square of the quantum Green's function, $|(H-E +
\mathrm{i}\varepsilon)^{-1}(x,y)|^2$. The analog of this Green's
function in the $\mathrm{H}^{2|2}$ model is the two-point correlation
function
\begin{equation}
    C_{xy} = \left\langle \mathrm{e}^{t_x} s_x \, \mathrm{e}^{t_y}
    s_y \right\rangle ,
\end{equation}
where the expectation is given by the full functional integral
defined in Sections \ref{sect:fullmodel}, \ref{sect:horo}. After
integration over the fields $\bar\psi$, $\psi$, and $s$, we have
\begin{equation}\label{eq:C-xy}
    C_{xy} = \big\langle \mathrm{e}^{t_x + t_y} D_{\beta,
    \varepsilon}(t)^{-1} (x,y) \big\rangle_{\Lambda,\beta,
    \varepsilon}\equiv \big\langle \tilde{D}_{\beta,\varepsilon}
    (t)^{-1} (x,y) \big\rangle_{\Lambda,\beta,\varepsilon} \;,
\end{equation}
where $\tilde{D} = \mathrm{e}^{-t} D \circ \mathrm{e}^{-t}$.
Note that $C_{xy}$ is positive both pointwise and
as a quadratic form. A simple
calculation shows that
\begin{equation}\label{eq:Dtilde}
    \tilde{D}_{\beta,\varepsilon}(t) = - \beta \Delta + \beta
    V(t) + \varepsilon\, \mathrm{e}^{-t} \;,
\end{equation}
where $V(t)$ is a diagonal matrix (or `potential') given by
\begin{displaymath}
    V_{jj}(t) = \sum\nolimits_{|i-j|=1} (\mathrm{e}^{t_i - t_j} - 1)
\end{displaymath}
(sum over nearest neighbors) and $\mathrm{e}^{-t}$ is the diagonal
matrix with $(\mathrm{e}^{-t})_{jj} = \mathrm{e}^{-t_j}$. In Appendix
\ref{app:symmetry} we establish the sum rule $\varepsilon \sum_{y \in
\Lambda} C_{xy} = 1$, reflecting conservation of probability for the
quantum dynamics generated by a Hamiltonian $H$.

Note that if $t$ were bounded, then $D(t)$ (given by \eqref{eq:1.1})
would be uniformly elliptic and we could establish good diffusive
bounds on the two-point function $C$ \eqref{eq:C-xy}. However,
Theorems \ref{thm:1} and \ref{thm:2} only say that large field values
are unlikely. To get optimal bounds on $C$ we would need to prove
uniform ellipticity on a percolating set. The set on which $|t_j +
t_{j'}| < M$, is presumably a percolating set but this does not
readily follow from our estimates.

Our next theorem states a quasi-diffusive estimate on $C$. More
precisely let $G_0 = (-\beta \Delta +\varepsilon )^{-1}$ be the
Green's function for the discrete Laplacian (with a regularization
term $\varepsilon $) and $\tilde{G}_0 = (-\beta\Delta +\varepsilon/2
)^{-1} $.  In 3 dimensions $G_\varepsilon (x,y) \leq \beta^{-1}  (1+
|x-y|)^{-1}$ (and the same is true for $\tilde{G}_0$). Then we have
\begin{theorem}\label{thm:new}
Let $f : \, \Lambda \to \mathbb{R}$ be non-negative. Then assuming
the hypotheses of Theorems \ref{thm:1} and \ref{thm:2} we have
\begin{equation}\label{eq:upperbound}
    \frac{1}{K'} [\tilde{f}; G_0 \tilde{f}] \leq [f;Cf] =
    \sum_{ij} C_{ij}\, f(i)f(j) \leq K [f;\tilde{G}_{0} f] \;,
\end{equation}
where $\tilde{f}(j) = (1+|j-x|^{\alpha})^{-1} f(j)$, $x\in \Lambda $
is any fixed point, and $K$ and $K'$
are constants independent of $f$.
\end{theorem}
\paragraph{Remark.} In this paper we always use periodic boundary
conditions on $\Lambda \subset \mathbb{Z}^3$. The distance $|x-y|$
between two points is always the distance on $\Lambda$ with periodic
boundary conditions.

\subsection{Saddle point}

One may try to gain a crude understanding of the behavior of the
$\mathrm{H}^{2|2}$ sigma model via a simple saddle-point analysis.
Let $t^{(0)}$ be the configuration of $t = \{t_j\}$ which minimizes
the effective action $F_{\beta,\varepsilon}(t)$ defined in
\eqref{eq:1.2}. In Appendix \ref{app:minimum} we prove that $t^{(0)}$
is unique and $t_j^{(0)} = t^\ast$ independent of $j$. For large
$\beta$ we find
\begin{equation}\label{eq:saddle-1D2D}
    \text{1D:} \quad \varepsilon\, \mathrm{e}^{-t^*}
    \simeq \beta^{-1}, \qquad \text{2D:} \quad \varepsilon\,
    \mathrm{e}^{-t^\ast} \simeq \mathrm{e}^{-\beta} \;,
\end{equation}
in one and two dimensions, respectively. Thus in 1D or 2D the saddle
point depends sensitively on the regularization parameter
$\varepsilon$. The value of $t^\ast$ suggests a strong asymmetry of
the field favoring negative values of $t$. On the other hand, in 3D
at low temperatures, we find $t^* = 0$ independent of $\varepsilon$.
Our estimates \eqref{eq:1.7-TS} confirm this value by controlling
fluctuations about the saddle. For $\beta$ small, in 3D, the saddle
$t^\ast$ is again strongly $\varepsilon$-sensitive, suggesting
localization.

The bias to negative values of the field $t$ is expected to be
closely related to localization. Note that since $- \Delta + V(t)
\geq 0\,$, the additional term $\varepsilon\, \mathrm{e}^{-t}$ makes
$\tilde{D}_{\beta,\varepsilon}$ strictly positive at the saddle
suggesting that $C_{xy}$ decays roughly like $\mathrm{e}^{- m |x-y|}$
with $m^2 = \varepsilon\, \mathrm{e}^{ -t^\ast} / \beta = \beta^{-2}$
and $\mathrm{e}^{-\beta}$ in 1D and 2D respectively. There are
important fluctuations away from this saddle but we do not expect
them to spoil the exponential decay. For the 1D chain this has been
proved \cite{toymodel}.

\subsection{Edge reinforced random walk}

A number of mathematicians (Kozma, Heydenreich, Sznitmann) have noted
that our random walk looks similar to a linearly edge reinforced
random walk (ERRW). ERRW is a history-dependent walk which prefers to
visit edges it has visited in the past. Let $n(e)$ denote the number
of times the walk has visited the edge $e$. Then the probability that
the walk at vertex $v$ will visit a neighboring edge $e$ equals $
(a+n(e)) / S_a(v)$ where $S$ is the sum of $a+n(e')$ over all the
edges $e'$ touching $v$. The parameter $a$ is analogous to our
$\beta$. Coppersmith and Diaconis \cite{diaconis} proved that this
history-dependent walk can be expressed as a random walk in a random
environment; see also more recent work by Merkl and Rolles
\cite{merkl} in which recurrence of the walk is established on a 2D
lattice for small $\beta$. This is analogous to localization in our
model. The environment of ERRW is very similar to the environment in
$\mathrm{H}^{2|2}$. In fact, both environments have nonlocal actions
arising from the square root of a determinant. Although the two
models do not seem to be identical, they may have similar properties.

\subsection{Outline of the paper}

The remainder of this paper is organized as follows. In the next
section we give a precise definition of the full $\mathrm{H}^{2|2}$
model and introduce the horospherical coordinate system. The
effective action defined in \eqref{eq:1.2} is then derived by
integration of the field $s$ and the Grassmann fields $\bar\psi$ and
$\psi\,$. Section \ref{sect:origin} provides a physical motivation
for the study of this model. In Section \ref{sect:syms} we explain
the symmetries of the model and briefly discuss its perturbative
renormalization group flow. The basic Ward identities we shall need
are given in Section \ref{sect:ward} and are derived in Appendix
\ref{app:Ward}. Section \ref{sect:ward} ends with a rough outline of
our proof and a description of the remaining sections of this paper.

\paragraph{Acknowledgments.} It is our pleasure to thank S.\ Varadhan
and J.\ Lebowitz for discussions and suggestions related to this
paper. Very special thanks go to D.\ Brydges for sharing his many
insights on the model and for many comments on an early version
of this paper. We wish to thank the Newton Institute (Cambridge) 
for its suppport and hospitality during the completion of this article.  

\section{Definition of the model}\label{sect:defs}\resetequ

We now fill in the details of the definition the $\mathrm{H}^{2|2}$
model and derive the free energy $F_{\beta,\varepsilon} (t)$ given
above.

\subsection{Full supersymmetric model}\label{sect:fullmodel}

As in Section \ref{sect:1.2}, let $\Lambda \subset \mathbb{Z}^{d}$ be
a cube of size $L$. For each lattice site $j \in \Lambda$ we
introduce a supervector $u_j \in \mathbb{R}^{3|2}$,
\begin{equation}
    u_j = (z_j, x_j, y_j, \xi_j, \eta_j) \;,
\end{equation}
with 3 real components $x_j, y_j, z_j$ and 2 Grassmann variable
components $\xi_j, \eta_j\,$. We then define an inner product on
$\mathbb{R}^{3|2}$ by
\begin{equation}\label{eq:inprod}
    (u,u') = - z z' + x x' + y y' + \xi \eta' - \eta \xi'
\end{equation}
and constrain $u_j$ by the quadratic equation
\begin{equation}
    \forall j \in \Lambda \;: \quad (u_j\, , u_j) = - 1 \;,
\end{equation}
which is solved by
\begin{equation}
    z_j = \pm \sqrt{1 + x_j^2 + y_j^2 + 2 \xi_j \eta_j} \;.
\end{equation}
In the following we take the positive square root for all $j \in
\Lambda\,$. This singles out a choice of connected subspace,
$\mathrm{H}^{2|2}$, parametrized by two bosonic variables $x_j , y_j$
and two fermionic variables $\xi_j, \eta_j\,$.

On the product space $(\mathrm{H}^{2|2})^{|\Lambda|}$ we introduce a
`measure' (more accurately, a Berezin superintegration form)
\begin{equation}\label{eq:dmu}
    D\mu_\Lambda = \prod_{k\in \Lambda} (2\pi)^{-1} dx_k dy_k\,
    \partial_{\xi_k} \partial_{\eta_k} \circ (1 + x_k^2 + y_k^2
    + 2\xi_k \eta_k)^{-1/2} \;.
\end{equation}
We use the notation $\partial_\xi \equiv \partial / \partial \xi$ for
the partial derivative. The statistical measure then is of the Gibbs
form $D\mu_\Lambda \,\mathrm{e}^{-A_{\beta, \varepsilon}}$ with
action
\begin{align}
    &A_{\beta,\varepsilon} = \frac{\beta}{2}\mathop\sum\limits_{i,j}
    J_{ij}\,(u_i-u_j\,,u_i-u_j) + \varepsilon \mathop\sum\limits_{k
    \in \Lambda} (z_k-1) \label{eq:fullaction} \\
    &= \beta \mathop\sum\limits_{i,j} J_{ij}
    \left(z_i z_j -(1+x_i x_j +y_i y_j +\xi_i \eta_j -\eta_i \xi_j)
    \right) + \varepsilon \mathop\sum\limits_{k \in\Lambda} (z_k-1)
    \;. \nonumber
\end{align}
Here $J_{ij} = 1$ if $i,j$ are nearest neighbors (NN) and $J_{ij} =
0$ otherwise. As will be discussed later, the action $A_{\beta,0}$ is
$\mathrm{SO}(1,2)$-invariant. The $\varepsilon$-term breaks this
noncompact symmetry and makes the integral $\int D\mu_\Lambda\,
\mathrm{e}^{-A_{\beta,\varepsilon}}$ converge.

\subsection{Horospherical coordinates}\label{sect:horo}

As with \cite{SZ}, it is very helpful to switch to horospherical
coordinates --- it is only in this coordinate system that we can
obtain the probabilistic interpretation of Section \ref{sect:1.2}. We
thus use the following parametrization of the supermanifold:
\begin{equation}\label{eq:horoc}
    x = \sinh t - \mathrm{e}^t \left(  {\textstyle{\frac{1}{2}}}
    s^2 + \bar\psi\psi \right)\ , \quad y = \mathrm{e}^t s \ ,\quad
    \xi = \mathrm{e}^t \bar\psi\;, \quad \eta = \mathrm{e} ^t \psi\;,
\end{equation}
where $t$ and $s$ range over the real numbers. Note that $(t,s\,
;\bar\psi,\psi)$ are globally defined coordinates and
\begin{displaymath}
    (t,s\,;\bar\psi,\psi) = (0,0;0,0) \Leftrightarrow
    (x,y\,;\xi,\eta) = (0,0;0,0) \;.
\end{displaymath}
The expression for the action in them is
\begin{equation}\label{eq:action}
    A_{\beta,\varepsilon} = \beta \mathop\sum\nolimits_{(ij)}
    (S_{ij}-1) + \varepsilon \mathop\sum\nolimits_{k\in\Lambda}
    (z_k - 1) \;,
\end{equation}
where $(ij)$ are NN pairs and
\begin{align}
    S_{ij} &= B_{ij} + (\bar\psi_i - \bar\psi_j)(\psi_i -
    \psi_j)\, \mathrm{e}^{t_i + t_j}\;, \label{eq:def-S}\\
    B_{ij} &= \cosh (t_i-t_j) + {\textstyle{\frac{1}{2}}}
    (s_i - s_j)^2\, \mathrm{e}^{t_i + t_j}\;, \label{eq:Bdef}\\
    z_{k} &= \cosh t_k +\left({\textstyle{\frac{1}{2}}}s_k^2 +
    \bar\psi_k \psi_k\right)\mathrm{e}^{t_k} \label{eq:def-z}\;.
\end{align}
We also need the expression for the measure $D\mu_\Lambda$ in
horospherical coordinates. By applying Berezin's transformation
formula \cite{berezin} for changing variables in a (super-)integral,
one finds that
\begin{equation}
    D\mu_{\Lambda} = \prod\nolimits_{j\in\Lambda} (2\pi)^{-1}
    \mathrm{e}^{-t_j} dt_j ds_j \, \partial_{\bar\psi_j}
    \partial_{\psi_j} \;.
\end{equation}
For any function $f$ of the lattice field variables $\{ t_j\,, s_j\,,
\bar\psi_j\,,\psi_j\}_{j\in\Lambda}$ we now define its expectation as
\begin{equation}\label{expectation}
    \left\langle f \right \rangle_{\beta,\varepsilon}  =
    \int D\mu_\Lambda \, \mathrm{e}^{-A_{\beta,\varepsilon}} f \;,
\end{equation}
whenever this integral exists.

\subsection{Effective bosonic field theory}\label{sect:2.2}

Since the action \eqref{eq:action} is quadratic in the fields
$\bar\psi$, $\psi$, and $s$, each with covariance $D_{\beta,
\varepsilon}(t)^{-1}$, we know from standard free-field calculus that
integration over $s$ yields a factor of $\mathrm{Det}^{-1/2}
(D_{\beta,\varepsilon} (t))$ while integration over $\bar\psi,\psi$
yields $\mathrm{Det}\, D_{ \beta ,\varepsilon}(t)$. By performing
these integrations, we arrive at the nonlocal free energy functional
$F_{\beta,\varepsilon}(t)$ given by \eqref{eq:1.2}. Moreover, the
basic two-point functions are
\begin{align}
    \big\langle s(v)^2 \big\rangle &= + \big\langle [v\,;
    D_{\beta,\varepsilon}(t)^{-1} v] \big\rangle \;, \cr
    \big\langle \bar\psi(v)\psi(v) \big\rangle &=
    - \big\langle [v\,; D_{\beta,\varepsilon}(t)^{-1} v]
    \big\rangle \;, \label{eq:2.Xa-TS}
\end{align}
where
\begin{displaymath}
    s(v) = \sum\nolimits_{j \in \Lambda} s_j \, v(j) \;, \quad
    \psi(v) = \sum\nolimits_{j \in \Lambda} \psi_j \, v(j) \;,
\end{displaymath}
and the expectations on the left-hand and right-hand side are defined
by \eqref{expectation} and \eqref{eq:1.5}, respectively. We will
often use the formula \eqref{eq:2.Xa-TS} as well as its
generalization
\begin{equation}\label{eq:2.15-TS}
    \big\langle \mathrm{e}^{\sum_{\lambda=1}^n \bar\psi(v_\lambda)
    \psi(v_\lambda)} \big\rangle = \big\langle \mathrm{Det}(1 -
    \mathcal{A}) \big\rangle \;,
\end{equation}
where $\mathcal{A}$ is the $n \times n$ matrix given by
\begin{equation}
    \mathcal{A}_{\lambda \lambda^\prime}(t) = [v_\lambda\,;
    D_{\beta,\varepsilon}(t)^{-1} v_{\lambda^\prime}] \;.
\end{equation}
\paragraph{Remark 2.1.}
If the Grassmann fields $\bar\psi, \psi$ were absent, then
$\mathrm{Det}^{1/2}$ in \eqref{eq:1.2} would be replaced by
$\mathrm{Det}^{-1/2}$ (and $\prod_k \mathrm{e}^{-t_k} dt_k$ by
$\prod_k \mathrm{e}^{t_k} dt_k$) and $Z_\Lambda$ would be the
partition function of the hyperbolic sigma model studied in
\cite{SZ}.
\paragraph{Remark 2.2.}
If we integrate only over the fields $\bar\psi,\psi$ (but not over
$s$) we produce a positive integrand depending on $t$ and $s$. The
square root of the determinant is then replaced by $\mathrm{Det} \,
D_{\beta,\varepsilon}(t) > 0\,$.
\paragraph{Remark 2.3.} The logarithm of $\mathrm{Det}\, D_{\beta,
\varepsilon}(t)$ is convex in $t$.
\paragraph {Proof (D.\ Brydges).} By the matrix tree theorem we have
\begin{equation}
    \mathrm{Det}\, D_{\beta,\varepsilon}(t) = \sum_\mathcal{F}
    \beta^{|\Lambda|-|R|}\, \varepsilon^{|R|} \prod_{\ell \in
    \mathcal{F} } \mathrm{e}^{t_{j_{\ell}}+t_{j'_{\ell }}}
    \prod_{k\in R} \mathrm{e}^{t_k} \;,
\end{equation}
where $\mathcal{F}$ denotes the spanning rooted forests, $R$ the set
of roots, $|R|$ the cardinality of this set, and $\ell = (j_{\ell
},j'_{\ell})$ denotes an edge in the forest. The proof is now
immediate since any positive sum of exponentials in $t$ is log
convex.

Note that the logarithm of $\mathrm{Det}\, D_{\beta,\varepsilon}(t)$
competes with the other factor, $\mathrm{e}^{-\beta \sum_{(ij)}
\cosh(t_i - t_j)}$, which is log concave.

\section{Microscopic origin of the model}\label{sect:origin}\resetequ

In this subsection we use the language and heuristic ideas of physics
to sketch the origin of our field theory model from a microscopic
model of disorder. Consider real symmetric random band matrices, $H$,
say with independent Gaussian distributed entries, of band width $W$
in $d$ dimensions. (Such a band matrix model possesses a
time-reversal symmetry and belongs to symmetry class $A$I --
traditionally referred to as the Wigner-Dyson class of orthogonal
symmetry -- of the 10-way classification of disordered fermion
systems \cite{HHZ}).

Now suppose that we wish to compute the disorder average of
\begin{equation}\label{eq:expression}
    \sqrt{\mathrm{Det}(E + \mathrm{i}\eta - H) /
    \mathrm{Det}(E + \mathrm{i}\varepsilon - H)} \times
    \big| (E + \mathrm{i}\varepsilon - H)^{-1}(x,y) \big|^2
\end{equation}
for real energy parameters $E$ and $\varepsilon, \eta > 0$. The
unconventional feature here is that the square $| (E + \mathrm{i}
\varepsilon - H)^{-1}(x,y)|^2$ of the Green's function is weighted by
the square root of a ratio of one determinant taken at energy $E +
\mathrm{i}\varepsilon$ and another one at energy $E+\mathrm{i}\eta$.
Although one might think that the presence of these extra factors
complicates the problem, quite the opposite is true; it will actually
lead to simplifications when $\eta$ is taken to be large.

First of all, the combination
\begin{displaymath}
    (E + \mathrm{i}\varepsilon - H)^{-1}(x,y) \;
    \mathrm{Det}^{-1/2}(E + \mathrm{i}\varepsilon - H)
\end{displaymath}
can be generated by Gaussian integration over a single real boson
field, $\phi_1^+\,$. Second, writing the complex conjugate
$\overline{(E + \mathrm{i} \varepsilon - H)^{-1}(x,y)}$ of the
Green's function as a Gaussian integral requires two real boson
fields $\phi_\alpha^-$ and two anticommuting fields $\psi_\alpha^-$
($\alpha = 1, 2$). Third, to express the square root of $\mathrm{Det}
(E+\mathrm{i}\eta -H)$ as a Gaussian integral, we need another real
boson $\phi_2^+$ and two more anticommuting fields $\psi_\alpha^+$.
Altogether, we then have four bosonic fields $\phi_\alpha^\sigma$ and
four fermionic fields $\psi_\alpha^\sigma$ ($\sigma = \pm$, $\alpha =
1, 2$).

Now assume for the moment that $\eta = \varepsilon$, in which case
the two determinants in \eqref{eq:expression} cancel each other. If
the band width $W$ is large enough, then the standard steps of
disorder averaging followed by Hubbard-Stratonovich transformation
and elimination of the massive modes, take us to Efetov's nonlinear
sigma model for systems with orthogonal symmetry (class $A$I).

Physically speaking, the order parameter fields of retarded ($+$) and
advanced ($-$) type acquire different expectation values:
\begin{displaymath}
    \langle \phi_\alpha^\sigma \phi_\beta^\sigma \rangle =
    \delta_{\alpha\beta} \langle G^\sigma \rangle \;, \quad
    \langle \psi_\alpha^\sigma \psi_\beta^\sigma \rangle =
    \epsilon_{\alpha\beta} \langle G^\sigma \rangle \qquad
    (\sigma = \pm \;; ~ \alpha , \beta = 1, 2) \;,
\end{displaymath}
where we are using the abbreviations $\langle \phi_\alpha^+ (x)
\phi_\beta^+ (x) \rangle = \langle \phi_\alpha^+ \phi_\beta^+
\rangle$,
\begin{displaymath}
    G^+ = (E + \mathrm{i}\varepsilon - H)^{-1}(x,x) \;, \quad
    G^- = \overline{G^+} \;,
\end{displaymath}
and $\epsilon_{\alpha\beta} = - \epsilon_{\beta\alpha}$ is the
antisymmetric tensor for two degrees of freedom. In the region of
nonzero average density of states, where $\langle G^+ \rangle \not=
\langle G^- \rangle$, these expectation values break a continuous
symmetry of the Gaussian integrand at $\varepsilon = 0$. The
components of Efetov's sigma model field have the physical meaning of
being the Goldstone modes associated with this broken symmetry. There
are $4$ bosonic Goldstone modes due to the symmetry breaking $\langle
\phi_\alpha^+ \phi_\alpha^+ \rangle \not= \langle \phi_\beta^-
\phi_\beta^- \rangle$ and four more such modes due to $\langle
\psi_1^+ \psi_2^+ \rangle = -  \langle \psi_2^+ \psi_1^+ \rangle$ not
being equal to $\langle \psi_1^- \psi_2^- \rangle = - \langle
\psi_2^- \psi_1^- \rangle$. There also exist $8$ fermionic Goldstone
modes due to the breaking of the odd symmetries connecting the
boson-boson sector $\langle \phi_1^\sigma \phi_1^\sigma \rangle =
\langle \phi_2^\sigma \phi_2^\sigma \rangle$ with the fermion-fermion
sector $\langle \psi_1^\tau \psi_2^\tau \rangle = - \langle
\psi_2^\tau \psi_1^\tau \rangle$ of opposite type $\tau = -\sigma$.
All these modes organize into a supermanifold with tangent space
$\mathbb{R}^{8|8}$ over a symmetric space $(\mathrm {H}^2 \times
\mathrm{H}^2) \times \mathrm{S}^4$.

Now let $\eta \gg \varepsilon > 0$, so that the two determinants in
the expression \eqref{eq:expression} no longer cancel. The difference
$\eta - \varepsilon \approx \eta$ then acts as a mass term for the
Goldstone modes connecting the advanced sector $(-)$ with the $\eta$
retarded sector $\langle \phi_2^+ \phi_2^+ \rangle = \langle \psi_1^+
\psi_2^+ \rangle = - \langle \psi_2^+ \psi_1^+ \rangle$. By a
Thouless-type argument, these massive Goldstone modes do not affect
the renormalized physics at length scales much greater than the
length $L^\prime$ determined by the equation
\begin{displaymath}
    \eta = 2\pi\hbar D / {L^\prime}^2 \;,
\end{displaymath}
where $D \propto W^2$ is the bare diffusion constant of the system.

Thus at large length scales $L \gg L^\prime$ we may simply drop the
massive Goldstone modes from the theory or, in a more careful
treatment, integrate them out perturbatively. What we are left with,
then, are the $2 + 2 = 4$ massless bosonic and fermionic Goldstone
modes connecting the retarded component $\langle \phi_1^+ \phi_1^+
\rangle$ of the order parameter with its four components $\langle
\phi_1^- \phi_1^- \rangle = \langle \phi_2^- \phi_2^- \rangle =
\langle \psi_1^- \psi_2^- \rangle = - \langle \psi_2^- \psi_1^-
\rangle$ in the advanced sector. These four residual Goldstone modes
organize into a supermanifold with tangent space $\mathbb{R}^{2|2}$
and base manifold $\mathrm{H}^2$ --- we thus arrive at the field
space $\mathrm{H}^{2|2}$ of the model we are going to study.

\section{Symmetries and their consequences}
\label{sect:syms}\resetequ

As an effective theory derived by reduction from an underlying sigma
model, the statistical mechanics problem posed by
\eqref{eq:1.1}--\eqref{eq:1.5} enjoys a number of symmetries. First
among these is a hidden supersymmetry which ensures that the
partition function is always equal to unity,
\begin{displaymath}
    Z_\Lambda (\beta,\varepsilon) = 1 \;,
\end{displaymath}
independent of the inverse temperature $\beta$ and regularization
parameter $\varepsilon$. Thus the reduced statistical measure
$\mathrm{e}^{-F_{\beta,\varepsilon}} d\mu_\Lambda$ can be regarded as
a probability measure, and the physical observables of the model are
given as expectations
\begin{displaymath}
    \langle f \rangle = \int f\, \mathrm{e}^{- F_{\beta,\varepsilon}}
    d\mu_\Lambda \;.
\end{displaymath}
In the following subsection we provide some background to the
normalization property $Z_\Lambda(\beta,\varepsilon) = 1$.

\subsection{$Q$-symmetry}

We start by observing that, for any $\varepsilon$, the full action
$A_{\beta, \varepsilon}$ defined in \eqref{eq:fullaction} is
invariant under transformations that preserve the short inner product
\begin{equation}\label{eq:osp}
    x_i x_j + y_i y_j + \xi_i \eta_j - \eta_i \xi_j
\end{equation}
for all $i, j \in \Lambda$. Such transformations are given, at the
infinitesimal level, by even and odd derivations (i.e., first-order
differential operators) with the property that they annihilate the
expression \eqref{eq:osp} for all $i,j$ and their coefficients are
linear functions of the coordinates $x_k, y_k, \xi_k, \eta_k\,$.
These differential operators form a representation of the
orthosymplectic Lie superalgebra $\mathfrak{osp}_{2|2}\,$. An
important example of an odd operator $Q\in \mathfrak{osp}_{2|2}$ is
\begin{equation}\label{eq:defnQ}
    Q = \sum\nolimits_{j\in\Lambda} \left(x_j \partial_{\eta_j}
    - y_j \partial_{\xi_j} + \xi_j \partial_{x_j} + \eta_j
    \partial_{y_j} \right) \;.
\end{equation}
Since $\prod_j dx_j dy_j\, \partial_{\xi_j} \partial_{\eta_j}$ is the
Berezin superintegration form given by the inner product
\eqref{eq:osp}, it is immediate that $D\mu_\Lambda$ is
$\mathfrak{osp}_{2|2}$-invariant, which implies that $\int
D\mu_\Lambda \, Q f = 0$ whenever the function $f$ is differentiable
and $Q f$ is integrable.

For present use, let us record here the explicit expression for the
$\mathfrak {osp}_{2|2}$ generator $Q$ in horospherical coordinates: a
straightforward computation starting from \eqref{eq:defnQ} gives $Q =
\sum_{j \in \Lambda} q_j$ with single-site generator (index $j$
omitted)
\begin{equation}\label{eq:horoQ}
   q = \bar\psi \partial_t + (\psi - s \bar\psi)\partial_s
    - s \partial_{\bar\psi}+ {\textstyle{\frac{1}{2}}}(1 -
    \mathrm{e}^{-2t} - s^2 - 4 \bar\psi \psi) \partial_\psi \;.
\end{equation}

Now consider any differentiable integrand $f$ which is invariant by
$Q$, i.e., $Q f = 0\,$. This invariance property has strong
consequences for the integral of $f$ (if it exists): in Appendix
\ref{app:Ward}, Proposition \ref{prop:SUSY}, we prove that the
integral of such $f$ equals $f$ evaluated on the zero-field
configuration (i.e., on $t_j = s_j = \bar\psi_j = \psi_j = 0$ or
equivalently, $x_j = y_j = \xi_j = \eta_j = 0\,$, for all $j \in
\Lambda$):
\begin{equation}\label{eq:4.6M}
    \int D\mu_\Lambda\, f = f(o) \;.
\end{equation}
The idea of the proof is easy to state: one shows that the integral
of $f$ remains unchanged by the replacement $f \to \mathrm{e}^{- \tau
h} f$ with $h = \sum_{j \in \Lambda} (x_j^2 + y_j^2 + 2\xi_j \eta_j)$
and $\tau \geq 0\,$, and then deduces the result \eqref{eq:4.6M} by
sending the deformation parameter $\tau \to +\infty$ to localize the
integral at the zero-field configuration.

Using the explicit expression \eqref{eq:horoQ} it is easy to check
that the action $A_{\beta,\varepsilon}$ is $Q$-invariant. Since the
differential operator $Q$ is of first order, one directly infers the
relation $Q\, \mathrm{e}^{ - A_{\beta,\varepsilon}} = 0$. Therefore,
as a particular consequence of \eqref{eq:4.6M} and $A_{\beta,
\varepsilon} (o) = 0$ it follows that the partition function equals
unity,
\begin{equation}\label{eq:partfunc}
    Z_\Lambda (\beta,\varepsilon) = \int D\mu_\Lambda \,
    \mathrm{e}^{- A_{\beta,\varepsilon}} =
    \mathrm{e}^{- A_{\beta,\varepsilon}(o)} = 1 \;,
\end{equation}
for all values of $\beta \geq 0$ and $\varepsilon > 0$.

Further consequences of \eqref{eq:4.6M} will be elaborated below.

\subsection{Hyperbolic symmetry}

While $Q$ is a symmetry of our action $A_{\beta,\varepsilon}$ for
\emph{all} values of $\varepsilon$, further symmetries emerge in the
limit of vanishing regularization $\varepsilon \to 0+$. Relegating a
more detailed discussion to Appendix \ref{app:symmetry}, we here
gather the crucial facts.

The model \eqref{eq:1.1}--\eqref{eq:1.5} for $\varepsilon \to 0+$
acquires a global symmetry by the Lorentz group $\mathrm {SO}(1,2)$
-- the isometry group of the hyperbolic plane $\mathrm{H}^2$ viewed
as a noncompact symmetric space $\mathrm{H}^2 \simeq \mathrm{SO}(1,2)
/ \mathrm{SO}(2)$. This global symmetry entails a number of conserved
currents and associated Ward identities. Of these let us mention here
the most important one,
\begin{equation}\label{eq:sumrule}
    \sum_{y \in \Lambda} \left\langle \mathrm{e}^{t_x +
    t_y} D_{\beta,\,\varepsilon}(t)^{-1}(x,y) \right\rangle
    = \frac{1}{\varepsilon} \;,
\end{equation}
which is the sigma model version of the quantum sum rule
\begin{displaymath}
    \sum_{y \in \Lambda} \big\langle \left| (E-\mathrm{i}
    \varepsilon - H)^{-1}(x,y) \right|^2 \big\rangle =
    \frac{1}{\varepsilon}\, \big\langle\mathrm{Im}\, (E -
    \mathrm{i} \varepsilon - H)^{-1}(x,x) \big\rangle =
    \frac{\pi}{\varepsilon} \rho(E) \;,
\end{displaymath}
where $\rho(E)$ is the mean local density of states. In the sigma
model approximation one sets $\pi \rho(E) = 1$. The above relation
reflects the unitarity of the quantum theory. Its classical
interpretation is conservation of probability.

Notice that the right-hand side of \eqref{eq:sumrule} diverges in the
limit of vanishing regularization $\varepsilon \to 0\,$. For an
infinite lattice $\Lambda$ there exist two principal scenarios
\cite{mckanestone} by which to realize this divergence. In the first
one, the correlation function $C_{xy} = \langle \mathrm{e}^{t_x +
t_y} D^{-1}_{\beta,\varepsilon}(t)(x,y) \rangle$, while bounded in
$\varepsilon$, becomes of long range and thus ceases to be summable
in the limit $\varepsilon \to 0$. In this case the $\mathrm{SO}(1,2)$
symmetry is spontaneously broken and the system is in a phase of
extended states. On the other hand, $C_{xy}$ may already diverge for
any \emph{fixed} pair of lattice sites $x,y$, signaling strong field
fluctuations and restoration of the noncompact symmetry $\mathrm{SO}
(1,2)$ as $\varepsilon \to 0\,$. Exponential decay of $C_{xy}$ with
distance $|x-y|$ then corresponds to exponential localization of the
energy eigenstates. Thus the question of extended versus localized
states of the disordered quantum system translates to the question of
the Lorentzian symmetry $\mathrm{SO}(1,2)$ of the statistical
mechanical model with free energy (\ref{eq:1.1}) being spontaneously
broken or not.

At this stage, a remark is called for. Niedermaier and Seiler have
recently shown \cite{ns1,ns2} for a large class of sigma models that
if the symmetry group of the sigma model is non-amenable -- this
includes in particular the case of the Lorentz group $\mathrm{SO}
(1,2)$ -- then spontaneous symmetry breaking occurs in \emph{all}
dimensions $d \ge 1$ and for all $\beta > 0\,$. It must therefore be
emphasized that, although our sigma model does acquire the
non-amenable symmetry $\mathrm{SO}(1,2)$ in the limit $\varepsilon
\to 0$, it does \emph{not} belong to the class of models where the
arguments of \cite{ns1,ns2} apply. (The culprit is the nonlocal part
of the free energy due to integration over the Grassmann fields.) In
fact, the non-amenable symmetry $\mathrm{SO}(1,2)$ of our model is
known to be unbroken in $d = 1$. This follows from the work of
\cite{toymodel} where the conductance of the one-dimensional system
was shown to exhibit exponential decay with increasing length of the
system.

\subsection{Perturbative renormalization group}\label{sect:RG}

We now sketch a perturbative result from Wilsonian renormalization
theory by which our model is expected to be in a symmetry-unbroken
phase also for $d = 2$ and all values of the inverse temperature
$\beta$, and thus to exhibit Anderson localization of all electronic
states.

This result follows from Friedan's work \cite{friedan} on
renormalization for the general class of nonlinear sigma models.
According to it, the RG flow of the temperature $T = \beta^{-1}$ with
increasing renormalization scale $a$ is given by
\begin{equation}\label{eq:RG}
    a \frac{dT}{da} = (2 - d)T + R\, T^2 + \mathcal{O}(T^3) \;,
\end{equation}
where $R$ is the target space curvature -- more precisely, the
multiplicative constant $R$ by which the Ricci tensor of the target
space differs from its metric tensor. For both the $\mathrm{H}^{2|2}$
model and Efetov's sigma model of class $A$I a quick computation
shows the curvature $R$ to be positive. In contrast, $R = 0$ and
$R<0$ for Efetov's sigma models of class $A$ (broken time-reversal
symmetry) and class $A$II (spin-orbit scattering), respectively.

According to \eqref{eq:RG}, a positive value of $R$ implies that a
small initial value of the temperature $T$ increases under
renormalization in dimension $d = 2$. By extrapolation, one therefore
expects the existence of a mass gap (or, equivalently, localization
of all states) in this case. For the localization length $\xi =
\xi(a,T(a))$, which is a physical observable and hence a
renormalization group invariant, one obtains the formula
\begin{displaymath}
    \xi \propto a\, \mathrm{e}^{1/(R\,T)} \quad (d = 2)
\end{displaymath}
by direct integration of the RG equation \eqref{eq:RG}.

In dimension $d = 3$, equation \eqref{eq:RG} predicts the localizing
tendency of positive target space curvature to become irrelevant at
small enough temperatures and hence the RG flow to be attracted to
the fixed point $T = 0$ corresponding to extended states. As was
remarked above, the Lorentzian symmetry $\mathrm{SO}(1,2)$ is
spontaneously broken at this fixed point.

With increasing temperature $T$ (or decreasing field stiffness
$\beta$) the model in $d = 3$ is expected to undergo an Anderson-type
transition to the phase of unbroken symmetry. This phase transition
was studied numerically in \cite{td}, where the critical value of
$\beta$ was found to be $\beta_c \approx 0.04$. The transition has
also been investigated in detail using the Migdal-Kadanoff
renormalization scheme \cite{migdal-kad}.

\section{Ward identities and outline of proof}\label{sect:ward}
\resetequ

In order to control fluctuations of the field $t$ at low temperatures
$T = \beta^{-1}$ we rely on a family of Ward identities due to the
internal supersymmetries of the model. These Ward identities are
naturally expressed in terms of both the real variables $t_j\,$,
$s_j$ and the Grassmann variables $\bar\psi_j\,$, $\psi_j$. In order
to obtain probabilistic information we integrate out the Grassmann
variables using \eqref{eq:2.Xa-TS} and \eqref{eq:2.15-TS}, thereby
producing a Green's function.

As was already mentioned, our partition function always equals unity
even when the temperature varies in space. By using this fact, we
show that gradients of the field $t$ between neighboring sites are
strongly suppressed for small $T$. There also exist Ward identities
at larger scales, and information may be extracted from them by using
information on previous length scales.

In addition to Ward identities, there are two other crucial
ingredients of our proof. The first one is a basic estimate on
Green's functions which are non-uniformly elliptic. The second one is
the use of SUSY characteristic functions, which help to control
large-scale field fluctuations.

A more detailed outline of our proof is given below, where the
notation and the needed Ward identities are explained. Once the Ward
identities are established, most of our proof is very classical.

\subsection{Ward identities due to $Q$-symmetry}\label{sect:2.4}
\resetequ

We recall the formula \eqref{eq:4.6M} for the integral of a
$Q$-invariant function. It is easy to check that the functions
$S_{ij}$ and $z_k$ given in \eqref{eq:def-S} and \eqref{eq:def-z}
satisfy the invariance conditions $Q S_{ij} = 0$ and $Q z_k = 0\,$.
Therefore, using $S_{ij} (o) = 1$ and $z_k(o) = 1$ we have the
identity
\begin{equation}\label{eq:var-par}
    \int D\mu_\Lambda\; \mathrm{e}^{-\beta \sum_{x,y} J_{xy}
    (S_{xy}-1) - \sum_{x \in \Lambda} \varepsilon_ x (z_x-1)} = 1
\end{equation}
for all values of $\beta\geq 0$, $J_{xy}\geq  0$, and $\varepsilon_x
> 0$. Note that in order for this statement to be true, $J_{xy}$ does
not have to be nearest neighbor. Moreover, for $m\in\mathbb{R}$ and
any pair $x, y \in \Lambda$ we have
\begin{equation}\label{eq:WI0}
    1 = \langle S_{xy}^m \rangle_{\beta,\varepsilon} = \left\langle
    B_{xy}^m + m B_{xy}^{m-1} (\bar\psi_x - \bar\psi_y) (\psi_x -
    \psi_y)\, \mathrm{e}^{t_x + t_y}\right\rangle_{\beta,\varepsilon}\;,
\end{equation}
where the expectation $\langle\cdot\rangle_{\beta,\varepsilon}$ was
defined in \eqref{expectation}, and we used the nilpotency $(\psi_x -
\psi_y)^2 = (\bar\psi_x - \bar\psi_y)^2 = 0$. By integrating over the
Grassmann fields $\bar\psi$ and $\psi$ as in \eqref{eq:2.Xa-TS} we
obtain our basic identity,
\begin{equation}\label{eq:WI1}
    1 = \langle S_{xy}^{m} \rangle_{\beta,\varepsilon} = \langle
    B_{xy}^{m}\, \big(1 - m G_{xy} \big)\rangle \;.
\end{equation}
The last expectation is taken with respect to the effective
action for the fields $t$ and $s$, and the Green's function $G_{xy}$
is
\begin{equation}\label{eq:Gf}
    G_{xy} = \frac{\mathrm{e}^{t_x+t_y}}{B_{xy}} \left[
    (\delta_x - \delta_y); \, D_{\beta,\varepsilon}(t)^{-1}
    (\delta_x - \delta_y) \right]_\Lambda \;.
\end{equation}

More generally if $(x_i , y_i)$ are $n$ pairs of points, then
\begin{equation}\label{eq:WI2}
    1 = \left\langle \prod\nolimits_{i=1}^{n} S_{x_i y_i}^m
    \right\rangle_{\beta,\varepsilon} = \left\langle
    \prod\nolimits_{i=1}^n B_{x_i y_i}^m \mathrm{Det} \big(1
    - m \mathcal{G}\big) \right\rangle\;,
\end{equation}
where $\mathcal{G}$ is an $n \times n$ matrix of Green's functions
\begin{equation}
    \mathcal{G}_{ij} = [g_i\, ; D_{\beta,\varepsilon}(t)^{-1}
    g_j]_\Lambda
\end{equation}
and
\begin{equation}
    g_i = B_{x_i y_i}^{-1/2}\, \mathrm{e}^{(t_{x_i}
    + t_{y_i})/2} (\delta_{x_i} - \delta_{y_i}) \;.
\end{equation}
The matrix $\mathcal{G}$ is real symmetric and positive. It will be
important later that we are choosing $g_i$ to be orthogonal to the
zero mode (i.e., the constant functions).

\subsection{Outline of proof}\label{sect:outline}

Our proof of Theorem \ref{thm:1} relies on the Ward identity
\eqref{eq:WI1}, \eqref{eq:WI2} and an induction on length scales. The
basic idea is quite simple: suppose $m>0$ and we had a uniform bound
$|G_{xy}| \le C/\beta < 1/m$ on the Green's function \eqref{eq:Gf},
for all configurations of $t$. Then we could conclude from
\eqref{eq:WI1} that
\begin{equation}\label{eq:cosh}
    \langle \cosh^m(t_x - t_y) \rangle \leq \langle B_{xy}^{m}
    \rangle \leq (1 -  m C/\beta)^{-1} \;,
\end{equation}
and this would imply Theorem \ref{thm:1}.

In Section \ref{sect:4} we prove that if $|x-y| = 1$ then indeed $0
\le G_{xy} \le 1/\beta\,$, and we establish an even stronger version
of \eqref{eq:cosh}. This proves that nearest neighbor fluctuations of
the field $t$ are very unlikely for large $\beta$ (see Lemma
\ref{lem:4}).

\begin{figure}
\centerline{\psfig{figure=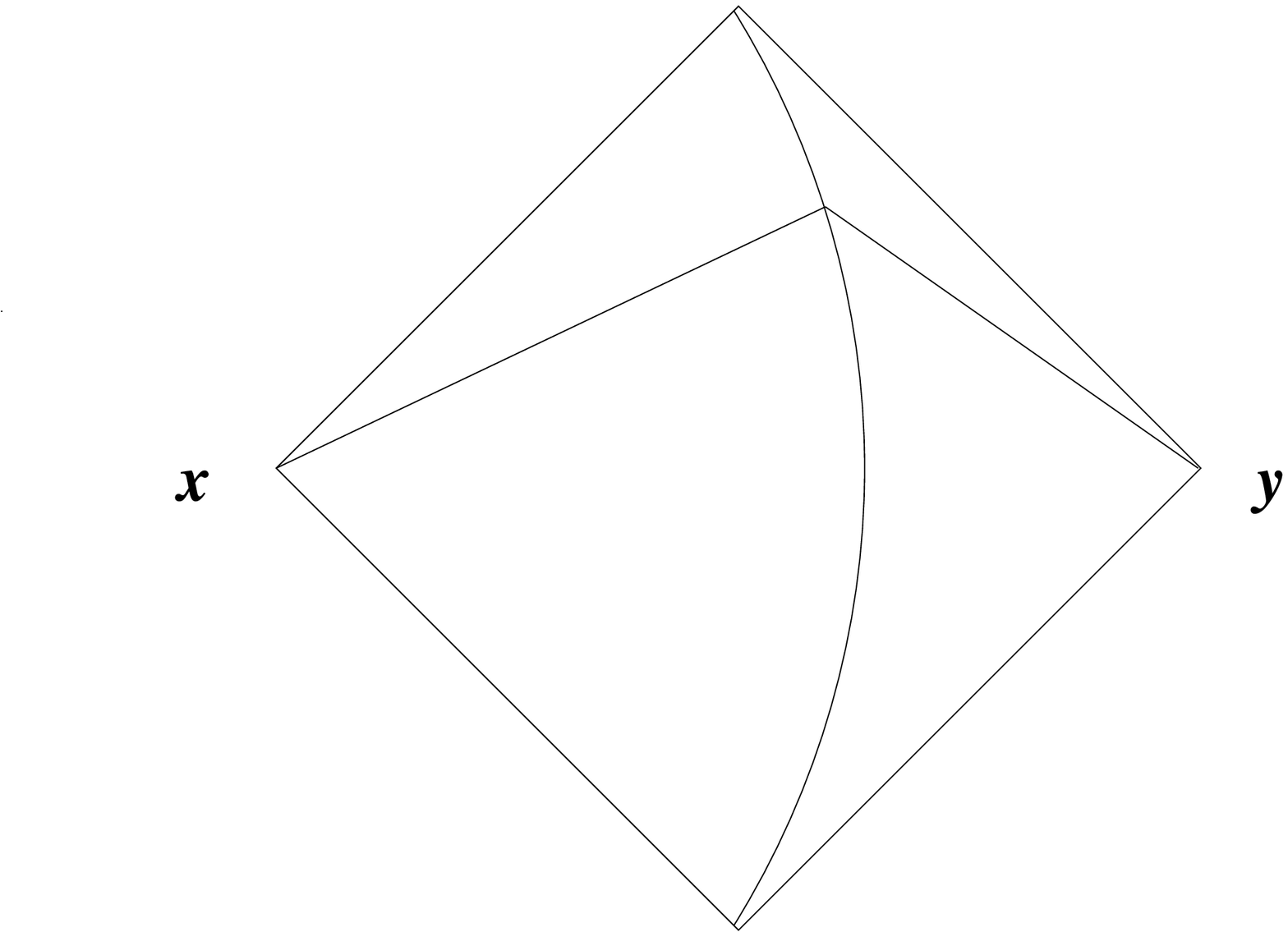 ,width=6cm}}
\caption{ ''diamond'' region: a double cone in 3 dimensions}
\label{fig:case0}
\end{figure}

For distances $|x - y| > 1$, however, there is no uniform bound on
$G_{xy}\,$. In Section \ref{sect:5} we study the Green's function
\eqref{eq:Gf} and establish sufficient conditions on the field $t$ to
obtain the desired bound on $G_{xy}\,$. In 3D these conditions are
roughly given as follows (where $|j-x| \geq 1$):
\begin{equation}\label{eq:5.9-mrz}
    \cosh (t_j-t_x) \leq B_{j x} \leq a\,
    |j-x|^\alpha\;,\quad 0 < \alpha < 1/2\;,
\end{equation}
and the same for $\cosh(t_j - t_y)$. The number $a$ is a constant, say
$a > 10$. It will turn out that these estimates are needed only for
the sites $j$ in a 3D diamond-type region, $R_{xy}\,$, containing $x$
and $y\,$; see Fig.\ \ref{fig:case0}. Notice that since the exponent
$\alpha$ is positive, we are allowing larger fluctuations at larger
scales. The probability that such a condition is violated will be
shown to be small by induction.

Section \ref{sect:6} uses the conditions described above to prove
conditional estimates on the fluctuations of the field $t$ at all
scales. These conditions are initially expressed in terms of
$Q$-invariant characteristic functions $\chi$. Later we show that the
nilpotent (or Grassmann) part of $\chi$ is not important, so we may
think of $\chi$ in the usual classical sense.

The remaining problem is to obtain unconditional estimates on the
fluctuations and thereby prove Theorem \ref{thm:1}. This is first
done for short scales in Section \ref{sect:7}. For larger scales we
use induction. Our induction hypothesis is formulated in Section
\ref{sect:8}. Roughly speaking it asserts that
\begin{equation}
    \left\langle  \prod\nolimits_{i=1}^n  B_{x_i y_i}^m
    \right\rangle \le 2^n
\end{equation}
holds under the assumption that the diamond-type regions $R_{x_i
y_i}$  (Fig.\ \ref{fig:case0}) associated with $i = 1, \ldots, n$
have disjoint interiors. The induction is in $\ell$, defined as the
maximal separation $|x_i - y_i|$ in the product over $i = 1, \ldots,
n$. For $\ell = 1$ this hypothesis was verified in Section
\ref{sect:4}.

Section \ref{sect:9} contains the technical core of our paper. There
we prove unconditional estimates on the fluctuations and thus obtain
Theorem \ref{thm:1}. The main idea is to consider a site $b$ in
$R_{xy}$ closest to $x$ or $y$ such that condition \eqref{eq:5.9-mrz}
is violated for $j = b$. We shall then prove by induction that the
probability for such an event to occur is small. The inequality
$B_{xy}^m < 2^m B_{x c}^m B_{c y}^m$ (see Lemma \ref{lem:Bf} below)
is used for a point $c$ near $b$. Since the distances $|x-c|$ and
$|c-y|$ are less than $|x-y|$, induction can be applied. The factor
$2^m$ is offset by the small probability of the event when $\beta$ is
large.

Theorem \ref{thm:2} is proved in Section \ref{sect:10}. Here we must
estimate the contribution of the zero mode and at this stage
$\varepsilon > 0$ plays a key role. Finally, Theorem \ref{thm:new}
follows from the estimates of Theorem \ref{thm:2}; its proof is given
in Section \ref{sec:new}.

\subsection{Two simple lemmas}\label{sect:3.3}

We conclude this section with two simple lemmas which will be
frequently used below. The first lemma is useful for estimates on
Green's functions. To state it, let $V$ be a finite-dimensional
Euclidean vector space with scalar product $[\, ; \,]_V$.
\begin{lemma}\label{lem:K}
Let $M : \, V \to V$ be a positive real symmetric operator. Then for
any set of $n$ vectors $v_i \in V$ we have
\begin{equation}\label{eq:Mv}
    M - \sum_{i=1}^n v_i\, [ v_i \,; \,\cdot\, ]_V \ge 0
\end{equation}
if and only if the symmetric $n \times n$ matrix $K$ with matrix
elements
\begin{equation}
    K_{ij} = [ v_i \,; M^{-1} v_j ]_V
\end{equation}
satisfies $0 \leq K \leq \mathrm{Id}$.
\end{lemma}
\paragraph{Proof.}
Letting $w_i = M^{-1/2} v_i \in V$ we observe that $K_{ij} = [ w_i
\,; w_j ]_V$ and \eqref{eq:Mv} is equivalent to
\begin{displaymath}
    \mathrm{Id} - \sum_{i=1}^n w_i \, [w_i\,; \,\cdot \,]_V \geq 0 \;.
\end{displaymath}
By evaluating this quadratic form at $w = \sum \lambda_i w_i$ for any
real numbers $\lambda_i$ we see that \eqref{eq:Mv} is equivalent to
\begin{equation}
    \sum_{i,j=1}^n \lambda_i \lambda_j\, (K - K^2)_{ij} \ge 0\;,
\end{equation}
or $0 \leq K^2 \leq K$, from which our assertion follows. \qed

The second lemma will be used in our induction process of Section
\ref{sect:9}.
\begin{lemma}\label{lem:Bf}
If $B_{xy\,}$, $B_{cy}\,$, $B_{xc}$ are defined as in \eqref{eq:Bdef}
then
\begin{equation}\label{Bbound}
    B_{xy} < 2 B_{xc} B_{cy} \;.
\end{equation}
\end{lemma}
The inequality \eqref{Bbound} can be verified by direct computation
(proof omitted).
\paragraph{Remark.} The {\it raison d'etre} behind \eqref{Bbound} is
easy to state: $B_{xy}$ has an interpretation as the hyperbolic
cosine of the geodesic distance on $\mathrm{H}^2$. Therefore, if
$x,y,c$ are three points on $\mathrm{H}^2$, then since the geodesic
distance $\mathrm{dist}(x,y)$ is the minimal length of any curve
connecting $x$ and $y$, the triple of geodesic distances satisfy the
triangle inequality $\mathrm{dist}(x,y)\le \mathrm{dist}(x,c) +
\mathrm{dist}(c,y)$. Given this, the inequality \eqref{Bbound}
follows by taking the hyperbolic cosine of both sides and using that
$\cosh(a + b) < 2\cosh(a) \cosh(b)$ holds for any two real numbers $a
, b$.

\section{Bounds on NN fluctuations}\label{sect:4}\resetequ

As was already mentioned, for nearest neighbor (NN) pairs we can
obtain a result stronger than \eqref{eq:cosh}. Recall that we have
now fixed $J_{xy} = 1$ for all $xy$ that are NN pairs, and $J_{xy} =
0$ otherwise. This fact is essential in the next lemma.
\begin{lemma}\label{lem:nnbound}
Let $x,y$ be an NN pair and suppose that $0 < \gamma < 1$. Then
\begin{equation}\label{eq:bound1}
    \left\langle \mathrm{e}^{\beta\gamma\,
    (B_{xy}-1)} \right\rangle \leq (1 - \gamma)^{-1} \;.
\end{equation}
More generally, if $(x_j , y_j)$, $j = 1,\dotsc, n$ is a set of $n$
different NN pairs, then
\begin{equation}\label{eq:bound2}
    \left\langle \mathrm{e}^{\beta\gamma \sum_{j=1}^n
    (B_{x_j y_j} - 1)} \right\rangle \leq (1-\gamma )^{-n} \;.
\end{equation}
\end{lemma}
This shows that NN fluctuations are strongly suppressed.
\paragraph{Remark.} Since $J_{x_{j}y_{j}} = 1$ and $\gamma < 1$
the integrals in \eqref{eq:bound1}--\eqref{eq:bound2} are well
defined. This would not be true if $x_{j},y_{j}$ were not NN, or if
$\gamma > 1$, or if two or more NN pairs were allowed to be identical
without further restrictions on the value of $\gamma$.
\paragraph{Proof.}
For $x,y$ an NN pair let
\begin{equation}\label{eq:6.3-TS}
    F_{xy}(j) = \mathrm{e}^{(t_x + t_y)/2} (\delta_x(j) -
    \delta_y(j)) \;,
\end{equation}
and introduce the Green's function
\begin{equation}
    G_{xy}^0 (t) = [ F_{xy}\,; D^{-1}_{\beta,\varepsilon}(t)\,
    F_{xy} ] = B_{xy} G_{xy} \;.
\end{equation}
Since $S_{xy}$ is $Q$-invariant, Proposition \ref{prop:SUSY} of
Appendix \ref{app:Ward} implies
\begin{align}
    \mathrm{e}^{\beta\gamma} &= \big\langle \mathrm{e}^{
    \beta\gamma S_{xy}} \big\rangle = \big\langle \mathrm{e}^{
    \beta\gamma (B_{xy} + \bar\psi(F_{xy}) \psi(F_{xy}) )}
    \big\rangle \cr &= \big\langle \mathrm{e}^{\beta\gamma B_{xy}}
    (1 + \beta\gamma\,\bar\psi(F_{xy}) \psi(F_{xy})) \big\rangle
    \cr &= \big\langle \mathrm{e}^{\beta\gamma B_{xy}}
    (1 - \beta\gamma\, G^0_{xy}) \big\rangle \;, \label{eq:6.5-TS}
\end{align}
where we have used $\psi^2 = \bar\psi^2 = 0$ and \eqref{eq:2.Xa-TS}.
Now from \eqref{eq:1.1} we have
\begin{equation}\label{eq:star-TS}
    [v\,; D_{\beta,\,\varepsilon} (t)\, v] = \beta \sum\nolimits_{(ij)}
    [v\,; F_{ij}]^2 + \varepsilon \sum\nolimits_k \mathrm{e}^{t_k}
    v_k^2 \geq \beta\, [ v\, ; F_{xy}]^2 \;.
\end{equation}
Therefore Lemma \ref{lem:K} implies that $0 \leq \beta\,G_{xy}^0 (t)
\leq 1$ for all $t$, and \eqref{eq:bound1} follows.

Similarly, for $n > 1$ we have
\begin{align}\label{eq:4.7}
    \mathrm{e}^{n \beta\gamma} = \left\langle \mathrm{e}^{
    \beta\gamma \sum_{j=1}^n S_{x_j y_j}} \right\rangle &=
    \left\langle \mathrm{e}^{\beta\gamma \sum_{j=1}^n B_{x_j y_j}}
    \mathrm{e}^{\beta\gamma \sum_{j=1}^n \bar\psi(F_{x_j y_j})
    \psi(F_{x_j y_j})} \right\rangle \cr &= \left\langle
    \mathrm{e}^{\beta\gamma \sum_{j=1}^n B_{x_j y_j}} \mathrm{Det}
    (1 - \gamma K) \right\rangle ,
\end{align}
where $K$ is the $n \times n$ matrix
\begin{equation}
    K_{ij} = \beta\, [ F_{x_j y_j}\,;
    D_{\beta,\varepsilon}(t)^{-1} \, F_{x_i y_i} ]
\end{equation}
given by $n$ different NN pairs $x_i , y_i\,$. From
\eqref{eq:star-TS} and Lemma \ref{lem:K} it follows that $\Vert K
\Vert \leq 1$. This implies $| \mathrm{Det}(1 - \gamma K) | \geq (1 -
\gamma)^n$ and the lemma follows. \qed

As a corollary, since $1 \le B_{xy}^m \le \mathrm{e}^{m(B_{xy}-1)}$
for $m \ge 0$, we have the bound
\begin{equation}\label{eq:4.12}
    \left\langle \prod\nolimits_{j=1}^n
    B_{x_j y_j}^m \right\rangle \leq (1 - m/\beta)^{-n} \leq 2^n
\end{equation}
for any $m$ in the range $m \leq \beta/2\,$.

A first important consequence of Lemma \ref{lem:nnbound} is the
following statement.
\begin{lemma}\label{lem:4}
Let $x_j,y_j$ be a set of $n$ different nearest neighbor pairs. Then
\begin{equation}
    \mathrm{Prob}\left(\forall j = 1,\ldots, n:\; B_{x_j y_j} > 1 +
    \delta\right)\le (1-\gamma )^{-n}\mathrm{e}^{-n (\beta\gamma)\delta}
\end{equation}
for any $0 < \gamma < 1$.
\end{lemma}
\paragraph{Proof.}
Let $n = 1$. By the Chebyshev inequality
\footnote{Actually, the Chebyshev inequality states that for any
random variable $X$ with average $X_{0}$ we have $ \mathrm{Prob}
[(X-X_{0})^{2}> a^{2}] \leq a^{-2}\left\langle (X-X_{0})^{2}
\right\rangle$. Here we are using the same principle.}
,
\begin{equation}
    \mathrm{Prob}\left(B_{xy} > 1 + \delta \right) =
    \left\langle \chi (B_{xy} > 1 + \delta )\right\rangle
    \le \mathrm{e}^{-\beta\gamma(1+\delta)} \left\langle
    \mathrm{e}^{\beta\gamma\,B_{xy}}\right\rangle \;,
\end{equation}
where $\chi (B_{xy} > 1 + \delta)$ is the characteristic function for
$B_{xy} > 1 + \delta$ to hold. The desired inequality for $n = 1$ now
follows directly from Lemma \ref{lem:nnbound}.

The proof for $n$ pairs is no different. \qed

\section{Conditional estimates on Green's functions}
\label{sect:5} \resetequ

For general $x,y$ (not NN) we do not have the option of considering
$\langle \mathrm{e}^{\beta\gamma\,S_{xy}}\rangle$, as the underlying
integral need not exist. Nevertheless, $\langle S_{xy}^{m} \rangle$
does exist and from \eqref{eq:WI1} we have
\begin{equation}
    1 = \left\langle B_{xy}^{m}\,\big(1-m G_{xy}\big)\right\rangle\;,
\end{equation}
with $G_{xy}$ defined by \eqref{eq:Gf},
\begin{displaymath}
    G_{xy} = \frac{\mathrm{e}^{t_x+t_y}}{B_{xy}} \left[
    (\delta_x - \delta_y)\, ; D_{\beta,\varepsilon}(t)^{-1}
    (\delta_x - \delta_y) \right] \;.
\end{displaymath}
Now, as was explained in Section \ref{sect:outline}, \emph{if we
knew that} $G_{xy} \leq C / \beta$ for all configurations of $t$,
then we could conclude that
\begin{equation}
    \langle B_{xy}^{m} \rangle \leq  (1 - m C/\beta)^{-1}.
\end{equation}
While we have seen that this estimate is true for $|x-y| = 1$ (with
$C = 1$), it is \emph{false} in general, as there are rare
configurations with large negative $t$ surrounding $x$ or $y$.
Nonetheless, in 3D we can get an upper bound on $G_{xy}$ by
estimating the local `conductance' at an edge $(ij)$ from below. This
conductance is
\begin{align}
    A_{xy} (ij) &\equiv B_{xy}\, \mathrm{e}^{- t_x - t_y}
    \mathrm{e}^{t_i + t_j} \cr &\geq  {\textstyle{\frac{1}{2}}}
    \, \mathrm{max} \left( \mathrm{e}^{t_i + t_j - 2t_x} ,
    \mathrm{e}^{t_i + t_j - 2t_y} \right) , \label{localC}
\end{align}
where we have used $B_{xy} \geq \cosh (t_x - t_y)$. It will suffice
to estimate the expression \eqref{localC} for NN pairs $(ij)$ in a
region $R_{xy}$ which is like a 3D double cone with vertices at $x$
and $y$. Note that Neumann boundary conditions increase $G_{xy}$ and
$\delta_x - \delta_y$ is orthogonal to the zero mode. We will have to
require that $R_{xy}$ be essentially three-dimensional in the
following sense:
\begin{defin}\label{def1}
A region $R_{xy} \subset \Lambda$ containing $x$ and $y$ is called
$\delta$-admissible if it is connected by nearest-neighbor bonds and
the two one-parameter families of intersections $R_{z}(r) \equiv
R_{xy} \cap B_z^r$ with the ball $B_z^r$ of radius $r$ centered at $z
= x , y$ satisfy
\begin{displaymath}
    \mathrm{vol}\{ R_{z} (r) \} \geq r^3 \delta \quad
    \text{for} \quad r\leq |x-y|/\sqrt{2} \qquad (z = x, y) \;.
\end{displaymath}
In addition we require that the following Poincar\'e inequality:
\begin{displaymath}
    \sum_{j \in R_{z}(r)} f(j)^2 \leq \mathrm{const}\;
    r^2 \sum_{j \in R_{z}(r)} (\nabla f)^2 (j)\;,
\end{displaymath}
holds for all functions $f : \Lambda \to \mathbb{R}$ subject to the
condition $\sum_{j \in R_{z}(r)} f(j) = 0\,$.
\end{defin}
We observe that by the choice of maximal radius $r = |x-y|/\sqrt{2}$
the scaling of volume is monitored up to the full side length of a
rectangular diamond $R_{xy}$ (or a double cone $R_{xy}\,$, see Fig.\
\ref{fig:case0}) with opposite corners placed at $x$ and $y$.

\begin{figure}
\centerline{\psfig{figure=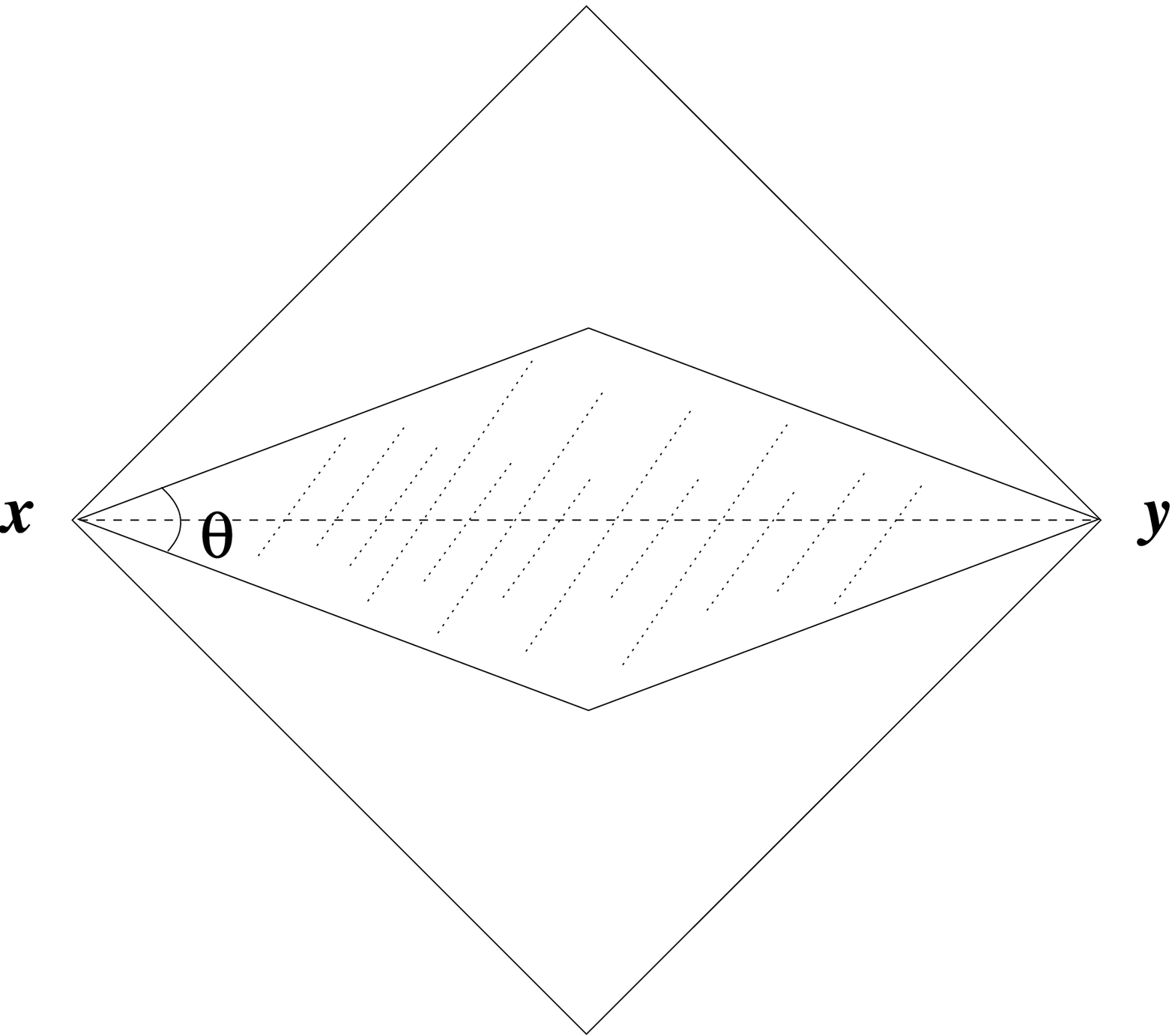,width=5cm}
\hspace{2cm} \psfig{figure=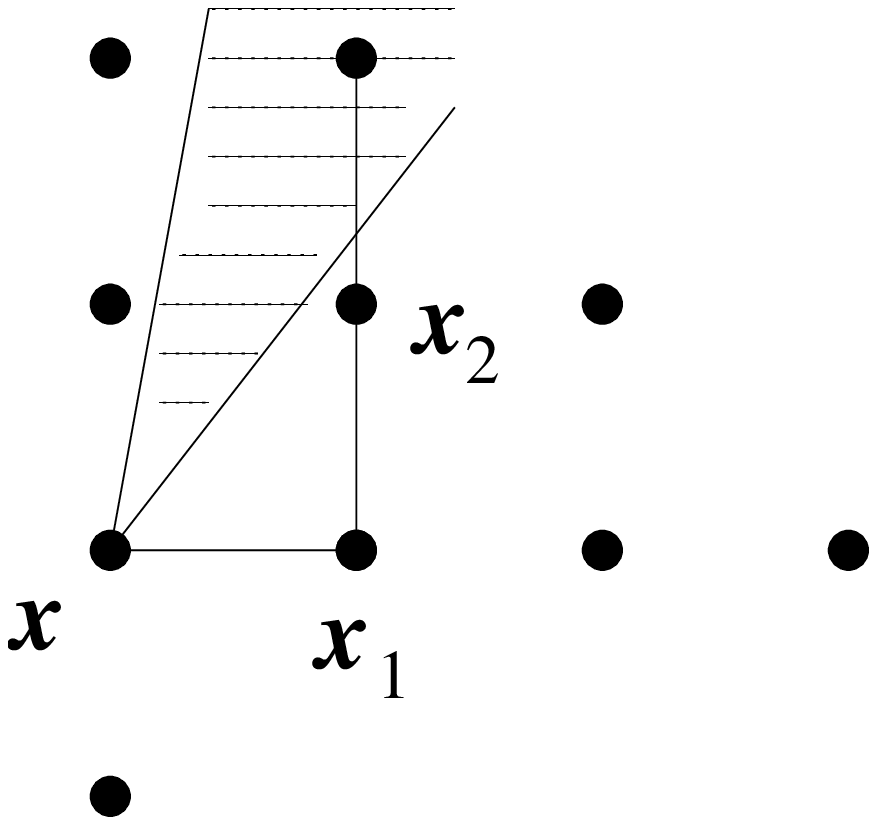,width=4cm}}
\caption{a) The region $R_{xy}$ in the continuum limit. b) On the
lattice, the points $x_1$ and $x_2$ must be added to ensure
connectedness.}
\label{fig:cutdiamond}
\end{figure}

In the continuum limit this definition is satisfied by a double cone
obtained by rotating (around the line $\overline{xy}$ connecting $x$
and $y$) a 2D diamond with vertices on $x$ and $y$ and angle $\theta
\geq \theta_{0} (\delta)\geq \pi/10 $ (see Fig.\ \ref{fig:cutdiamond}a).
Since we are on a lattice we may have to add a few lattice points near $x$
and $y$ to ensure connectedness (see Fig.\ \ref{fig:cutdiamond}b).
The Poincar\'e inequality is straightforward to prove in such
convex regions.
\begin{defin}\label{def:2}
Given a $\delta$-admissible region $R_{xy}\,$, we define the regions
$R_{xy}^z$ for $z = x$ and $z = y$ by
\begin{equation}\label{Rdef}
    R^z_{xy} = \{j \in R_{xy} \mid 1\leq |j-z| \leq |x-y|/\sqrt{2}\}\;.
\end{equation}
\end{defin}
For the case of a diamond,
$R_{xy}^x \cup R_{xy}^y = R_{xy} \setminus \{ x,y \}\,$.
\paragraph{Remark 7.1.}
The values of the field $t$ outside the region $R_{xy}$ are not
important, as we can use Neumann boundary conditions to eliminate the
exterior of $R_{xy}\,$. Indeed, in the subspace orthogonal to the
constant functions the Laplacian on $R_{xy}$ with Neumann boundary
conditions is bounded (by the Poincar\'e inequality) from below by
some number, say $c\,$, times the inverse square of the linear size
$L$ of $R_{xy}\,$. By this token, since the vector $\delta_{x} -
\delta_{y}$ used in the definition of $G_{xy}$ lies in that subspace,
we may utilize the bound on the inverse of the Neumann Laplacian by
$c^{-1} L^2$ and in this way eventually obtain an upper bound on
$G_{xy}$ (see Lemma \ref{lem:5}).

\paragraph{Remark 7.2.}
From \eqref{localC} we have
\begin{equation}\label{eq:7.5-TS}
    A_{xy}(ij)^{-1} \leq 8 \cosh(t_i - t_z) \cosh(t_j - t_z)
\end{equation}
for both $z = x$ and $z = y$. The main result of this section is that
we can get an upper bound on $G_{xy}$ even without imposing a uniform
upper bound on $A_{xy} (ij)^{-1}$, as long as some growth restriction
on the fluctuations of $t$ is met for a $\delta$-admissible region
$R_{xy}:$
\begin{lemma}\label{lem:5}
Fix two constants $a > 1$ and $1/2 > \alpha > 0\,$. If $R_{xy}$ is a
$\delta$-admissible region in the sense of Def.\ \ref{def1} and the
statement
\begin{equation}\label{conditions}
    \forall j \in R^{z}_{xy} \; :
    \quad \cosh(t_j - t_z) \leq \ a\, |j-z|^\alpha
\end{equation}
holds for both $z = x$ and $z = y$, then we have
\begin{equation}
    0\leq  G_{xy} \leq G_{xy}^N \leq C(a,\alpha,\delta) / \beta \;,
    \label{eq:lemmabound}
\end{equation}
where $C(a,\alpha,\delta)$ is some constant depending on the
parameters $a$, $\alpha$ and the geometry of the region $R_{xy}$
(encoded in the parameter $\delta$). The notation $G^N$ means Neumann
boundary conditions on $\partial R_{xy}\,$.
\end{lemma}
\paragraph{Proof.}
The following is a variation on an argument presented in \cite{SZ}.
For each $k \in \mathbb{N}$ consider two cubes of side $2^k$ centered
at $x$ and $y$. (For concreteness, imagine the edges of the two cubes
to be parallel to the vector $x-y$.) Let $R_x^k\, , R_y^k$ denote the
corresponding intersections with $R_{xy}$ and let $I_k\,$,
$\tilde{I}_k$ be the indicator functions of $R_x^k$ and $R_y^k\,$,
respectively, normalized so that for each $k$
\begin{equation}
    \sum\nolimits_j I_k(j) = 1 = \sum\nolimits_j \tilde{I}_k(j) \;.
\end{equation}
We observe that
\begin{equation}\label{eq:km}
    \forall k \geq k_m \; : \quad R_x^k = R_y^k = R_{xy} \;,
\end{equation}
where $k_m$ is the smallest number $k \in \mathbb{N}$ such that $2^{k
-1} \geq |x-y|$. Since $R_{xy}$ is $\delta$-admissible, $R_z^k$ has
the same properties as $R_{z}(r = 2^k)$ in Def.\ \ref{def1} and we
therefore have
\begin{equation}\label{Rvol}
    \mathrm{vol}\, R_x^k \geq 2^{k d}\delta
    \qquad (k \leq k_m -1 \;, \quad d = 3)
\end{equation}
for all $2^k > 10$. For $2^k\leq 10$ this is not true (see Def.\
\ref{def1}) but the corresponding volume is no less than unity, as
$R_{xy}$ is connected.

Now we express $\delta_x - \delta_y$ as a telescopic sum:
\begin{equation}
    \delta_x - \delta_y = \sum_{k = 1}^{k_m}
    \left( \rho_k - \tilde{\rho}_k \right) \;,
\end{equation}
where $\rho_k = I_{k-1}-I_k\,$, $\tilde{\rho}_k = \tilde{I}_{k-1} -
\tilde{I}_k\,$, $I_{0} = \delta_x\,$, and $\tilde{I}_0 = \delta_y\,$.
This sum terminates at $k_m$ because by \eqref{eq:km} we have $I_k =
\tilde{I}_k$ for $k\ge k_m\,$. Note that $\rho_k\,$, $\tilde{\rho}_k$
are orthogonal to the constant functions: $\sum_{j\in\Lambda} \rho_k
(j) = 0 = \sum_{j \in \Lambda} \tilde{\rho}_k(j)$.

Next we put the telescopic sum to use by the following computation:
\begin{align}
    &\left[\delta_x -\delta_y\,; D_{\beta,\varepsilon}(t)^{-1}
    (\delta_x -\delta_y) \right] = \sum_{k,l=1}^{k_m} \left[\rho_k
    -\tilde{\rho}_k \,; D_{\beta,\varepsilon}(t)^{-1} (\rho_{l} -
    \tilde{\rho}_l) \right] \nonumber \\ &\leq \left(\sum_{k=1}^{k_m}
    \Big( \left[\rho_k\,; D_{\beta,\varepsilon}(t)^{-1} \rho_k
    \right]^{1/2}+\left[\tilde{\rho}_k\,;D_{\beta,\varepsilon}(t)^{-1}
    \tilde{\rho}_k \right]^{1/2} \Big) \right)^2 \;, \label{eq:telsum}
\end{align}
where the Cauchy-Schwarz inequality was employed. Hence we need to
estimate $[\rho_k\,; D_{\beta,\varepsilon}(t)^{-1} \rho_k ]$ and $[
\tilde{\rho}_k\,; D_{\beta,\varepsilon}(t)^{-1} \tilde{\rho}_k]$.
This is done, say for the former, by the inequality
\begin{equation}\label{eq:5.13}
    [ \rho_k\,; D_{\beta,\varepsilon}(t)^{-1}\,  \rho_k ] \leq
    \| D^{-1}_{R_x^k} \| \, \| \rho_k \|_2^2 \;,
\end{equation}
where $D_R$ (for a region $R$) stands for the operator \eqref{eq:1.3}
with Neumann boundary conditions on $R$. In view of $\sum \rho_k(j) =
0$ the operator norm is to be taken on the orthogonal complement of
the constant functions.

The square of the $L^2$-norm of $\rho_{k+1}$ is bounded by
$(\mathrm{vol}\, R_x^k)^{ -1}$. Thus by \eqref{Rvol}
\begin{equation}\label{Ibound}
    \| \rho_{k+1} \|_2^2 \leq (\mathrm{vol}\, R_x^k)^{-1}
    \leq 2^{-kd} \delta^{-1} \;.
\end{equation}
The corresponding inequality also holds for $\tilde{\rho}_{k+1}\,$.

We must still bound the operator norm $\| D^{-1}_{R_x^k} \|$. For
this we observe that the conditions \eqref{eq:7.5-TS} ensure that
\begin{displaymath}
    A_{xy}(jj')^{-1} \leq 2^3 (a\, |j-x|^\alpha)^2
    \leq 8 a^2 2^{2k\alpha}
\end{displaymath}
for all $j,j'\in R_x^k$ and $k\leq k_m - 1$, since in that case
$jj'\in R_{xy}^x$ and we apply \eqref{conditions} for $z = x$. For $k
= k_m$ we are looking at pairs that belong to $R_{xy}^y$ but not to
$R_{xy}^x$. In that case we apply \eqref{conditions} for $z = y$ and
still have $A_{xy}(jj')^{-1} \leq 8 a^2 2^{2k\alpha}$. Therefore,
since $R_{xy}$ is $\delta$-admissible and the lowest nonzero
eigenvalue of the Neumann Laplacian on $R_x^k$ is of the order of
$(2^k)^{-2}$, we obtain
\begin{equation}\label{Dnorm}
    \mathrm{e}^{t_x + t_y} B_{xy}^{-1} \| D^{-1}_{R_x^k}\|
    \leq c(\delta)\, a^2 \beta^{-1} 2^{2k} 2^{2k\alpha}
\end{equation}
for some $c(\delta)$ and all $k \leq k_m\,$. For $2^k \leq 10$ the
connectedness of $R_{xy}$ ensures that  $\| D^{-1}_{R_x^k} \| \leq
\mathrm{const}$. The same bounds apply for $\| D^{-1}_{R_y^k} \|$.

Finally, by combining \eqref{eq:telsum} with \eqref{eq:5.13},
\eqref{Ibound}, and \eqref{Dnorm}, we arrive at
\begin{displaymath}
    G_{xy} = \frac{\mathrm{e}^{t_x + t_y}}{B_{xy}} \left[
    \delta_x - \delta_y \,; D_{\beta,\varepsilon}(t)^{-1}
    (\delta_x - \delta_y) \right] \leq c(\delta)\,
    \frac{a^2}{\beta}\left( 2 \sum_{k=1}^{k_m}
    \sqrt{2}^{\,k  (2\alpha + 2 - d)} \right)^2 .
\end{displaymath}
For $2\alpha < d - 2 = 1$ the value of this sum is bounded uniformly
in $k_m\,$. \qed
\paragraph{Remark.}
The bound \eqref{eq:lemmabound} also applies when the definition of
$R_{xy}^x$ and $R_{xy}^y$ is modified in the following way (for $z =
x,y$ as before):
\begin{equation}\label{Rdefbis}
    R_{xy}^z = \{ j \in R_{xy} : \, |j-z| \leq |x-y|\, f_z \,\} \;,
\end{equation}
where $f_x\,, f_y$ are a pair of positive numbers which add up to (at
least) unity and neither of which is too small. It is easy to see
that the relevant scales involved are the ones for $k$ near $k_{m}$
and we can get the same bound but with a change of overall factor.
This remark will become important in Section \ref{sect:9}, Lemma
\ref{remainder}, where we will need this estimate with $f_y \simeq
1/5\,$.

\section{Conditional estimates on fluctuations}\label{sect:6}
\resetequ

In this section we establish bounds on the fluctuations of the field
$t$ by bounding $\langle B_{xy}^m \, \bar\chi_{xy} \rangle$ where
$\bar\chi_{xy}$ has the property that $\bar\chi_{xy} = 0$ whenever
the hypothesis \eqref{conditions} of Lemma \ref{lem:5} fails.
\begin{defin}\label{def:3}
{\bf (characteristic function)}: As before, fix two constants $a > 1$
and $1/2> \alpha >0$, and let $r_{j-k} := (a\,|j-k|^\alpha)^{-1}$ for
$j$, $k\in\Lambda$, $j\neq k$.
Let $\chi:\,\mathbb{R}_+ \to \mathbb{R} $ be the
characteristic function of the interval $[0, 1]$, i.e., $\chi(t) = 1$
for $0 \leq t \leq 1$ and $\chi(t) = 0$ for $t > 1$. Moreover, let
$R_{xy}$ be $\delta$-admissible and choose the regions $R_{xy}^x \,$,
$R_{xy}^y$ as in \eqref{Rdef}. In this setting we define
\begin{equation}\label{chidef1}
    \bar\chi_{xy} = \prod_{j\in R_{xy}^x} \chi_{xj}
    \prod_{j \in R_{xy}^y} \chi_{yj}\;, \qquad
    \chi_{zj} = \chi(r_{j-z} B_{z j}) \quad (z = x, y)\;.
\end{equation}
\end{defin}
Here the constants $a$, $\alpha$ are taken to coincide with those in
Lemma \ref{lem:5}.

With these definitions we have
\begin{lemma}\label{lem:6}
Let $R_{xy}$ be a $\delta$-admissible region, and let $C = C(a,\alpha
,\delta)$ be the constant that appears in Lemma \ref{lem:5}. Then for
$0\leq m < \beta / C$ we have
\begin{equation}\label{eq:lemma6}
    \left\langle  B_{xy}^m\, \bar\chi_{xy} \right\rangle \leq
    \left(1 - m C / \beta \right)^{-1} \;.
\end{equation}
\end{lemma}
\paragraph{Proof.}
Our proof uses the identity $\langle S_{xy}^m\,\chi_{xy}^S\rangle =1$
where $\chi_{xy}^S$ is a supersymmetric version of $\bar \chi_{xy}$
defined above. After integrating out the Grassmann fields we shall
show that this identity implies
\begin{displaymath}
    \big\langle B_{xy}^m \, \bar\chi_{xy}\, (1 - m G_{xy})
    \big\rangle \leq 1 \;.
\end{displaymath}
Lemma \ref{lem:5} and the presence of $\bar\chi_{xy}$ then yield
\eqref{eq:lemma6}.

More precisely, let
 $\chi_\gamma\in\mathrm{C}^\infty(\mathbb{R}_+)$ with
$\frac{d}{dt} \chi_\gamma(t) \leq 0$ and
\begin{displaymath}
    \chi_\gamma (t) = \left\{ \begin{array}{ll} 1 &\quad t \leq
    1-\gamma \;, \\ 0 &\quad t \geq 1 \;, \end{array} \right.
\end{displaymath}
be a smooth regularization of $\chi = \lim_{\gamma\to 0} \chi_\gamma
\,$. We fix a small value of $\gamma > 0$ and write $\tilde\chi
\equiv \chi_\gamma$ for short. Then, recalling the definition
\eqref{eq:6.3-TS} of $F_{xy}$ we introduce
\begin{equation}\label{eq:chiS}
    \chi_{x j}^S = \tilde\chi (r_{j-x}\, S_{x j} ) = \tilde\chi_{x j}
    + r_{j-x}\,\tilde{\chi}'_{xj}\, \bar\psi(F_{x j}) \psi(F_{x j})\;,
\end{equation}
where $\tilde\chi_{xj} = \tilde\chi(r_{j-x} B_{x j})$. Since $\chi_{x
j}^S$ is $Q$-invariant and $\chi_{x j}^S (0) = 1$ we have
\begin{equation}\label{eq:6.6}
    1 = \left\langle S_{xy}^m \prod_{j\in R_{xy}^x} \chi^S_{xj}
\prod_{j\in R_{xy}^y} \chi_{yj}^S
    \right\rangle \equiv \left\langle S_{xy}^m \, \tilde\chi_{xy}^S
    \right\rangle \;.
\end{equation}
Now, we express
\begin{equation}\label{eq:8.5-TS}
    \big\langle S_{xy}^m \, \tilde\chi_{xy}^S \big\rangle =
    \big\langle S_{xy}^m \, \tilde\chi_{xy} \,
    \exp - [ \bar\psi \,; A \psi ] \big\rangle ,
\end{equation}
where the symmetric operator $A$ is given by
\begin{align}
    [f ; A f] &= - \sum_{j \in R_{xy}^x}
    \frac{r_{j-x}\tilde{\chi}'_{xj} }{\tilde{\chi}_{xj}}
    [f ; F_{x j}]^2 - \sum_{j \in R_{xy}^y} \frac{r_{j-y}
    \tilde{\chi}'_{yj}}{\tilde{\chi}_{yj}} [f;F_{y j}]^2 \;.
\label{eq:A0}\end{align}
Clearly $A \geq 0$ as a quadratic form since $\tilde\chi^\prime \leq
0\,$. The total $\bar\psi\psi$ contribution to \eqref{eq:8.5-TS}
including the fermionic part of the action is
\begin{displaymath}
    [ \bar\psi\,; (D_{\beta,\varepsilon} (t) + A) \psi ]
    - m B_{xy}^{-1} \bar\psi(F_{xy})\, \psi(F_{xy}) \;,
\end{displaymath}
where the second summand stems from $S_{xy}^m\,$, see
\eqref{eq:WI0}-\eqref{eq:WI1}. Thus, integration over the Grass\-mann
fields $\bar\psi, \psi$ gives $\mathrm{Det}(Q+A)$ where
\begin{displaymath}
    Q = D_{\beta,\varepsilon} (t) - m B_{xy}^{-1}
    F_{xy}\, [ F_{xy} \,; \,\cdot\, ] \,.
\end{displaymath}

Since we are taking $m$ to be less than $\beta / C$, the presence of
the factor $\tilde\chi_{xy}$ in \eqref{eq:8.5-TS} ensures (by Lemma
\ref{lem:5}) that $m G_{xy} < 1\,$. Now by Lemma \ref{lem:K} the
inequality $1 \geq m G_{xy} = m B_{xy}^{-1} [F_{xy} ; D_{\beta,
\varepsilon }(t)^{-1} F_{xy} ]$ is equivalent to $Q \geq 0\,$.
Therefore the result $\mathrm{Det}(Q + A)$ of integrating over
$\bar\psi, \psi$ is bounded from below by
\begin{displaymath}
    \mathrm{Det}(Q+A) \geq \mathrm{Det}(Q) = \mathrm{Det}
    (D_{\beta,\varepsilon} (t))\, (1 - mG_{xy}) \geq 0 \;,
\end{displaymath}
and we obtain the estimate
\begin{displaymath}
    1 = \big\langle B_{xy}^m \, \tilde\chi_{xy} \, \mathrm{e}^{-
    [ \bar\psi\,; (Q+A) \psi]} \big\rangle \geq \big\langle
    B_{xy}^m \, \tilde\chi_{xy} \, (1 - m G_{xy}) \big\rangle\;.
\end{displaymath}
We finally take the limit $\gamma \to 0$. The smooth function
$\tilde\chi_{xy}$ then converges to the characteristic function
$\bar\chi_{xy}\,$. Hence
\begin{displaymath}
    1 \geq \big\langle B_{xy}^m \, \bar\chi_{xy} \,
    (1 - m G_{xy}) \big\rangle \geq \big\langle B_{xy}^m
    \, \bar\chi_{xy} \big\rangle \, (1 - m C/\beta) \;,
\end{displaymath}
which is the desired result. \qed
\begin{lemma}
If all of the regions $R_{x_1 y_1}\,, R_{x_2 y_2}\,, \ldots, R_{x_n
y_n}$ are $\delta$-admissible and disjoint (meaning they have
disjoint interiors), then we still have
\begin{equation}
    \left\langle \prod_{j=1}^n B_{x_j y_j}^m \bar\chi_{x_j y_j}
    \right\rangle \leq (1 -m C / \beta)^{-n} \;.
\end{equation}
\end{lemma}
\paragraph{Proof.}
As before we use the fact that the supersymmetrized observable, which
here results from replacing $B_{x_j y_j}$ by $S_{x_j y_j}$, has
expectation one.

Consider first the simpler problem of computing the expectation of
the product $\prod S_{x_i y_i}^m \bar\chi_{x_i y_i}\,$. After
integrating over $\psi$ and $\bar\psi$ we see that
\begin{equation}
    \left\langle \prod_{i=1}^{n} S_{x_i y_i}^m \, \bar\chi_{x_i y_i}
    \right\rangle = \left\langle \prod_{i=1}^{n} B_{x_i y_i}^m \,
    \bar\chi_{x_i y_i}\, \mathrm{Det} (1 - m \mathcal{G})
    \right\rangle \;,
\end{equation}
where $\mathcal{G}$ is an $n\times n$ matrix of Green's functions
\begin{equation}
    \mathcal{G}_{ij} = \beta^{-1} [g_i\,;D_{\beta,\varepsilon}(t)^{-1}
    g_j] \;, \qquad g_i = B^{-1/2}_{x_i y_i}\, \mathrm{e}^{(t_{x_i} +
    t_{y_i})/2}\, (\delta_{x_i} - \delta_{y_i}) \;.
\end{equation}
The matrix $\mathcal{G}$ is positive as a quadratic form. In order to
reduce the problem to the previous case (of just a single region)
note that $\mathcal{G} \leq \mathcal{G}_N\,$, where the subscript
denotes Neumann boundary conditions on the boundaries of the disjoint
regions $R_{x_i y_i}\,$. The presence of the factors $\bar\chi_{x_i
y_i}$ implies bounds on the $G_{x_i y_i}$ so that
\begin{equation}
    \mathrm{Det} (1 - m \mathcal{G}) \geq
    \mathrm{Det} (1 - m \mathcal{G}_{N}) =
    \prod_{i = 1}^n  (1 - m  G_{x_i y_i})
    \geq (1 - m C/\beta)^{-n} \;.
\end{equation}

The proof of the lemma is completed by introducing the effects of
$\tilde{\chi}^\prime$ as before. Since there are no new aspects to
this argument, we omit it.
\paragraph{Remark.} From this lemma one obtains estimates for
conditional probabilities only. Yet, in order to bound $C_{xy}$ in
Theorem \ref{thm:new} we need probability estimates without any
conditions, which is why we now have to develop an inductive
argument.

\section{Unconditional estimates on fluctuations}
\label{sect:uncond}\resetequ

We are now going to remove the constraints enforced by insertion of
$\bar\chi$. In order to do so, we have to consider $\chi_{xj}^c = 1
-\chi_{xj}$ for $\chi_{xj}$ defined by \eqref{chidef1}. Short scales
(given by $0 < |j-x| < \beta^{1/4}$) will be treated separately by
monitoring, in Section \ref{sect:7}, only the size of nearest
neighbor gradients inside the region $R_{xy}\,$. At the very large
scales of $|j-x|\geq \beta^{1/4}$, however, looking only at NN
fluctuations is not enough. There, in order to remove the $\bar\chi$
constraints we will show by induction on the distance $|j-x|$ that
the corresponding contribution is small.

\begin{figure}
    \centerline{\psfig{figure=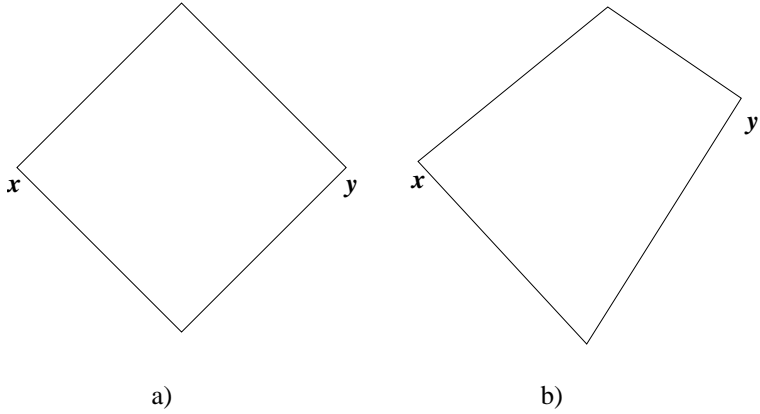 ,width=10cm}}
    \caption{a) diamond region, %
    b) $C$-admissible region (deformed diamond)}
    \label{fig:case1}
\end{figure}

We will distinguish between two types of geometry: diamonds and
deformed diamonds. For deformed diamonds we will quantify the bounds
given by \eqref{eq:lemmabound} and call such regions $C$-admissible.
\begin{defin}\label{def:C-adm}
Let $R_{xy} \subset \Lambda$ be $\delta$-admissible in the sense of
Def.\ \ref{def1}.
\begin{enumerate}
\item We call $R_{xy}$ a {\rm diamond} if it is the set of
    lattice points which is contained in a 3-dimensional double
    cone obtained in the following way: we take a 2-dimensional
    rectangular diamond with opposite vertices placed on $x$ and
    $y$ and edges of length $|x - y| / \sqrt{2}$ (see Fig.\
    \ref{fig:case1}) and rotate it around the line $\overline
    {xy}$. In order to ensure connectedness we may have to add a
    few lattice points near $x$ and $y$ (see Fig.\
    \ref{fig:cutdiamond} a,b).
\item We call $R_{xy}$ a $C$-{\rm admissible} region (or deformed
    diamond) if
\begin{align}
    0 \leq G_{xy}^N\, \bar\chi_{xy} \leq C/\beta & \quad \text{for}
    ~ |x-y| > \beta^{1/4}, \\ 0\leq G_{xy}^N {\prod}'_{pq} \chi_{pq}
    \leq C/\beta & \quad \text{for} ~ |x-y| \leq \beta^{1/4}
    \label{eq:9.2},
\end{align}
where $\bar\chi_{xy}$ is defined in \eqref{chidef1}, the
superscript $N$ stands for Neumann boundary conditions on
$R_{xy}$ and $\prod'$ denotes the product over all nearest
neighbor pairs in $R_{xy}\,$.
\end{enumerate}
\end{defin}
Note that for short scales, dealt with in \eqref{eq:9.2}, instead of
using $\bar\chi_{xy}$ we impose constraints on \emph{all} NN pairs in
the region $R_{xy}\,$.

With these definitions we can state the main result of this paper.
\begin{theorem}\label{thm:3}
Let $m=\beta^{1/8}$, and let $R_{x_i y_i}$ for $i = 1, \ldots, n_1$
be diamonds with disjoint interiors and $|x_i - y_i| > \beta^{1/4}$.
Then we have
\begin{equation}\label{Ih1}
    \left\langle \prod_{i=1}^{n_1} B_{x_i y_i}^m \right\rangle
    \leq 2^{n_1}
\end{equation}
for all $n_1 \geq 0$. Moreover if $p_j , q_j$ for $j = 1, \ldots,
n_2$ are such that $| p_j - q_j | > \beta^{1/4}$, the regions $R_{p_j
q_j}$ are $C$-admissible, have disjoint interiors and do not overlap
with any of the regions $R_{x_i y_i}\,$, then there exists a constant
$\rho \leq 1/2$ such that
\begin{equation}\label{Ih2}
    \left\langle  \prod_{i=1}^{n_1} B_{x_i y_i}^m \prod_{j=1}^{n_2}
    B_{p_j q_j}^{3m} \bar\chi_{p_j q_j} \right\rangle \leq 2^{n_1}
    (1+\rho)^{n_2}
\end{equation}
for all $n_1 \geq 0$ and $n_2 \geq 0$. Finally, let $r_k , s_k$ for
$k = 1, \ldots, n_3$ be such that $| r_k - s_k | \leq \beta^{1/4}$,
$R_{r_k s_k}$ are $C$-admissible, have disjoint interiors and do not
overlap with any of the regions $R_{x_i y_i}$ or $R_{p_j q_j}\,$.
Then for all $n_1 \geq 0$, $n_2 \geq 0$ and $n_3 \geq 0$ we have
\begin{equation}\label{Ih3}
    \left\langle \prod_{i=1}^{n_1} B_{x_i y_i}^m \prod_{j=1}^{n_2}
    B_{p_j q_j}^{3m} \bar\chi_{p_j q_j} \prod_{k=1}^{n_3} B_{r_k
    s_k}^{3m} \right\rangle \leq 2^{n_1} (1+\rho)^{n_2}\, 2^{n_3}
\end{equation}
with $\rho \leq 1/2$.
\end{theorem}
The proof of the theorem is carried out in Sections
\ref{sect:7}--\ref{sect:9}. We will need to distinguish between three
situations, which we refer to as \emph{classes}.

\smallskip\noindent{\bf Class 1.}
$|x-y| > \beta^{1/4}$ and the pair is not protected by a factor of
$\bar\chi_{xy}\,$. In this case we need an inductive argument on
scales to prove a bound on the expectation of $B_{xy}^m\,$. The
induction will be done on $\ell = \max_j |x_j - y_j|$ and is carried
out in Sections \ref{sect:8} and \ref{sect:9}. We will need to
inductively select non-overlapping smaller diamonds inside the region
$R_{xy}$ while making sure that these remain $\delta$-admissible. To
arrange for all geometrical details to work out, we take $R_{xy}$ to
be a perfect diamond.

\smallskip\noindent{\bf Class 2.}
$|x-y| > \beta^{1/4}$ but the pair is protected by a factor of
$\bar\chi_{xy}\,$. In this case we can apply the results of Section
\ref{sect:6}, thereby obviating the need for any induction. $R_{xy}$
is then allowed to be a deformed diamond and the bound we can get is
stronger than in Class 1 (power $3m$ instead of $m$). %

\smallskip\noindent{\bf Class 3.}
$|x-y|\leq \beta^{1/4}$. This includes short scales and the NN case,
which was already treated in Section \ref{sect:4}. We will show in
Section \ref{sect:7} that these scales do not require any factor of
$\bar\chi_{xy}$ to ensure a good bound. No induction is needed, and
we can therefore take $R_{xy}$ to be a deformed diamond. \smallskip

Note that the larger exponent $3m$ appearing in \eqref{Ih2} and
\eqref{Ih3} is important for the inductive proof to go through. The
enlarged exponent can be handled either because of the presence of
$\bar \chi$ or because the pair is of Class 3.

\subsection{Fixing the different parameters}\label{sect:fix}

We have introduced a certain number of parameters: $m$, $a$, $\rho$,
$\delta$, $C$, $\alpha$. Before going on, we briefly review why they
appeared and how to choose their values.
\begin{enumerate}
\item The parameter $m$ is ubiquitous in this paper as the power
    of $B_{xy}\,$. Since the probability of large deviations will
    be bounded by $K^{-m}$ with $K > 1$, we want $m$ to be as
    large as possible. On the other hand, to apply the SUSY
    argument of Section \ref{sect:6} we must have $(1- 3 m C /
    \beta)^{-1} <1$, where the factor $3m$ in this inequality
    comes from the power of $B$ in \eqref{Ih2}, \eqref{Ih3}.
    Therefore the magnitude of $m$ is limited by $\beta$. To
    arrange for all the conditions to be met, we fix $m = \beta^{
    1/8}$. The factors $m$ and $3m$ will be kept fixed in the
    whole course of proof.
\item The constants $C$ and $\delta$ appearing in the definition
    of the region $R_{xy}$ (see Def.\ \ref{def1}) are not subject
    to any special requirements, but their values do constrain
    the other parameters. They will be fixed throughout.
\item To prove the induction hypothesis we need $0 \leq \rho \leq
    1$. More precisely (see Eq.\ \eqref{eq:12.6}) we need $\rho +
    \mathcal{R}(x,y)\leq 1$. Since we prove $\mathcal{R}(x,y)
    \leq 1/2$ we will take $\rho \leq 1/2$.
\item The constant $a$ in Lemma \ref{lem:5} in Section \ref{sect:5} plays 
    a key role in
    bounding the entropy for small scales; see Section
    \ref{sect:9}, Case 1, Eq.\ \eqref{eq:case1}. It will become
    clear there that $a > 10$ is sufficient.
\item We need to take $\alpha > 0$ in Lemma \ref{lem:5} in order
    to control entropy factors for large deviations (see Section
    \ref{sect:9}). On the other hand, the result of Theorem 2
    would be optimal for $\alpha = 0$. Therefore we wish to make
    $\alpha$ as small as possible. We will see in Section
    \ref{sect:9} (Case 2b eq. \eqref{eq:2b} and Case 2c eq. \eqref{eq:2c}) 
    that $\alpha \geq O (1 / \ln \beta)$
    is a requirement for our analysis to go through.
\end{enumerate}

\section{Short-scale fluctuations}\label{sect:7}\resetequ

We now prove Theorem \ref{thm:3} for $\ell \leq \beta^{1/4}$, i.e.,
for Class 3 pairs. These estimates will follow from the bounds on NN
fluctuations established in Section \ref{sect:4}.
\begin{lemma}
There is a constant $\beta_0$ such that for $\beta \geq \beta_0\,$,
$|x - y| = \ell \leq \beta^{1/4}$ and $3m \leq \beta^{1/8}$, we have
\begin{equation}
    \left\langle B_{xy}^{3m} \right\rangle \leq 2 \;.
\end{equation}
More generally let $(x_1, y_1), \ldots, (x_n , y_n)$ be $n$ pairs
with $|x_j - y_j |\leq \ell$ for all $j$, and let the interiors of
the corresponding $C$-admissible regions $R_{x_1 y_1}$, $R_{x_2 y_2},
\ldots, R_{x_n y_n}$ be disjoint. Then
\begin{equation}
    \left\langle\prod_{j=1}^n B_{x_j y_j}^{3m} \right\rangle\leq 2^n.
\end{equation}
\end{lemma}
\paragraph{Proof.}
As in Def.\ \ref{def:3}, let $\chi$ be the characteristic function of
the interval $[0,1]$ and let (with a parameter $\delta$ to be defined
shortly)
\begin{equation}
    \chi_{pq} = \chi( (1+\delta)^{-1} B_{pq} )\;,\quad |p-q| = 1 \;,
\end{equation}
and $\chi_{pq}^c = 1 - \chi_{pq}\,$. Using $\chi_{pq} \leq 1$ we have
$1 \leq \prod_{(pq)} \chi_{pq} + \sum_{(pq)} \chi_{pq}^c$ and
\begin{equation}\label{dec}
    \big\langle  B_{xy}^{3m} \big\rangle \leq \big\langle B_{xy}^{3m}
    \prod\nolimits_{(pq)} \chi_{pq}\big\rangle + \sum\nolimits_{(pq)}
    \big\langle  B_{xy}^{3m} \, \chi_{pq}^c \big\rangle \;,
\end{equation}
where the product and the sum are over all nearest neighbor pairs
$(pq)$ in $R_{xy}\,$.

We estimate the first term on the right-hand side of \eqref{dec} by
applying the strategy of the proof of Lemma \ref{lem:6} to show that
\begin{equation}\label{eq:10.6-TS}
    \big\langle  B_{xy}^{3m} \big(1 - 3m\, G_{xy}\big)
    \prod\nolimits_{(pq)} \chi_{pq} \big\rangle \leq 1\;.
\end{equation}
To bound $G_{xy}\,$, note that on the support of $\chi_{pq}$ we have
\begin{equation}
    0 \leq {\textstyle{\frac{1}{2}}} (t_p - t_q)^2 \leq
    \cosh (t_p - t_q) - 1 \leq \delta \;.
\end{equation}
Thus $|t_p - t_q| \leq \sqrt{2\delta}$ and $|t_z - t_j| \leq \ell
\sqrt{2\delta}$ for $z = x, y$ and all $j \in R_{xy}\,$. Now let us
require
\begin{equation}\label{eq:10.7-TS}
    \ell \sqrt{2\delta} = 1 \;, \quad \text{or} \quad
    \delta = {\textstyle{\frac{1}{2}}} \beta^{-1/2} \;,
\end{equation}
since $\ell \leq \beta^{1/4}$. Thus we have a uniform lower bound on
the conductance \eqref{eq:7.5-TS}. It then follows that $0\leq G_{xy}
\leq C/\beta$ with $C$ independent of $\beta$, and \eqref{eq:10.6-TS}
gives
\begin{equation}\label{eq:10.8-TS}
    \big\langle B_{xy}^{3m} \prod\nolimits_{(pq)}\, \chi_{pq}
    \big\rangle \leq (1 - 3m\, C / \beta)^{-1} \leq 3/2
\end{equation}
for $\beta$ large.

For the second summand of \eqref{dec} we use
\begin{equation}\label{eq:10.9-TS}
    \chi_{pq}^c = 1 - \chi\big( (1+\delta)^{-1} B_{pq}\big)
    \leq \left( \frac{B_{pq}}{1+\delta} \right)^{\beta/2} \;.
\end{equation}
The factor $B_{xy}^{3m}$ is estimated by repeated application of
\eqref{Bbound}:
\begin{equation}\label{eq:10.10-TS}
    2 B_{xy} \leq \prod\nolimits_j 2 B_{p_j q_j} \;,
\end{equation}
where the product ranges over a set of NN pairs connecting $x$ and
$y$. By combining \eqref{eq:10.9-TS} and \eqref{eq:10.10-TS} and then
using the result \eqref{eq:4.12} for NN pairs we have
\begin{align}
    \left\langle B_{xy}^{3m}\, \chi_{pq}^c \right\rangle &\leq
    \frac{2^{3m(\ell-1)}}{(1+\delta)^{\beta/2}} \left\langle
    B_{pq}^{\beta/2} \prod\nolimits_j B_{p_j q_j}^{3m}\right\rangle
    \cr &\leq \frac{2^{3m(\ell-1)}}{(1+\delta)^{\beta/2}}\,
    \big( {\textstyle{\frac{1}{2}}} - 3 m /\beta \big)^{-1}
    (1 - 3m /\beta)^{-\ell} \cr &\leq \mathrm{e}^{3m\ell}
    \mathrm{e}^{- \beta \delta / 3} \;. \label{eq:10.11-TS}
\end{align}
Since $3m\ell \leq \beta^{1/8}\beta^{1/4}$ by hypothesis, and $\delta
= \frac{1}{2} \beta^{-1/2}$ by \eqref{eq:10.7-TS}, we see that the
expression \eqref{eq:10.11-TS} is less than $\exp( \beta^{3/8} -
\beta^{1/2} / 6)$.

Combining our estimates on the two terms on the r.h.s.\ of
\eqref{dec} we have
\begin{displaymath}
    3/2 + 3 \ell^3 \mathrm{e}^{\beta^{3/8} - \beta^{1/2}/6} \leq 2
\end{displaymath}
for large enough $\beta$. The factor $3 \ell^3 \leq 3 \beta^{3/4}$
comes from the sum over all NN pairs in $R_{xy}\,$.

When several disjoint regions are present, the bounds over disjoint
regions factor, and we can get the same result using the same
argument. \qed

\section{Induction hypothesis and some preliminary estimates}
\label{sect:8}\resetequ

The argument in the last section cannot be repeated for all values of
$\ell$. In order to control all scales we need an inductive argument.
\paragraph{Induction Hypothesis:}
Let $x_i , y_i$ ($i = 1, \ldots , n_1$), $p_j , q_j$ ($j = 1, \ldots,
n_2$), and $r_k , s_k$ ($k = 1, \ldots, n_3$) be pairs of Class 1, 2,
resp.\ 3, in the sense of Theorem 3. Then the bounds \eqref{Ih1},
\eqref{Ih2}, \eqref{Ih3} hold when $|x_i - y_i| \leq \ell$ for all $i
\leq n_1\,$.\qed

The induction is on $\ell = \max_i |x_i - y_i|$. The said bounds were
already established for $\ell = 1$ (NN case, Section \ref{sect:4})
and $\ell \leq \beta^{1/4}$ (Section \ref{sect:7}). Assuming that the
Induction Hypothesis holds up to scale $\ell$, we shall prove (in
Section \ref{sect:9}) that it holds up to scale $\ell+1$. This will
complete the proof of Theorem \ref{thm:3} and, as an immediate
consequence, Theorem \ref{thm:1}.

The idea of the proof is the same as in Section \ref{sect:7}. If the
pair $xy$ is protected by a $\bar\chi_{xy}$ factor (Class 2), then we
apply Lemma \ref{lem:6} in Section \ref{sect:6}.

To get the unconditional estimates we must study the situation when
$\bar\chi_{xy}$ is violated. This violation may happen at any scale
from $1$ up to $\ell $. To quantify this we introduce the following
definition.
\begin{defin}\label{def:good}
A point $x \in \Lambda$ is called $n$-good if
\begin{equation}
    B_{xy} \leq a\, |x-y|^{\alpha}
%
%
\end{equation}
for all $y \in \Lambda$ with distance $1 \leq |x-y| \leq 4^n$ from
$x$.
\end{defin}
\begin{defin}\label{def:ngp}
For a cube $R_n$ of side $4^n$ we define $\chi_{R_n}^c$ to be the
indicator function of the event that there exists no $n$-good point
in $R_n\,$.
\end{defin}
Our goal in the present section is to bound the expectation of the
indicator function $\chi_{R_n}^c$. In brief we will achieve this by
estimating $\chi_{R_n}^c$ by a sum of products of factors of $B_{xy}$
and then using \eqref{Ih3}. The details are as follows.

\begin{figure}
    \centerline{\psfig{figure=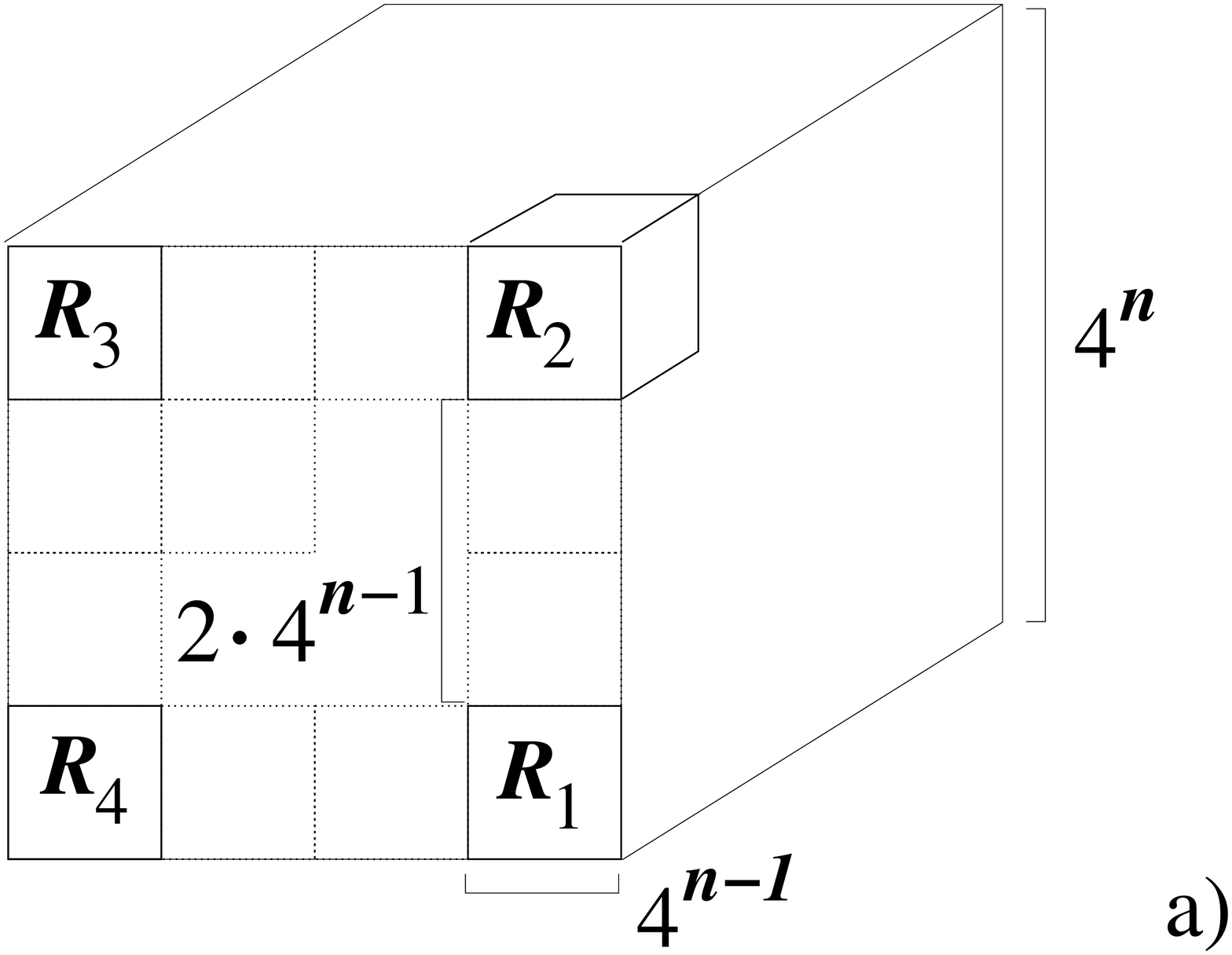,width=4cm}\hspace{0.8cm}
    \psfig{figure=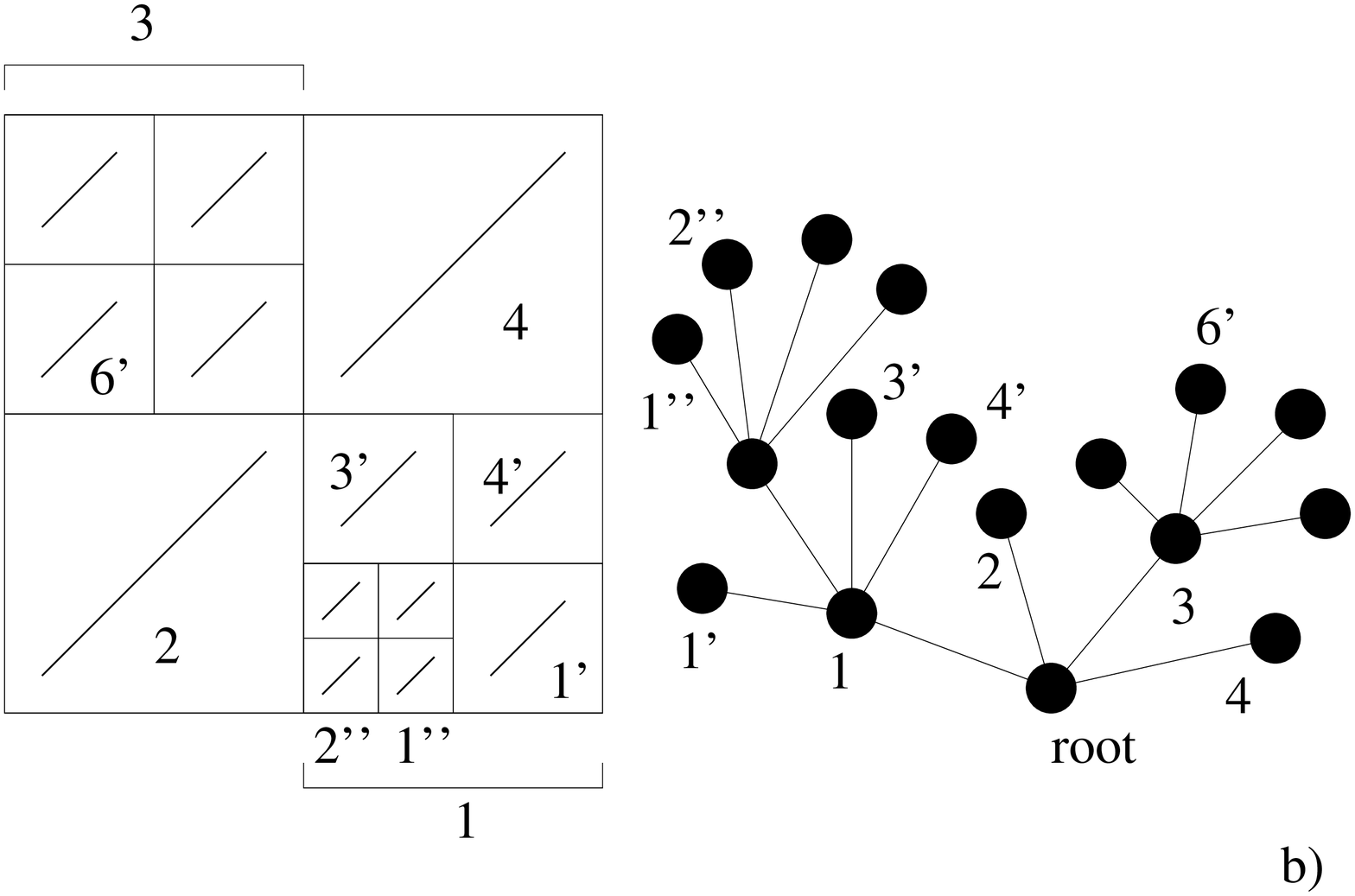,width=7cm} }
    \caption{a) In a 3D cube of side $4^n$ we select 8 cubes of
    side $4^{n-1}$. b) Here we see an example of a rooted tree (on a
    2D square) with coordination number 5 or 1 at each vertex,
    and the corresponding set of subsquares. The root corresponds
    to the large square.}\label{cubes}
\end{figure}

A 3D cube $R_n$ of side $4^n$ can be expressed as a union of $4^3$
disjoint subcubes of side $4^{n-1}$. It is clear by inspection of
Fig.\ \ref{cubes} that we can select $2^3 = 8$ of these subcubes, say
$R_{n-1}^i$ ($i = 1, \ldots, 8$), so that dist$(R_{n-1}^i,R_{n-1}^j)
> 4^{n-1}$ ($i \neq j$). Our approach now rests on the following
simple observation: if there is no $n$-good point in $R_n\,$, then
there is either no $(n-1)$-good point in any of the 8 subcubes
$R_{n-1}^i\,$, or else there exists at least one bad pair $(x,y) \in
R_n \times \Lambda$ at scale $4^{n-1} < |x-y| \leq 4^n$. Thus,
$\chi_{R_n}^c$ is bounded by the inequality
\begin{equation}\label{chidec}
    \chi_{R_n}^c \leq S_{R_n}^c + \prod_{i=1}^8 \chi_{R_{n-1}^i}^c ,
\end{equation}
where
\begin{equation}\label{Scdef}
    S_{R_n}^c = \sum_{\begin{array}{c} x \in R_n ,\, y \in \Lambda
    \\ 4^{n-1} < |x - y| \leq 4^n\end{array}} \chi_{xy}^c \;.
\end{equation}
We iterate \eqref{chidec} inside each cube $R_{n-1}^j\,$, thus
selecting $8^2$ subcubes of side $4^{n-2}$, and we keep repeating
this procedure until we reach cubes of side $4^0 = 1$ (i.e.\ points).
We denote by $\tilde{R}_{n-k}$ ($k = 0, \ldots, n$) the set of $8^k$
cubes of side $4^{n-k}$ obtained in this way. $\tilde{R}_n = \{R_n\}$
is the starting cube. Moreover let $\tilde{R}=\cup_{k=0}^n
\tilde{R}_{n-k}$. In this way we bound $\chi_{R_n}^c$ by a
positive sum of products of $\chi_{x y} ^c\,$, which in turn are
bounded by $B_{xy}^m/(a\,|x-y|^{\alpha})^m$. The resulting expression
can be organized as a sum over rooted trees picturing the hierarchy
of inclusion relations of the subcubes.


The following set of definitions serves to prepare the statement of
Lemma \ref{lem:tree} below.
Let $V$ be an abstract set of vertices such that 
$|V|=|\tilde{R}|$. We associate by a fixed 
bijective map each vertex $v$ in $V$ to a cube $R_v\in\tilde{R}$.
We denote by $k_v$ the scale of the corresponding cube: 
$R_v\in \tilde{R}_{n-k_v}$. The vertex associated with the largest
cube $R_n$ is denoted by $r$ (root). 
 Finally let ${\cal A}(v)$ (ancestor of $v$)
be the unique vertex in $V$ such that $R_v\subset R_{{\cal A}(v)}$ 
(see vertex 1 and 1' in Fig.\ \ref{cubes}b). 

With these definitions we can introduce $\mathcal{T}_n$ the set of
labelled rooted trees on some subset of $V$ with root $r$, such that 
the root has coordination number $d_r = 8$ or $d_r =
0$ (in which case the tree is reduced to a single vertex) and the
other vertices have coordination number $d_v = 9$ or $d_v = 1$.
Moreover if $v$ belongs to the tree then there must be 
a tree line connecting $v$ to its ancestor ${\cal A}(v)$. The
maximal distance of a vertex from the root is $n$. Let $L_T$ denote
the set of vertices in $T$ with $d_v = 1$ (the leaves) or $d_r = 0$
(then $L_T$ contains only the root). Let $V_k$ be the set of vertices
in $T$ at distance $k$ from the root.
With these definitions the tree is completely fixed by  the 
leaves $L_T$ (or equivalently by the choice of the coordination
numbers for each vertex).
See Fig.\ \ref{cubes}b for an example in the case of $d_v = 5$ instead of 9.
\begin{lemma}\label{lem:tree}
With the definitions above we have the inequality
\begin{equation}\label{eq:tree}
    \chi_{R_n}^c \leq  \sum_{T \in \mathcal{T}_n}
    \prod_{v \in L_T} S_{R_v}^c \;,
\end{equation}
where $R_v$ is a cube of side $4^{n - d_v}$  and $S_{R_v}^c$ is
defined as in \eqref{Scdef}, with $n$ replaced by $n - d_v\,$.
\end{lemma}
\paragraph{Proof.}
Our trees $T \in \mathcal{T}_n$ are constructed by iterating
\eqref{chidec}. In each iteration we get to choose between the first
and second term of the r.h.s.\ of \eqref{chidec}.

The construction starts with the root of the tree. In the first step
of the iterative scheme, if we pick the first term of \eqref{chidec}
then the construction ends and we have produced nothing but the
trivial tree (the root). If we pick the second term, we have a
product of 8 different indicator functions $\chi_R^c\,$, one for each
subcube in $\tilde{R}_{n-1}\,$. We represent them in the tree by
attaching 8 vertices to the root. Each vertex is then associated to a
subcube by lexicographical order (see Fig.\ \ref{cubes}b). Now we
repeat this procedure at the end of each branch and, continuing in
this way, construct a tree. Whenever we pick the first term in
\eqref{chidec}, the corresponding branch of the tree terminates and
we produce a terminal vertex or leaf. If we pick the second term, we
generate a vertex of coordination number 9. The iteration stops when
we reach scale $n$. \qed

We are now in a position to state and prove the main result of this
section.
\begin{proposition}\label{lastcase}
Let $d = 3$ and let the parameters $a, m,\alpha$ be chosen such that
$m \alpha \geq 4d$ and $a \geq 2^{3\alpha + (d+2)/m}$. Assume that
the induction hypothesis \eqref{Ih3} holds  up to scale $\ell=
4^{n+1}$. Then
\begin{equation}\label{sc}
    \left\langle \chi_{R_n}^c \right\rangle
    \leq 2^{-(n+1)\alpha m} \;,
\end{equation}
and if $R_{n(k)}$, $k = 1, \ldots, N,$ denotes a family of cubes of
side $4^{n(k)} \leq \ell$ such that
\begin{equation}
    \mathrm{dist} (R_{n(k)},R_{n(k')}) \geq \max (4^{n(k)}, 4^{n(k')})
\end{equation}
then
\begin{equation}\label{sc1}
    \left\langle\prod\nolimits_{k=1}^N \chi_{R_{n(k)}}^c \right\rangle
    \leq \prod\nolimits_{k=1}^N 2^{-(n(k)+1) \alpha m}\;.
\end{equation}
\end{proposition}
\paragraph{Proof.}
Note that the average of \eqref{Scdef} is bounded by
\begin{align*}
    \langle S_{R_n}^c\rangle &\leq \sum_{\begin{array}{c} x \in R_n
    ,\, y \in \Lambda \\ 4^{n-1} < |x - y| \leq 4^n\end{array}}
    \frac{\langle B_{xy}^m \rangle}{a^m |x-y|^{\alpha m}} \\ &\leq
    (4^n)^d (4^n 2)^d \frac{2}{a^m 4^{(n-1)\alpha m}} \leq
    {\textstyle{\frac{1}{2}}}\, 2^{-(n+1) \alpha m}
\end{align*}
for all $n\geq 1$. For $n = 0$ the sum is over $y \in \Lambda$ at
distance $|x - y| = 1$ and we obtain the same bound (see below).

First, consider the minimal scale $n = 0\,$. In this case $R_n = R_0$
is just a single point, the sum over $y$ is a sum over $2d$ nearest
neighbors, and we simply have
\begin{displaymath}
    \langle \chi_{R_0}^c \rangle \leq \langle S_{R_0}^c \rangle
    \leq 2d \, \frac{2}{a^m} \leq 2^{-\alpha m} \;.
\end{displaymath}
Thus \eqref{sc} holds for $n = 0$. Now let $n \geq 1$. By
\eqref{chidec} we have
\begin{displaymath}
    \langle \chi_{R_n}^c \rangle \leq \langle S_{R_n}^c \rangle
    + \left\langle \prod\nolimits_{i=1}^8 \chi_{R_{n-1}^i}^c
    \right\rangle  \;.
\end{displaymath}
Assuming that $\langle \chi_{R_{n-1}}^c \rangle \leq \;2^{-n \alpha
m}$ holds, we conclude that
\begin{displaymath}
    \langle \chi_{R_n}^c \rangle \leq  {\textstyle{\frac{1}{2}}}\,
    2^{-(n+1)\alpha m} + (2^{-n\alpha m})^8 < 2^{-(n+1)\alpha m}\;.
\end{displaymath}
Hence the result is true for all $n$ by induction.

To prove \eqref{sc1} we apply \eqref{eq:tree} to each term $
\chi_{R_{n(k)}}^c$. Then we can again apply \eqref{Ih3} as long as
the corresponding diamonds are disjoint -- this is ensured by the
procedure for choosing subcubes and by the constraint $\mathrm
{dist}(R_{n(k)},R_{n(k')}) \geq \max (4^{n(k)}, 4^{n(k')})$. This
concludes the proof of Proposition \ref{lastcase}. \qed

\section{Proof of the Induction Hypothesis}
\label{sect:9}\resetequ

In this section we shall establish the induction hypothesis of
Section \ref{sect:8} at scale $\ell$ assuming that it holds up to
scale $\ell-1$. Since our regions $R_{x_i y_i}$ are disjoint by
assumption, we will be able to re-express each factor $B_{x_i y_i}$
as a sum over non-overlapping regions where our induction hypothesis
applies. To simplify the notation let $x_i = x$ and $y_i = y$. We
will assume that $\ell = |x-y|$ is large, i.e., $|\ell| \geq
\beta^{1/4}$. (The case of small $\ell < \beta^{1/4}$ was dealt with
in Section \ref{sect:7}.)

For $z = x, y$ we recall the meaning of the regions $R_{xy}^z$ from
Def.\ \ref{def:2}, Section \ref{sect:5}. To ensure that the new
regions produced by the analysis below remain inside the original
region $R_{xy}$ we need to introduce the following subsets.

\begin{defin}\label{def:5}
Let $R_{xy}$ be a diamond region as described in Def.\
\ref{def:C-adm}. Then we define $\tilde{R}^{z}_{xy}$ for $z=x,y$ as
\begin{equation}\label{eq:12.1}
    \tilde{R}^{z}_{xy} = \{j \in R^{z}_{xy} : \; \angle(jz,xy)
    \leq \pi / 8 \; \mathrm{for}\; |j-z|>10 \} \;,
\end{equation}
where $\angle(jz,xy)$ is the angle between the lines $\overline{jz}$
and $\overline{xy}$.
\end{defin}
This definition roughly selects (at distances larger than 10) a
double cone which is obtained by rotating around $\overline{xy}$ a 2D
diamond with vertices on $x$ and $y$ and opening angle $\theta =
\pi/8$ (see Fig.\ \ref{fig:cutdiamond}a). The condition $|j-k| > 10$
ensures that $\tilde{R}^{z}_{xy} \cup \{ z \}$ is connected.

We also define
\begin{equation}
    u_{xy} = \prod_{j \in \tilde{R}^{x}_{xy}} \chi_{xj}
    \prod_{j \in \tilde{R}^{y}_{xy} } \chi_{yj}
\end{equation}
for $\chi_{xj}\,$, $\chi_{yj}$ as defined in \eqref{chidef1}. Note
that $\tilde{R}^x_{xy} \cup \tilde{R}^y_{xy}\cup\{x,y\}$ is a
$\delta$-admissible region in the sense of Def.\ \ref{def1} (Section
\ref{sect:5}), so Lemma \ref{lem:5} and Lemma \ref{lem:6} can be
applied to give
\begin{equation}
    \langle B_{xy}^m \, u_{xy} \rangle \leq (1 - mC/\beta)^{-1} \;.
\label{eq:12.3}\end{equation}

Now let $\chi_{xj}^c = 1-\chi_{xj}$ and $\chi_{yj}^c = 1 - \chi_{yj}
\,$. The next lemma is nothing but a combinatorial identity based on
the following partitions of unity:
\begin{equation}\label{part1}
    1 = \prod_{j \in \tilde{R}_{xy}^x} (\chi_{xj} + \chi_{xj}^c)\;,
    \qquad
    1 = \prod_{j \in \tilde{R}_{xy}^y} (\chi_{yj} + \chi_{yj}^c)\;,
\end{equation}
where $\tilde{R}_{xy}^x$ and $\tilde{R}_{xy}^y$ are the regions
defined above.
\begin{lemma}\label{lem:11mz}
The identity function can be written as
\begin{equation}\label{part2}
    1 = u_{xy}  + \sum_{b\in \tilde{R}_{xy}^x} \chi_{xb}^c
    \prod_{j,|j-x| < |b-x|} \chi_{xj} + \sum_{b\in \tilde{R}_{xy}^y}
    \chi_{yb}^c \prod_{i \in \tilde{R}_{xy}^x} \chi_{xi}
    \prod_{j,|j - y| < |b - y|} \chi_{yj} \;.
\end{equation}
\end{lemma}
\paragraph{Proof.}
We start from \eqref{part1} and expand the first product over $j$,
beginning with small $|j-x|$. For each factor $\chi_{xj}+\chi_{xj}^c$
we have two possibilities: either we pick $\chi_{xj}\,$, in which
case we proceed to the next factor and repeat, or else we pick
$\chi_{xj}^c$ and then we stop expanding and leave all the other
factors (with larger $|j-x|$) in summed form $\chi_{xj}+ \chi_{xj}^c
= 1$. In the resulting sum there is the term $\prod_{j \in
\tilde{R}_{xy}^x} \chi_{xj}\,$. This we multiply by the other product
(over $j \in \tilde{R}_{xy}^y$) in \eqref{part1}, which we expand in
the same way.

In total, we have either picked a factor $\chi$ for all $j \in
\tilde{R}^{x}_{xy}$ and $j \in \tilde{R}^{y}_{xy}$ (this results in
the term $u_{xy}$), or we have picked a term $\chi_{xj}^c$ or
$\chi_{yj}^c$ somewhere during the course of the expansion process
(this gives all the other terms). The point where we stopped is
denoted by $b$ (where $b$ stands for `bad') because $\chi_{xb}^c > 0$
or $\chi_{yb}^c > 0$ means that there is a large deviation at that
point. \qed

Using the equality \eqref{part2} of Lemma 
\ref{lem:11mz} we can rewrite $B_{xy}^m$ as
\begin{equation}\label{eq:12.6}
    B_{xy}^m  = B_{xy}^m \, u_{xy}  + \mathcal{R}(x,y) \;,
\end{equation}
where $\mathcal{R}(x,y)$ is defined as
\begin{eqnarray}
    &&\mathcal{R}(x,y) = \sum_{b\in  \tilde{R}_{xy}^x} B_{xy}^m
    \, \chi_{xb}^c \prod\nolimits_{j:\, |j-x| < |b-x|} \chi_{xj}
    \nonumber\\ &&\ \ + \sum_{b\in  \tilde{R}_{xy}^y} B_{xy}^m\,
    \chi_{yb}^c \prod\nolimits_{i\in \tilde{R}_{xy}^x}\chi_{xi}\;
    \prod\nolimits_{j: \, |j-y|<|b-y|} \chi_{yj} \;. \label{rem}
\end{eqnarray}
We have $\langle B_{xy}^m \, u_{xy} \rangle \leq (1 - mC /
\beta)^{-1}$ by \eqref{eq:12.3}, without any need for an inductive
argument. Thus if we can prove that $\langle\mathcal{R}(x,y)
\rangle\leq 2 - (1 - mC/\beta)^{-1}$ our proof will be complete. The
desired statement is formulated in the next lemma.
\begin{lemma}\label{remainder}
For large $\beta$ the remainder \eqref{rem} is bounded in average by
\begin{equation}\label{eq:12.8}
    \langle\mathcal{R}(x,y)\rangle \leq 1/2 \;.
\end{equation}
Moreover, $\mathcal{R}(x,y)$ can be written as a sum over products of
$B_{x'y'}$ with $|x' - y'| \leq \ell-1$ in such a way that the
corresponding regions $R_{x'y'}$ are disjoint.
\end{lemma}

Now, using this lemma  and  arranging for $(1-mC/\beta)^{-1}$ not to
exceed $3/2$, we have
\begin{equation}
    \langle B_{xy}^m \rangle \leq 3/2 + 1/2 = 2 \;,
\end{equation}
thus completing the proof of the Induction Hypothesis for the case
$n_1 = 1$ and $n_2\, , n_3 = 0\,$. The general case -- \eqref{Ih3} --
is done in the same way. \medskip

Before starting the proof of Lemma \ref{remainder} we provide some
orientation and motivation. To bound the expectation of $B_{xy}^m \,
\chi_{xb}^c$ we could try to use \eqref{Bbound}:
\begin{equation}
    B_{xy}^m \leq 2^m B_{xb}^m B_{by}^m \;,
\end{equation}
while $\chi_{xb}^c = 1 - \chi_{xb}$ can be bounded by
\begin{equation}\label{eq:cheb}
    \chi_{xb}^c \leq B_{xb}^p \, a^{-p} |b-x|^{-\alpha p}
\end{equation}
where we used Def.\ \ref{def:3} of Section \ref{sect:6}. Since both
$|b-x|$ and $|b-y|$ are smaller than $\ell$ it is natural to try to
apply the Induction Hypothesis. However we face at least two
problems:
\paragraph{[A].}
The Induction Hypothesis does not cover $B_{xb}^{m+p}$ when $p>0\,$.
Indeed, a factor $\bar\chi_{xb} = \prod \chi_{xj} \prod \chi_{bj}$
must be present in order for \eqref{Ih2} to apply in such a case with
$0 < p \leq 2m$. Notice, however, that while we have no immediate
control without the missing factors $\chi_{bj}$, the factors
$\chi_{xj}$ for $j\in \tilde{R}_{xy}^x$ and $|j-x| < |b-x|$ are
already in place. To overcome the problem, we shall introduce the
needed factors $\prod \chi_{bj}$ by  the same partition of unity
scheme that was used  above.

\begin{figure}
\centerline{\psfig{figure=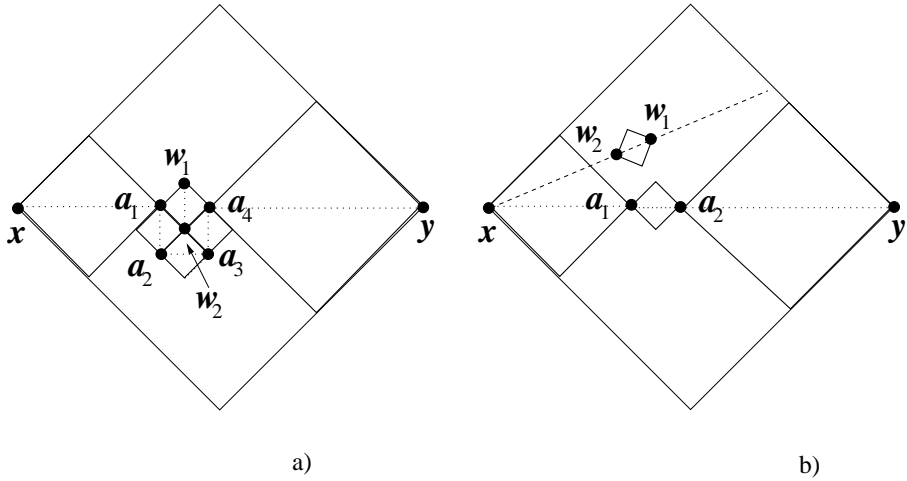, width=12cm}}
    \caption{a) If the pair $w_1 w_2$ is right in the middle, then
    we need to add four intermediate points $a_1, \ldots, a_4\,$.
    b) Even if the pair $w_1 w_2$ is located on the boundary of
    $\tilde{R}_{xy}^{x}\,$, the region $R_{w_1 w_2}$ still lies
    inside $R_{xy}\,$.} \label{fig:gbound1}
\end{figure}

Before embarking on that scheme, let us quickly evaluate the
situation which emerges after insertion of $\bar\chi_{xb}\,$. We can
then choose $p = 2m\,$, the induction \eqref{Ih2} applies, and we get
a small contribution
\begin{align}
    \langle B_{xy}^m \chi_{xb}^c \bar\chi_{xb} \rangle &\leq 2^m
    a^{-p} |b-x|^{-\alpha p} \langle B_{xb}^{m+p} \bar\chi_{xb}
    B_{by}^m \rangle \nonumber \\ &\leq 4 \cdot  2^m  a^{-2m}
    |b-x|^{-2\alpha m}  \label{eq:A}
\end{align}
for $|b-x|$ large. Here we used \eqref{Ih2} since $xb$ is of Class 2
and $by$ is of Class 1. Note that the expression \eqref{eq:A} is
summable in $b$ for $m \alpha$ large. Moreover, the factor $a^{-2m}$
ensures that also the contributions for $|b-x| = O(1)$ are small.
\paragraph{[B].}
The second problem is that, in order for \eqref{Ih2} to apply we must
make sure that we can find inside $R_{xy}$ two non-overlapping
regions $R_{xb}$ and $R_{by}$ of which the former is $C$-admissible
and the latter of diamond type. Moreover, since we have $\prod
\chi_{xj}$ only for $j \in \tilde{R}_{xy}^x$ we must ensure that
$R_{xb}$ is inside the reduced region $\tilde{R}_{xy}^x \cup
\tilde{R}_{xy}^y\cup \{x,y\}$ in $R_{xy}\,$. Since this might have to
be repeated many times at smaller and smaller scales, we must be sure
that all regions remain $\delta$-admissible (or, put differently, we
do not want $\delta$ to be scale-dependent). We will see in the next
lemma that this can be arranged.
\begin{lemma}\label{geomb}
For a diamond $R_{xy}$ consider the subsets $R_{xy}^x$ and $R_{xy}^y$
of \eqref{Rdef}.

[1] Let $w$ be any point in $\tilde{R}_{xy}^x\,$. Then we can always
find a point $a \in R_{xy}$ and regions $R_{xw}\,$, $R_{wa}\,$, and
$R_{ay} \,$, such that $R_{wa}$ and $R_{ay}$ are diamonds inside
$R_{xy}\,$, $R_{xw}$ lies inside $\tilde{R}_{xy}^x\cup\{x\}$ and is
$C$-admissible, and the three regions have disjoint interiors (see
Fig.\ \ref{fig:gbound}).

\begin{figure}
\centerline{\psfig{figure=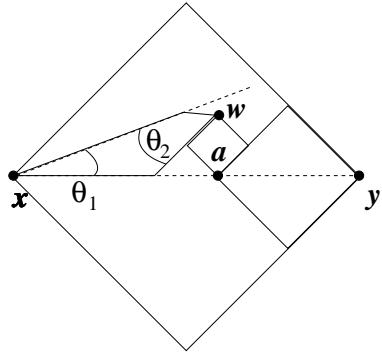, width=5cm}}
    \caption{We need one intermediate point $a$. The two angles
    $\theta_{1}$ and $\theta_{2}$ are never smaller than $\pi/8$.}
\label{fig:gbound}
\end{figure}

[2] Let $w$ be any point in $\tilde{R}_{xy}^x$ such that $|w-x| >
\beta^{1/4}$ ($xw$ not of Class 3). Let $w_1 w_2$ be a pair in
$\tilde{R}_{xy}^x \cup \tilde{R}_{xy}^y\cup\{x,y\}$ such that $|w_1 -
w| \leq |w - x|^{1/2}$ and $|w - x|^{1/2} \leq |w_1 - w_2| < |w -
x|/5\,$ (see \eqref{eq:case2a}). Let $R_{w_1 w_2}$ be the
corresponding diamond region. Then we can always find 4 points $a_i
\in R_{xy}$ ($i = 1, \ldots, 4$), such that all of the regions $R_{x
a_1}$, $R_{a_j a_{j+1}}$ ($j = 1, \ldots, 3$), and $R_{a_4 y}\,$, are
diamonds with disjoint interiors and do not overlap with $R_{w_1 w_2
}$ (see Fig.\ \ref{fig:gbound1}). The same can be done for $x \in
R_{xy}^y\,$.
\end{lemma}
\paragraph{Proof.}
The most dangerous situations are shown in Figures \ref{fig:gbound}
and \ref{fig:gbound1}. It is a simple geometrical argument to see
that the region $R_{xw}$ in Fig.\ \ref{fig:gbound} is $C$-admissible,
as the angles $\theta_{1}$ and $\theta_{2}$ are never smaller than
$\pi/8\,$, see \eqref{eq:12.1}. In the cases shown in Figs.\
\ref{fig:gbound1}a and \ref{fig:gbound1}b one has to check that the
diamonds do not transgress $R_{xy}\,$. This never happens since
$|w_{1}-x| \leq |x-y| /\sqrt{2}$ and $|w_{1} - w_{2}|\leq |w_{1} -
x|/5$. \qed

\paragraph{Proof of Lemma \ref{remainder}.}
We split the sum over bad points $b$ in
\eqref{rem} into several groups of terms.
\paragraph{Case 1.}
The bad point $b$ is located close to $x$, i.e., $|b-x|\leq
\beta^{1/4}$. Then we can bound $B_{xy}^m$ by \eqref{Bbound} and
$\chi_{xb}^c$ by \eqref{eq:cheb}, which gives
\begin{equation}
     B_{xy}^m \chi_{xb}^c\ \leq\  2^m a^{-p}
    |b-x|^{-\alpha p}  B_{xb}^{m+p} B_{by}^m \;.
\end{equation}
To apply the Induction Hypothesis we need to select inside the
diamond $R_{xy}$ two  regions $R_{xb}$ and $R_{by}\,$. The first one,
$R_{xb\,}$, need only be $C$-admissible (since $xb$ is of Class 3,
see Section \ref{sect:uncond}), so it may be a deformed diamond
(Fig.\ \ref{fig:case1}b). On the other hand, $R_{by}$ has to be
diamond-shaped, since $|b-y|>\beta^{1/4}$ (Class 1). To make the
requirement of diamond shape conform with our constrained geometry,
we must add an intermediate point $a$ as in Fig.\ \ref{fig:gbound}
with $w = b$, and use
\begin{equation}
    B_{by}^m \leq 2^m B_{ba}^m B_{ay}^m \;.
\end{equation}
We have seen in Lemma \ref{geomb} that we can always find such a
point $a$, so the induction \eqref{Ih2} does apply. Note that since
$|b-x| < \beta^{1/4}$ there will be no additional induction on
$R_{xb}\,$. Therefore there is no risk that the region might get more
and more deformed by the induction steps and $\delta$-admissibility
might finally be lost. Thus we have
\begin{equation}\label{eq:12b1}
    B_{xy}^m \chi_{xb}^c \leq 2^{2m} a^{-p}
    |b-x|^{-\alpha p} B_{xb}^{m+p} B_{ba}^m B_{ay}^m \;.
\end{equation}
The situation for $b$ near $y$ is analogous. Summing the
contributions from $b$ near $x$ or $y$ we obtain
\begin{align}
    &\sum_{z = x,y}\; \sum_{|b-z| \leq \beta^{1/4}} \left\langle
    B_{xy}^m\, \chi_{zb}^c \prod\nolimits_{j: \,|j-z|<|b-z|}
    \chi_{zj} \right\rangle \cr &\leq \sum_{z = x,y}\; \sum_{|b-z|
    \leq \beta^{1/4}} 4^m a^{-2m}  |b-x|^{-2\alpha m}\, 2^2
    (1+\rho)\cr &\leq 2^3 (1 + \rho)\, \frac{4^{m}}{a^{2m}} K_1
    \sum_{|b-x| = 1}^{\beta^{1/4}} |b-x|^{2-2\alpha m} \leq
    (4 / a^2)^m K'_1 \leq \frac{\rho}{10} \;, \label{eq:case1}
\end{align}
where in the second line  we used \eqref{Ih2} and $p = 2m\,$. We can
accommodate $m + p = 3m > m$ without any protection factor $\bar\chi$
since $bx$ is Class 3. In the third line, $K_1|b-x|^2$ is the entropy
factor for the 3D sum over bad points at distance $|b-x|$, the factor
$K'_1$ is a constant of order unity, and we used that $4 / a^2 < 1$
and $m > 4d / \alpha$ is large. We bounded the expression by
$\rho/10$ for convenience; since both $a$ and $m$ are large, the
factor $(4/a^2)^m K_1^\prime$ is in fact very small.
\paragraph{Case 2.}
The first bad point $b$ is far from $x$ (i.e., $|b - x| > \beta^{
1/4}$) and also far from $y$. Let us consider the case $b \in
\tilde{R}_{xy}^x$ for definiteness. (The other case, $b \in
\tilde{R}_{xy}^y\,$, is treated in the same way.) Again, we have to
estimate
\begin{equation}
    \left\langle B_{xy}^m \, \chi_{xb}^c \prod\nolimits_{
    j:\, |j-x| < |b-x|} \chi_{xj} \right\rangle \;.
\end{equation}
As was observed above, if we succeeded in promoting the last product
in the average to a complete factor $\bar\chi_{xb}\,$, then we could
apply the Induction Hypothesis as in \eqref{eq:A}. In order to
satisfy the hypothesis of Lemma \ref{lem:5}, Eq.\ \eqref{conditions},
we should have a constraint $\chi_{jb}$ for all $|j-b| \leq |b-x| /
\sqrt{2}$. Actually, from the remark after the proof of that lemma we
only require $\chi_{jb}$ for $|j-b| < |b-x| / 5$ since we know that
all $t_x - t_j$ fluctuations are good up to $|j-x| \approx|b-x|$.

Guided by the idea of partition of unity (cf.\ \eqref{part1}--\eqref{part2}), 
we will first check whether there is some
large fluctuation $\chi^c > 0$ at large scale near $b$. If no such
event occurs, we proceed to the step of checking fluctuations at
intermediate distance scales. Then either all intermediate distance
fluctuations are good too (and we have the desired factor
$\bar\chi$), or there must be some bad event at intermediate scale.
In this last case we will see that many bad events must happen. We
will now make this more precise.
\paragraph{Case 2a.}
The nearest bad point $b$ is far from $x$ (and $y$), $|b-x| >
\beta^{1/4}$, and there is a large scale bad event near $b$. This
means that $B_{j k} \geq a \, |j - k|^\alpha$ for some pair $j , k
\in \tilde{R}_{xy}^{x}\cup\tilde{R}_{xy}^{y}$ such that
\begin{equation}
    |j-b|\leq |b-x|^{1/2} \quad \text{and} \quad
    |b-x|^{1/2} \leq |j-k| \leq |b-x|/5 \;.
\label{eq:case2a}
\end{equation}

Now, using \eqref{Bbound} and \eqref{eq:cheb},
\begin{equation}\label{eq:12b2}
    B_{xy}^m \chi_{jk}^c \leq 2^{4m} B_{x a_1}^m
    \prod_{i=1}^{3} B_{a_{i} a_{i+1}}^m
    B_{a_4 y}^m B_{j k}^m \; |j-k|^{-\alpha m} a^{-m} \;.
\end{equation}

To apply the Induction Hypothesis the corresponding regions must all
be diamonds (all pairs are Class 1). By the assumptions made on the
pair $jk$, Lemma \ref{geomb} guarantees that we can choose the four
intermediate points $a_i \in R_{xy}$ ($i = 1, \ldots, 4$) so that all
of the regions $R_{x a_1}$, $R_{a_j a_{j+1}}$ ($j = 1, \ldots, 3$),
and $R_{a_4 y}\,$, are diamonds with disjoint interiors and do not
overlap with $R_{j k}$ (see Fig.\ \ref{fig:gbound1}). Since the
regions are non-overlapping and $\ell > |j-k| \geq |b-x|^{1/2}$, our
induction hypothesis yields
\begin{displaymath}
    \left\langle  B_{xy}^m \chi_{jk}^c \right\rangle
    \leq 2^{4m} 2^6 |b-x|^{-\alpha m/2} a^{-m} \;.
\end{displaymath}
For large $m$ the value of the sum over $b$ is small.

To estimate the entropy factor, note that there are less than $|b -
x|^{d + d/2}$ pairs $jk$ satisfying \eqref{eq:case2a}. Altogether
then, the present partial sum of contributions from $r \equiv |b-x|
\geq \beta^{1/4}$ is bounded by
\begin{displaymath}
    (4/a)^m K_2 \sum_{r > \beta^{1/4}} r^{(d-1)+d+d/2
    -\alpha m/2} = O \big(\beta^{-1/4} \big)<\frac{\rho}{10}\;.
\end{displaymath}
Note that there is nothing special or optimal about the exponent
$1/4$ of $1/\beta$ -- it is just convenient.

\paragraph{Case 2b.}
We now suppose that $|b-x|\geq \beta^{1/4}$ and there is \emph{no}
large deviation near $b$, i.e., $B_{jk} \leq a\,|j-k|^{\alpha}$ holds
for all $j , k$ subject to \eqref{eq:case2a}. This implies that at
long scales $|j-k| \geq |b - x|^{1/2}$ we have $\chi_{jk} = 1$. It
remains to check whether $\chi_{jk}$ holds also at shorter scales
$|j-k|\leq |b-x|^{1/2}$.

First we consider the case of there being a point $g$ ($g$ stands for
good) in $R_{xy}$ with $|g-b| \leq |b-x|^{1/2}$ such that $\chi_{g h}
= 1$ holds for all $h$ with $|g-h| \leq |b-x|^{1/2}$. We then have in
particular that $\chi_{g b} = 1$, and so by Def.\ \ref{def:3}
\begin{displaymath}
    B_{g b} \leq a\,|b-g|^{\alpha} \leq a\, (|b-x|^{1/2})^{\alpha}
    = a\, |b-x|^{\alpha/2} \;.
\end{displaymath}
This inequality combined with the constraint $\chi_{xb}^c = 1$ and
\eqref{Bbound} yields
\begin{displaymath}
    2 B_{xg} \geq \frac{B_{xb}}{B_{gb}} \geq \frac{a\,
    |b-x|^{\alpha}}{a\,|b-x|^{\alpha/2}} = |b-x|^{\alpha/2} \;.
\end{displaymath}
Thus we have
\begin{equation}\label{eq:12b3}
   B_{xy}^m\, \chi_{xb}^c\, \bar\chi_{xg} \leq 2^m (B_{xg}^m
    \bar\chi_{xg}) B_{gy}^m\, \chi_{xb}^c \leq 2^{3m} (B_{xg}^{3m}
    \bar\chi_{xg}) B_{gy}^m \, |b-x|^{-\alpha m} \;.
\end{equation}
Now we have to be somewhat careful about the choice of the regions
$R_{xg}$ and $R_{gy}\,$, as they may not have the canonical diamond
shape. For $R_{xg}$ this is not a problem, because of the presence of
$\bar\chi_{xg}\,$ ($xg$ is of Class 2). All we need is that $R_{xg}$
be $C$-admissible. On the other hand, $R_{gy}$ is (as in Case 1)
slightly more delicate. To be sure that we deal with diamond-shaped
regions, we add an intermediate point $a$ as in Fig.\
\ref{fig:gbound} and use $B_{gy}^m \leq 2^m B_{ga}^m B_{ay}^m\,$. We
have seen in Lemma \ref{geomb} that it is always possible to find
such a point $a$.

The regions $R_{ga}$ and $R_{ay}$ are of diamond type, so induction
applies. Note that $R_{xg}$ comes with a $\bar\chi_{xg}$ factor ($xg$
is of Class 2) so  no additional induction is required for it.
Therefore, as in Case 1, there is no risk that the region might get
more and more deformed by the induction steps. It should be
emphasized, however, that $\bar\chi_{xg}$ is not exactly the same as
in \eqref{chidef1}, but rather is given by
\begin{displaymath}
    \bar\chi_{xg} = \prod_{\tilde{R}_{xy}^x \ni j \,:\,|j-x|<|b-x|}\chi_{xj}
    \prod_{\tilde{R}_{xy}^x \ni j \,:\, |j-g| \leq |b-x|/5} \chi_{gj} \;.
\end{displaymath}
Since $|b-g| \leq |b-x|^{1/2}$ and $|b-x| > \beta^{1/4}$ we have
$|g-x| \simeq |b-x|$ up to a correction factor of order $O( |b-x|^{
-1/2}) \leq O (\beta^{-1/8}) \ll 1$. Therefore $\bar\chi_{xg}$ is
equivalent to the following constraints:
\begin{align}
    &\forall j \in R_{xg}\, , \; |j-x| \leq |g-x| f_1 \;: \quad
    B_{xj} \leq a\, |j-x|^{\alpha}\;, \nonumber\\ \text{and} \quad
    &\forall j \in R_{xg}\, , \; |j-g| \leq |g-x| f_2 \;: \quad
    B_{gj} \leq a\, |j-g|^{\alpha}\;, \nonumber
\end{align}
with $f_2 = 1/5$ and $f_1 = 1 - O(\beta^{-1/8})$. From Remark 7.3 we
know that Lemma \ref{lem:5} and hence Lemma \ref{lem:6} still hold,
so we can apply the induction and
\begin{displaymath}
    \left\langle B_{xy}^m \chi^c_{xb} \bar\chi_{xg}\right\rangle
    \leq 2^{3m} (1+\rho)\, 2^2 \, |b-x|^{-\alpha m} .
\end{displaymath}
There are $O (|b-x|^{d/2})$ choices for $g$, so the sum over
these contributions is bounded by
\begin{equation}
    2^{3m} K_3 \sum_{r > \beta^{1/4}} r^{d-1+d/2-\alpha m} =
    O(\beta^{-1/4}) < \frac{\rho }{10}\;.
\label{eq:2b}
\end{equation}
\paragraph{Case 2c.}
The last case to consider is the situation where no such point $g$
exists. In that case we can always find a cube $R_n \subset \tilde
{R}_{xy}^x \cup\tilde{R}_{xy}^y$ which contains the point $b$ and has
side $4^n = |b-x|^{1/2}$ such that $R_n$ is at least at distance
$\beta^{1/4}\leq 4^n \ll \ell$ from the boundary of $R_{xy}$ and
contains no $n$-good point (see Def.\ \ref{def:good} in Section
\ref{sect:8}). Then by \eqref{eq:tree} and \eqref{eq:cheb} with $p =
m$, we have
\begin{displaymath}
    \chi_{R_n}^c \leq \sum_{T \in \mathcal{T}_n} \sum_{\{j_v k_v
    \}_{v \in L_T}} \prod\nolimits_{v \in L_T} \chi_{j_v k_v}^c
    \leq \sum_{T \in \mathcal{T}_n}\sum_{\{j_v k_v \}_{v \in L_T}}
 \prod\nolimits_{v \in L_T}    
\frac{B^m_{j_v k_v}}{a^m |j_v - k_v|^{\alpha m}} \;,
\end{displaymath}
where according to \eqref{Scdef} the sum over configurations of pairs
$(j_v , k_v) \in R_v \times \Lambda$ is constrained by $4^{n_v - 1} <
| j_v - k_v | \leq 4^{n_v}$, with $n_v$ the scale of the leaf $v$.
Since all cubes $R_v$ are inside the small region $R_n$ and all pairs
$j_v , k_v$ satisfy the conditions of Lemma \ref{geomb} for the pair
$w_{1}w_{2}$, we can proceed as in Case 2b and select 4 intermediate
points $a_j$, $j = 1, \dotsc, 4$ such that the corresponding regions
are diamonds and do not overlap with any $R_{j_v k_v}$ (see Fig.\
\ref{fig:gbound1}). Then by \eqref{Bbound} we have
\begin{equation}\label{eq:12b4}
    B_{xy}^m \chi_{R_n}^c \leq 2^{4m} B^{m}_{x a_1}
    \cdots B^{m}_{a_4 y} \sum_{T \in \mathcal{T}_n}
    \sum_{\{j_v k_v \}_{v \in L_T}}\prod\nolimits_{v \in L_T} 
    \frac{B^m_{j_v k_v}}{a^m |j_v - k_v|^{\alpha m}} \;,
\end{equation}
and we can apply the Induction Hypothesis. By Proposition
\ref{lastcase} in Section \ref{sect:8} we have
\begin{equation}
    \left\langle B_{xy}^m  \chi_{R_n}^c \right\rangle \leq
    2^{4m} 2^{- n \alpha m}\;, \quad n \approx \ln |b-x|
    \geq \ln(\beta^{1/4}) \;.
\label{eq:2c}
\end{equation} 
Therefore we have enough decay to control the entropy factors:
\begin{displaymath}
    2^{4m} K_4 \sum_{r>\beta^{1/4}} |b-x|^{-\alpha m}
    < O(\beta^{-1/4}) < \frac{\rho}{10} \;.
\end{displaymath}
This concludes the proof of \eqref{eq:12.8}. From Eqs.\
\eqref{eq:12b1} and \eqref{eq:12b2}--\eqref{eq:12b4}, we see that
$\mathcal{R}(x,y)$ can be written as a sum over products of such
$B_{x'y'}$ with $|x'-y'|\leq \ell -1$ that the corresponding regions
$R_{x'y'}$ are disjoint. This concludes the proof of Lemma
\ref{remainder} and the Induction Hypothesis. \qed

\section{Proof of Theorem 2}\label{sect:10}\resetequ

Now that we have estimated $\langle B_{xy}^m \rangle$ for all $x,y$
we need to estimate  $\langle \cosh^p (t_x) \rangle$ for moderate
values of $p \leq 10$. If we suppose that the field $t$ is pinned at
some point $j_0\,$, so that $t_{j_0} = 0\,$, then Theorem 2 follows
directly from Theorem 1:
\begin{displaymath}
    \langle \cosh^p t_x \rangle =
    \langle \cosh^p (t_x-t_{j_0}) \rangle \leq 2\;,
\end{displaymath}
for any $x$ in the lattice (since Theorem 1 does not require bounds
on $\varepsilon$). When the field is not pinned, we need $\varepsilon
> 0$ and some conditions on the volume. The rest of this section is
devoted to this case. As in the proof of Theorem \ref{thm:1} we will
first prove bounds on conditional expectations.

\begin{defin}\label{def:chibar}
A point $x\in \Lambda$ is called `good at all scales' if
    \begin{displaymath}
    \forall j \in \Lambda\setminus \{x \}\; :
    \quad B_{xj} \leq a\,|j-x|^{\alpha}
\end{displaymath}
(see also Def.\ \ref{def:good} in Section \ref{sect:8}). The
corresponding characteristic function is
\begin{equation}\label{eq:chibar}
    \bar\chi_x := \prod\nolimits_{j \in \Lambda\setminus \{x\} }
    \chi_{xj} \;,
\end{equation}
where the factors $\chi_{xj}$ are those of Def.\ \ref{def:3} (Section
\ref{sect:6}).
\end{defin}
\begin{lemma}\label{Ngoodcase}
Let $x$ be good at all scales, and let
\begin{equation}
    B_x = \cosh t_x + {\textstyle{\frac{1}{2}}}\mathrm{e}^{t_x}
    s_x^2\;.
\end{equation}
If $\beta \gg 1$ and $\varepsilon \geq 8 p a\, L^{-d + \alpha}$, then
for any  $0 < p \leq O (\beta)$ we have
\begin{equation}
    \langle B_x^p \, \bar\chi_x \rangle \leq 2 \;.
\end{equation}
\end{lemma}
\paragraph{Proof.}
The proof uses a combination of ideas already present in the proofs
of Lemma \ref{lem:5} (Section \ref{sect:5}) and Lemma \ref{lem:6}
(Section \ref{sect:6}). By supersymmetry (Proposition
\ref{prop:SUSY}, Appendix \ref{app:Ward}) we have
\begin{equation}
    1 = \left\langle z_x^p \prod\nolimits_{j\in\Lambda \setminus
    \{x\} } \chi^{S}_{xj} \right\rangle \;,
\end{equation}
where $z_{x}$ is defined in \eqref{eq:def-z} and $\chi_{xj}^S$ in
\eqref{eq:chiS}. Following exactly the same steps as in the proof of
Lemma \ref{lem:6}, we obtain the inequality
\begin{equation}\label{eq:11.4}
    1 \geq \left\langle B_x^p\, \bar\chi_x (1-p\, G_x)\right\rangle
    \;, \qquad G_x = \frac{\mathrm{e}^{t_x}}{B_x}[\delta_x\,;
    D_{\beta,\varepsilon}(t)^{-1} \delta_x] \;,
\end{equation}
if $p\, G_{x} < 1$. We must now bound the Green's function $G_x$
using the constraint $\bar\chi_{x}$ (as we did in Lemma \ref{lem:5}).
For this purpose define $\tilde{D} = \mathrm{e}^{-t_x} B_x
D_{\beta, \varepsilon}(t)$ by
\begin{displaymath}
    [v\,;\tilde{D}v] = \mathrm{e}^{t_x} B_x \beta\sum\nolimits_{(ij)}
    \mathrm{e}^{t_i+t_j-2t_x} (v_i-v_j)^2 + \varepsilon B_x
    \sum\nolimits_k \mathrm{e}^{t_k - t_x} v_k^2 \;,
\end{displaymath}
and note that $B_{xj} \leq a\,|j-x|^\alpha$ implies the bound
\begin{displaymath}
    \mathrm{e}^{t_j - t_x} \geq (2 a \, |j - x|^\alpha)^{-1} \;.
\end{displaymath}
We then follow the proof of Lemma \ref{lem:5} and introduce a
telescopic sum
\begin{equation}
    \delta_x = (\delta_x - I_1) + (I_1 - I_2) + \ldots
    + (I_{N-1}-I_N) + I_N = \sum_{n=0}^N \rho_n \;,
\end{equation}
where $I_n$ is the (normalized) indicator function of a cube of
center $x$ and side $2^n$, and $\rho_n = I_n - I_{n+1}\,$. There is
no need to introduce $\tilde{I}$ as we did in the proof Lemma
\ref{lem:5}, as we are now working not on $R_{xy}$ but on the whole
volume. The sum terminates on reaching the system size $2^N$. Note
that for $n < N$ we have $\sum_j \rho_n (j) = 0$ and  $\Vert \rho_n
\Vert_2^2\leq 2^{-nd} = 2^{-3n}$. The function $\rho_n$ for $n = N$
is constant: $\rho_N (j) = I_N (j) = |\Lambda|^{-1}$ for all $j \in
\Lambda$.

Now, by the Cauchy-Schwarz inequality,
\begin{equation}
    [\delta_x \,;\tilde{D}^{-1} \delta_x] \leq \left(\sum_{n=0}^N
    [\rho_n \,;\tilde{D}^{-1} \rho_n]^{1/2} \right)^2 \;.
\end{equation}
For $n < N$ we use the bound $[\rho_n\,; \tilde{D}^{-1} \rho_n] \leq
\Vert (\tilde{D}^{-1})_{\rho_n} \Vert\, \Vert \rho_n \Vert_2^2\,$,
where $\Vert (\tilde{D}^{-1})_{\rho_n} \Vert\leq 2^{2n+2n\alpha}
c_1\,$. Then for the sum of terms with $n < N$ we have
\begin{equation}
    \sum_{n=0}^{N-1} [\rho_n \,; \tilde{D}^{-1} \rho_n]^{1/2}
    \leq \frac{c_1}{\sqrt{\beta}} \sum_{n=0}^{N-1} \sqrt{2}^{\,
    -n(d-2-2\alpha)} < \frac{\gamma_1}{\sqrt{\beta}}
 \end{equation}
uniformly in $N$ since $d=3$ and $2 \alpha \ll 1$. For $n=N$, on the
other hand, we no longer have orthogonality to the constant functions
(`zero mode') and therefore must take recourse to the $\varepsilon
$-term in $\tilde{D}:$
\begin{equation}
    [\rho_N\,;\tilde{D}^{-1}\rho_N] \leq \big( \varepsilon
    |\Lambda| \min_{j\in \Lambda} \mathrm{e}^{t_j - t_x}
    \big)^{-1} \leq 2a \frac{L^\alpha}{\varepsilon L^d}
    \leq \frac{1}{4p} \;.\label{eq:13.9}
\end{equation}
Hence $G_x \leq (\gamma_1 / \sqrt{\beta} + 1/\sqrt{4p}\,)^2 < 1/
(2p)$ for $\beta \gg 1$. So,
\begin{equation}
    {\textstyle{\frac{1}{2}}} \langle B_x^p \, \bar\chi_x \rangle
    \leq \langle B_x^p \, \bar\chi_x (1 - p\,G_x) \rangle \leq 1
\end{equation}
by \eqref{eq:11.4}, and the lemma is proved.
\qed

With this lemma we can finally complete the proof of Theorem
\ref{thm:2}, i.e.\ the bound on the unconditional expectation of
$\cosh^p t_x\,$.
\paragraph{Proof of Theorem \ref{thm:2}.}
We recall from Def.\ \ref{def:good} (Section \ref{sect:8}) that a
point $x$ is said to be $n$-good if $B_{x j} \leq a\,|j-x|^\alpha$
for all $j \in \Lambda$ subject to $1 \leq |j-x| \leq 4^n$. A point
$x$ is good at all scales if $B_{x j}\leq a\,|j-x|^\alpha$ for all $j
\in \Lambda\setminus \{x \}$; we then say that $x$ is $N$-good.

We proceed as in Lemma \ref{remainder} (Section \ref{sect:9}):
\begin{equation}
    \langle \cosh^p t_x \rangle = \langle \bar\chi_x \cosh^p t_x
    \rangle + \langle \bar\chi_x^c \cosh^p t_x \rangle \;,
\end{equation}
where $\bar\chi_x$ ensures that the point $x$ is $N$-good. Then by
Lemma \ref{Ngoodcase} we have
\begin{equation}
    \langle \bar\chi_x \cosh^p t_x \rangle \leq 2\;.
\end{equation}
It remains to estimate the second term, $\langle \bar\chi_x^c \cosh^p
t_x \rangle$. We prove in Lemma \ref{lem:rem2} below that this term
is bounded by a constant. Once this has been accomplished, the proof
of Theorem 2 will be finished. \qed
\begin{lemma}\label{lem:rem2}
Let $\bar\chi_x^c = 1 - \bar\chi_x\,$, with $\bar\chi_x$ defined by
\eqref{eq:chibar}. Let $\beta \gg 1$ and $\varepsilon \geq \, 8 \cdot
4 \cdot 10\, a\, L^{\alpha - d}$. Then for any $0\leq p\leq 10$ we
have
\begin{equation}
    \langle \bar\chi_x^c \cosh^p t_x \rangle \leq 1/2 \;.
\end{equation}
\end{lemma}
\paragraph{Proof.}
While $\bar\chi_x^c$ means that $x$ is not $N$-good, it is still
possible for other points in $\Lambda$ to be $N$-good. If $g \not= x$
is the nearest such point (as seen from $x$), then none of the points
inside the ball $K_{|g-x|}^x$ of radius $|g-x|$ and center $x$ is
$N$-good. Denoting the indicator function for the latter event by
$\chi_{K_{|g-x|}^x}^c$ we have the identity
\begin{equation}
    \bar\chi_x^c = \sum_{\Lambda \ni g \not= x} \bar\chi_g \,
    \chi_{K_{|g-x|}^x}^c + \prod_{j \in \Lambda} \bar\chi_j^c \;,
\end{equation}
where the last term accounts for the possibility that there is no
$N$-good point in $\Lambda$ at all. Thus we obtain the decomposition
\begin{equation}\label{eq:11.7}
    \langle \bar\chi_x^c \cosh^p t_x \rangle = \sum_{g \not= x}
    \left\langle \bar\chi_g \, \chi_{K_{|g-x|}^x}^c \cosh^p t_x
    \right\rangle + \left\langle \prod\nolimits_j \bar\chi_j^c
    \cosh^p t_x \right\rangle \;.
\end{equation}
We will prove that both of these two terms are bounded by $1/4$.
\paragraph{1.}
We consider the first sum. Using $\cosh t_x \leq 2 \cosh (t_x-t_g)
\cosh t_g$ and applying the Cauchy-Schwarz inequality twice, we
obtain
\begin{align*}
    &\big\langle \chi_{K_{|g-x|}^x}^c \bar\chi_g \cosh^p t_x
    \big\rangle \leq \big\langle \chi_{K_{|g-x|}^x}^c \bar\chi_g
    \big\rangle^{1/2} \big\langle \bar\chi_g \cosh^{2p} t_x
    \big\rangle^{1/2} \cr
    &\leq \big\langle \chi_{K_{|g-x|}^x}^c \bar\chi_g
    \big\rangle^{1/2} \, 2^p \big\langle \bar\chi_g
    \cosh^{2p}(t_x - t_g) \cosh^{2p} t_g \big\rangle^{1/2} \cr
    &\leq 2^p \, \big\langle \chi_{K_{|g-x|}^x}^c \bar\chi_g
    \big\rangle^{1/2} \big\langle \bar\chi_g \cosh^{4p} t_g
    \big\rangle^{1/4}\big\langle \cosh^{4p}(t_x - t_g)
    \big\rangle^{1/4} \cr
    &\leq  2^p \,c_p\; \big\langle \chi_{K_{|g-x|}^x}^c \bar\chi_g
    \big\rangle^{1/2},
\end{align*}
where in the last step we used Lemma \ref{Ngoodcase} and Theorem
\ref{thm:1}, and we introduced $c_0 = 1$ and $c_p = 2^{1/2}$ for $p
\geq 1$. It remains to bound
\begin{equation}\label{eq:13.16}
    2^p c_p \sum_g \big\langle \chi_{K_{|g-x|}^x}^c
    \bar\chi_g \big\rangle^{1/2} = 2^p c_p \sum_{n\geq 0}
    \; \sum_{4^{n}\leq |g-x|< 4^{n+1}} \big\langle
    \chi_{K_{|g-x|}^x}^c \bar\chi_g \big\rangle^{1/2}\;.
\end{equation}
Let $R_{n}^{x}$ be the cube centered at $x$ of side $4^{n}$. Now
fixing a point $g$ with $4^{n} \leq |g-x| < 4^{n+1}$ we have
$R_{n}^{x}\subseteq K_{|g-x|}^x\;$, and  we distinguish between two
cases:
\paragraph{1a.}
The interior of $K_{|g-x|}^x$ is void not only of $N$-good points but
also of $n$-good points. Let $\chi_n^c$ denote the corresponding
indicator function. Then for $n\geq 1$, using Proposition
\ref{lastcase} we have
\begin{equation}
    \big\langle\chi_{K_{|g-x|}^x}^c \bar\chi_g\,\chi_n^c
    \big\rangle \leq \big\langle \chi_{R_{n}^{x}}^c \big\rangle
    \leq 2^{- (n+1)\alpha m} \;,
\end{equation}
where $ \chi_{R_n^x}^c$ is given in Def.\ \ref{def:ngp}, Section
\ref{sect:8}. For $n = 0$ the cube $R_0^x$ contains only the point
$x$, so
\begin{displaymath}
    \big\langle \chi_{R_0^x}^c \big\rangle \leq
    \sum_{|z-x|=1} \big\langle \chi_{xz} \big\rangle \leq
    \sum_{|z-x|=1} \frac{\big\langle B_{xz}^m \big\rangle}{a^m}
    \leq 2\frac{2d}{a^m} < 2^{-\alpha m} \;.
\end{displaymath}
\paragraph{1b.}
There is at least one $n$-good point $y$ inside $K_{|g-x|}^x\,$. Let
$\chi_n(y)$ be the corresponding indicator function. Because the
point $y$ cannot be $N$-good, there must be a first scale $q>n$ so that
$y$ is $q$-bad. Thus there exists a first point $b$ at distance
$|b-y| > 4^n$ with $B_{y b} > a\, |b-y|^\alpha$. It follows that
\begin{align*}
    \sum_{y \in K_{|g-x|}^x} \big\langle \bar\chi_g
    \chi_{K_{|g-x|}^x}^c \chi_n(y) \big\rangle &\leq
    \sum_{y \in K_{|g-x|}^x} \sum_{b:\, |b-y| > 4^n}
    \frac{\langle B_{yb}^m \rangle}{a^m |b-y|^{\alpha m}}\\
    &\leq \frac{4^{nd} k_1}{a^m} \sum_{r > 4^n} \frac{2\,
    r^{d-1} k_2}{r^{\alpha m}} \leq 2^{-n\alpha m} a^{-m}
    \leq  2^{- (n+1)\alpha m} \;,
\end{align*}
where the factor $4^{nd} k_1$ comes from the sum over $y$ and $r^{d
-1} k_2$ comes from the sum over $b$. Inserting these results into
\eqref{eq:13.16} we obtain
\begin{displaymath}
    2^p c_p \sum_{n \geq 0} \sum_{4^n \leq |g-x|\leq 4^{n+1}}
    \big\langle \chi_{K_{|g-x|}^x}^c \bar\chi_g \big\rangle^{1/2}
    \leq 2^p c_p \sum_{n\geq 0} 4^{(n+1)d} k_3 \big(2 \cdot
    2^{-(n+1)\alpha m} \big)^{1/2} \;,
\end{displaymath}
where $4^{(n+1)d} k_3$ comes from the sum over $g$. This will be no
greater than $1/4$ provided that $\alpha m$ is  large enough.
\paragraph{2.}
To complete the proof, we have to estimate the last term $\langle
\prod_j \bar\chi_j^c \cosh^p t_x \rangle$ in \eqref{eq:11.7}. By
Proposition \ref{lastcase} the probability for no $N$-good point to
be found in a cube $\Lambda$ of side $L = 4^N$ is bounded by $2^{-N
\alpha m} = L^{-\alpha m /2}$. Hence
\begin{displaymath}
    \big\langle \prod\nolimits_j \bar\chi_j^c \cosh^p t_x
    \big\rangle \leq \big\langle \cosh^{2p} t_x \big\rangle^{1/2}
    \big\langle \prod\nolimits_j \bar\chi_j^c \big\rangle^{1/2} \leq
    \big\langle \cosh^{2p} t_x \big\rangle^{1/2} L^{-\alpha m/4}\;.
\end{displaymath}
To get a bound on the expected value of $\cosh^{2p} t_x$ we once
again use supersymmetry (Proposition \ref{prop:SUSY}), as follows:
\begin{displaymath}
    \mathrm{e}^{\gamma \varepsilon} = \langle \mathrm{e}^{\gamma
    \varepsilon z_x}\rangle = \langle \mathrm{e}^{\gamma\varepsilon
    B_x} (1 - \gamma \varepsilon\, G'_x)\rangle \;,
\end{displaymath}
where we choose $0 < \gamma < 1/2$, and $G'_x = \mathrm{e}^{t_x}
[\delta_x\,;D_{\beta,\varepsilon}(t)^{-1} \delta_x]$. Since the
operator $D_{\beta,\varepsilon}(t) - \varepsilon\, \mathrm{e}^{t_x}
\delta_x [\delta_x\,; \cdot]$ is non-negative, by Lemma \ref{lem:K}
we have $\varepsilon\, G_x^\prime \leq 1$, so
\begin{displaymath}
    \langle \mathrm{e}^{\gamma\varepsilon (B_x - 1)} \rangle
    \leq (1-\gamma)^{-1} \;.
\end{displaymath}
Also, $\cosh^{2p} t_x \leq (2p)! \,(\gamma\varepsilon)^{-2p}\,
\mathrm{e}^{\gamma\varepsilon B_x}$ by an elementary computation, and
hence
\begin{displaymath}
    \langle\cosh^{2p} t_x \rangle \leq (2p)!\,(\gamma\varepsilon)
    ^{-2p} \langle \mathrm{e}^{\gamma\varepsilon B_x}\rangle \leq
    O(\varepsilon^{-2p}) \;.
\end{displaymath}
We thus finally obtain
\begin{equation}\label{eq:13.20}
    \big\langle \prod\nolimits_j \bar\chi_j^c \cosh^p t_x \big\rangle
    \leq \big\langle \cosh^{2p} t_x \rangle^{1/2} L^{-\alpha m / 4}
    \leq O(\varepsilon^{-p}) L^{-\alpha m/4} < 1/4\;,
\end{equation}
since $\alpha m$ is large and $\varepsilon \geq  L^{\alpha-d}$. This
concludes the proof of Lemma \ref{lem:rem2}. \qed

\paragraph{Remark.} In the proof of Theorem \ref{thm:2} the
$\varepsilon$ term (zero mode) appears only in two places:
\eqref{eq:13.9} of Lemma \ref{Ngoodcase} (the last term in the
telescopic sum) and \eqref{eq:13.20} (when no $N$ good point is
present). The inequality \eqref{eq:13.9} is the reason why we cannot
take $\varepsilon = O(L^{-d})$ but must take $\varepsilon = O(
L^{\alpha-d})$.

\section{Proof of Theorem \ref{thm:new}}\label{sec:new}\resetequ

Finally we can prove the bound on the Green's function $C_{xy}$ of
\eqref{eq:C-xy}. Let $f$ be such that $f(j) \geq 0$ for all $j \in
\Lambda $. We need to estimate
\begin{equation}\label{eq:Crep}
    [f ; Cf] = \left\langle  [\mathrm{e}^t f ;
    D_{\beta,\varepsilon}(t)^{-1} \mathrm{e}^t f] \right\rangle
    = \left\langle [W ; G_t W] \right\rangle \;,
\end{equation}
where $D_{\beta,\varepsilon}(t)^{-1} = G_t$ was defined in
\eqref{eq:1.1}, and $W(j) = \mathrm{e}^{t_j} f(j)$.

\subsection{Upper bound}

Let $L_0 = -\beta\Delta + \varepsilon$ and $G_0 = L_0^{-1}$ (as
defined in the statement of the Theorem). Now
\begin{align*}
    [W ; G_t W] &= [L_0 G_0 W ; G_t W] = \beta\, [\nabla (G_0 W)
    ; \nabla (G_t W) ] + \varepsilon\, [G_0 W; G_t W ] \\
    &= \beta \sum_{(jj')} \nabla_{jj'} (G_0 W) \nabla_{jj'}
    (G_t W) + \varepsilon \sum_j (G_0 W) (j) (G_t W) (j) \cr
    &= \beta \sum_{(jj')} \frac{\nabla_{jj'}(G_0 W)}
    {\mathrm{e}^{(t_j + t_{j'})/2}}\; \frac{\nabla_{jj'}(G_t W)}
    {\mathrm{e}^{-(t_j + t_{j'})/2}} + \varepsilon \sum_j \left(
    \frac{(G_0 W)(j)}{ \mathrm{e}^{+t_j / 2}} \right)
    \left( \frac{(G_t W)(j)}{ \mathrm{e}^{-t_j / 2}} \right) \;.
\end{align*}
Since $|a\cdot b + c\cdot d| \leq (a\cdot a + c\cdot c)^{1/2} (b\cdot
b + d\cdot d)^{1/2} $ we have
\begin{displaymath}
    [W ; G_t  W] \le \left( \beta \sum_{(jj')}
    \frac{|\nabla_{jj'}(G_0 W)|^2}{\mathrm{e}^{t_j + t_{j'}}}
    + \varepsilon \sum_j \frac{|(G_0 W)(j)|^2}{\mathrm{e}^{t_j}}
   \right)^{1/2} [W ; G_t W]^{1/2}.
\end{displaymath}
Therefore\footnote{We thank S.R.S.\ Varadhan for explaining the
inequality \eqref{eq:V} to us.}
\begin{equation}\label{eq:V}
    [W ; G_t W] \le \beta \sum_{(jj')} \frac{|\nabla_{jj'}
    (G_0 W)|^2}{\mathrm{e}^{t_j + t_{j'}}} + \varepsilon
    \sum_j \frac{|(G_0 W)(j)|^2}{\mathrm{e}^{t_j}} \;.
\end{equation}
Now
\begin{equation}\label{eq:14.3}
    |\nabla_{jj'}(G_0 W)| \leq \sum_k |(G_0(j,k) - G_0(j',k))|
    \, W(k) \leq \mathrm{const} \sum_k H_{jk} W(k) \;,
 \end{equation}
where we defined  $H_{jk} = \beta^{-1} (|j-k|^2+1)^{-1} \mathrm{e}^{
- \tilde\varepsilon |j-k|}$, $\tilde\varepsilon = (\varepsilon / 2
\beta)^{1/2}$, and we used
\begin{displaymath}
    |(G_0(j,k) - G_0(j',k))| \leq \mathrm{const}\, H_{jk}\;.
\end{displaymath}
By inserting \eqref{eq:14.3} into \eqref{eq:V} we get
\begin{align}
    [f ; C f] &\leq \mathrm{const}\, \beta\sum_{(j,j'),k,l}
    H_{jk} H_{j l} \, f(k)f(l)\, \big\langle \mathrm{e}^{(t_k +
    t_l - t_j - t_{j'})} \big\rangle\cr
    &+ \varepsilon \sum_{j,k,l} G_0(j,k)\, G_0(j,l)\, f(k) f(l)\,
    \big\langle \mathrm{e}^{t_k + t_l - t_j} \big\rangle \;.
\end{align}
By Theorems \ref{thm:1} and \ref{thm:2} the expectation over the
field $t$ is uniformly bounded. Now we can sum over $j:$
\begin{equation}
    \sum_j H_{jk} H_{j l} \leq \mathrm{const}\ \tilde{G}_0(k,l),
    \qquad \sum_j  G_0(j,k)\, G_0(j,l) = G_0^2 (k,l) ,
\end{equation}
where $\tilde{G}_0 = (-\beta \Delta +\varepsilon/2)^{-1}$. Note that
$G_0 \leq \tilde{G}_0\,$. We finally obtain
\begin{equation}
    [f ; C f] \le \mathrm{const}\ [f ; \tilde{G}_0 f ] + \varepsilon
    [f ; G_0^2 f] \leq 2\, \mathrm{const}\, [f ; \tilde{G}_0 f ]\;.
\end{equation}
This completes our proof of the upper bound. \qed

\subsection{Lower bound}

Let $\bar\chi_x$ be the characteristic function ensuring that $x \in
\Lambda$ is good at all scales (see \eqref{eq:chibar}). Recall that
if $\bar\chi_x > 0$ then
\begin{equation}\label{eq:bj}
    \mathrm{e}^{t_j - t_x}\geq (2a (1+|j-x|^{\alpha }))^{-1}
\end{equation}
for all $j \in \Lambda$. We have the inequality $1 = \bar\chi_x +
\bar\chi_x^c \geq \bar\chi_x\,$. Inserting it into \eqref{eq:Crep} we
obtain
\begin{align}
    \left\langle [W ; G_t W] \right\rangle \geq &\left\langle
    \bar\chi_x [W ; G_t W] \right\rangle = \sum_{jk} \big\langle
    \bar\chi_x W(j)W(k)\, G_t(j,k) \big\rangle \\
    \geq &\frac{1}{4 a^2} \sum_{jk} \tilde{f}(j) \tilde{f}(k)
    \big\langle \bar\chi_x \bar{D}_t^{-1} (j,k) \big\rangle
    = \frac{1}{4 a^2} \big\langle \bar\chi_x [ \tilde{f} ;
    \bar{D}_t^{-1}\tilde{f} ] \big\rangle\;,\nonumber
\end{align}
where $\tilde{f}(j) = (1+|j-x|^\alpha)^{-1} f(j)$ and $\bar{D}_t^{-1}
= \mathrm{e}^{2 t_x}G_t  = (\mathrm{e}^{-2 t_x} D_{\beta,\varepsilon}
(t))^{-1}$. In the first line we used the fact that $G_t$ is positive
as a quadratic form for each configuration of $t$. In the second line
we used the fact that this is a sum of positive terms since $W(j)
\geq 0$ and $G_t$ is pointwise positive. Furthermore, we applied
\eqref{eq:bj} to estimate $W(j)$. Now,
\begin{equation}
    \big\langle \bar\chi_x [\tilde{f} ; \bar{D}_t^{-1} \tilde{f} ]
    \big\rangle = \left\langle \bar\chi_x \right\rangle \mathrm{E}
    \big( [\tilde{f} ; \bar{D}_t^{-1} \tilde{f} ] \big) \geq
    \left\langle \bar\chi_x \right\rangle [\tilde{f} ; \mathrm{E}
    (\bar{D}_t)^{-1} \tilde{f}]
\end{equation}
where
\begin{displaymath}
    \mathrm{E}(\cdot) = \frac{\left\langle \bar\chi_x \,
    \cdot \right\rangle}{\left\langle \bar\chi_x \right\rangle}
\end{displaymath}
is a probability distribution and we used Jensen's inequality.

In order to complete the proof we need to estimate $\left\langle
\bar\chi_x \right\rangle$ and $\left\langle \bar\chi_x \bar{D}_t
\right\rangle$. From Lemma \ref{lem:rem2} (in the previous section)
with $p = 0$ we know that
\begin{displaymath}
    \left\langle \bar\chi_x \right\rangle = 1 -
    \left\langle \bar\chi_x^c \right\rangle \geq 1/2 \;.
\end{displaymath}
Moreover $\left\langle \bar\chi_x \bar{D} \right\rangle \leq
\left\langle \bar{D} \right\rangle$ as a quadratic form and for any
function $u$ we have
\begin{align}
    [u;\left\langle \bar{D} \right\rangle u] &= \beta \sum_{(jk)}
    (u(j) - u(k))^2 \left\langle\mathrm{e}^{t_j + t_k - 2t_x}
    \right\rangle + \varepsilon \sum_j u(j)^2 \left\langle
    \mathrm{e}^{t_j - 2t_x} \right\rangle\cr &\leq \beta c \sum_{(jk)}
    (u(j)- u(k))^2 +\varepsilon c' \sum_j u(j)^2 \leq c_1 [u ; G_0 u]\;,
\end{align}
where we applied Theorems \ref{thm:1} and \ref{thm:2}, and $c_1 =
\sup \{ c , c' \}$. Thus $\left\langle \bar{D} \right\rangle \leq c_1
G_0\,$. By applying these relations we see that $\mathrm{E}
(\bar{D}_t) \leq 2 c_1 G_0$ and hence
\begin{equation}
    [f ; C f] \geq \frac{1}{4a^2 c_1}\;
    [\tilde{f} ; G_0 \tilde{f}] \;.
\end{equation}
This concludes the proof of Theorem \ref{thm:new}. \qed

\paragraph{Remark.} If $W$ did not depend on $t$ we would have the
quadratic form estimate
\begin{displaymath}
    [W ; G_0 W]\, c_1 \leq \left\langle [W ; G_t W] \right\rangle
    \leq c_2\, [W ; G_0 W]
\end{displaymath}
with
\begin{displaymath}
    c_2 = \sup_{(jj'),k} \left(
    \big\langle \mathrm{e}^{- t_j - t_{j'}} \big\rangle\, ,\,
    \big\langle \mathrm{e}^{-t_k} \big\rangle \right) \, ,\quad
    c_2 = \sup_{(jj'),k} \left( \big\langle \mathrm{e}^{
    t_j + t_{j'}} \big\rangle\, ,\, \big\langle \mathrm{e}^{t_k}
    \big\rangle \right)\,.
\end{displaymath}
The upper bound follows directly from \eqref{eq:V}, the lower bound
from Jensen's inequality.

\vspace{1cm}\centerline{\Large Appendices}
\resetsect
\setcounter{section}{0}
\renewcommand{\thesection}{\Alph{section}}

\section{Minimum of the effective action}\label{app:minimum}
\resetequ

Let $j \mapsto t_j \equiv \bar{t} \in \mathbb{R}$ (for all $j \in
\Lambda$) be a constant field configuration. Evaluating the
statistical weight function on it we get
\begin{equation}\label{eq:A.1}
    \mathrm{e}^{-\varepsilon|\Lambda|(\cosh\bar{t}-1)}\,\mathrm{Det}
    ^{1/2} (-\beta \Delta + \varepsilon\, \mathrm{e}^{-\bar{t}})\;.
\end{equation}
Let $t^*$ be the number that maximizes this statistical weight. The
condition for the first derivative to vanish at $t^\ast$ is
\begin{equation}
    2 \sinh t^* = - \mathrm{e}^{- t^\ast} G_0(x,x) \qquad
    (x \in \Lambda) \;,
\end{equation}
where $G_0 \equiv (-\beta\Delta + \varepsilon\, \mathrm{e}^{
-t^*})^{-1} \geq 0\,$. Equivalently, $1 - \mathrm{e}^{2 t^*} =
G_{0}(x,x)$, and since $G_0$ is non-negative, it follows that $t^\ast
\leq 0\,$. We thus infer that
\begin{equation}
    0 \leq 1 - \mathrm{e}^{2 t^*} = G_{0}(x,x) \leq 1 \;.
\end{equation}

Next, we show that the constant field $t^*$ maximizes the integrand
over the full set of all field configurations $t = \{ t_j \}$. For
this, we recall the definition \eqref{eq:1.2} of the effective action
or free energy $F_{\beta,\varepsilon}$ in combination with
\eqref{eq:Dtilde}:
\begin{eqnarray*}
    F_{\beta,\varepsilon}(t) &=& \beta \sum\nolimits_{(ij)} (\cosh
    (t_i-t_j)-1) + \varepsilon \sum\nolimits_k (\cosh t_k - 1)
    \nonumber\\ &-& \ln \mathrm{Det}^{1/2}\big(-\beta\Delta +
    \beta V(t) + \varepsilon\, \mathrm{e}^{- t} \big) \;.
\end{eqnarray*}
Now we introduce $A := G_0^{1/2} \big(\beta V(t) + \varepsilon\,
(\mathrm{e}^{- t} - \mathrm{e}^{- t^* \mathrm{Id}}) \big)\,
G_0^{1/2}$ and write
\begin{displaymath}
    \mathrm{Det}\big( -\beta \Delta + \beta V(t) +
    \varepsilon\, \mathrm{e}^{- t} \big) = \mathrm{Det}
    (G_0^{-1})\, \mathrm{Det}(\mathrm{Id} + A) \;.
\end{displaymath}
Using $\ln \mathrm{Det}(\mathrm{Id} + A) \leq \mathrm{Tr}\, A$ we
then obtain
\begin{eqnarray*}
    F_{\beta,\varepsilon}(t) &\geq& -\ln\mathrm{Det}(G_0^{-1/2})
    + \varepsilon |\Lambda|\, (\cosh t^\ast - 1) \\ &+&\beta
    \sum\nolimits_{(ij)} (\cosh(t_i - t_j) - 1)- G_0(x,x)\,
    {\textstyle{\frac{1}{2}}} \sum\nolimits_k \beta V_{kk} \\ &+&
    \varepsilon \sum\nolimits_k \big( \cosh t_k -\cosh t^\ast
    - {\textstyle{\frac{1}{2}}} (\mathrm{e}^{-t_k} -
    \mathrm{e}^{-t^\ast}) \, G_0(x,x) \big) \;.
\end{eqnarray*}
The second line of the r.h.s.\ is non-negative by $\frac{1}{2} \sum_j
V_{jj} = \sum_{(ij)} (\cosh(t_i - t_j) - 1)$ and $G_0(x,x) \leq 1$,
and so is the third line by the identity $G_0(x,x) = 1 - \mathrm{e}^{
2t^\ast}$ and a trivial computation. This proves that $F_{\beta,
\varepsilon}(t)$ is bounded from below by $F_{\beta,\varepsilon}
(t^\ast) = - \ln \mathrm{Det}(G_0^{-1/2}) + \varepsilon |\Lambda|\,
(\cosh t^\ast - 1)$.

\section{Hyperbolic symmetry}\label{app:symmetry}
\resetequ

In Section \ref{sect:syms} we explained that the $\mathrm{H}^{2|2}$
nonlinear sigma model in the limit of vanishing regularization
$\varepsilon \to 0+$ acquires a global symmetry by the Lorentz group
$\mathrm{SO}(1,2)$. We will now exhibit the Ward identities due to
this Lorentzian symmetry $\mathrm{SO} (1,2)$. (Consequences due to
the supersymmetries of model will be explored in Appendix
\ref{app:Ward}.) To prepare the discussion, the reader is invited to
recall the expressions \eqref{eq:horoc} for the functions $x, y, \xi,
\eta$ in horospherical coordinates. We also recall that $z = \cosh t
+\mathrm{e}^t (\frac{1}{2}s^2 + \bar\psi \psi)$.

We now seek the first-order differential operator, $L_1$, generating
Lorentz boosts in the $zx$-plane, i.e.,
\begin{displaymath}
    L_1\, z = x \;, \quad L_1\, x = z \;, \quad
    L_1\, y = L_1\, \xi\, = L_1 \eta = 0 \;.
\end{displaymath}
It is easy to verify that the unique operator with these properties
is
\begin{equation}
    L_1 = \partial_t - \bar\psi \partial_{\bar\psi}
    - \psi \partial_\psi - s \partial_s \;.
\end{equation}
Similarly, the generator $L_2$ of Lorentz boosts in the $zy$-plane
and the generator $L_0$ of Euclidean rotations in the $xy$-plane, are
expressed by
\begin{eqnarray*}
    &&L_2 = s \left( \partial_t -
    \bar\psi \partial_{\bar\psi} - \psi \partial_\psi \right)
    + {\textstyle{\frac{1}{2}}} \left( 1 + \mathrm{e}^{-2t}
    - s^2 + 2\bar\psi \psi \right) \partial_s\;,\\
    &&L_0 = s \left( - \partial_t +
    \bar\psi \partial_{\bar\psi} + \psi \partial_\psi \right)
    + {\textstyle{\frac{1}{2}}} \left( 1 - \mathrm{e}^{-2t}
    + s^2 - 2 \bar\psi \psi \right) \partial_s \;.
\end{eqnarray*}
Being the generators of the Lie algebra $\mathfrak{so}_{1,2}$ of the
Lorentz group, the operators $L_0, L_1, L_2$ satisfy the commutation
relations:
\begin{displaymath}
    [L_0 , L_1] = - L_2 \;, \quad [L_0 , L_2] = L_1 \;,
    \quad [L_1 , L_2] = L_0 \;.
\end{displaymath}
In particular, the generator $L_0+L_2 = [L_1 , L_0+L_2]$ is the
generator of translations of the coordinate $s\,$.

So far, we have been concerned with the case of a single site. To
pass to a lattice $\Lambda$ with many sites, we take the sum
\begin{displaymath}
    L_a = \sum\nolimits_{j \in \Lambda} L_a(j) \quad (a = 0,1,2)
\end{displaymath}
of differential operators over all sites.

By construction, the $\mathfrak{so}_{1,2}$ operators $L_a = \sum
L_a(j)$ are symmetries of the Berezin measure $D\mu_\Lambda\,$.
Therefore, they give rise to Ward identities:
\begin{equation}
    0=\int D\mu_\Lambda\,L_a \left(\mathrm{e}^{-A_{\beta,\varepsilon}}
    F \right) = \left\langle L_a F - F L_a A_{\beta,\varepsilon}
    \right\rangle \quad (a = 0,1,2)\;,
\end{equation}
which hold for any observable $F$ as long as these expectations
exist. By computing the symmetry-breaking terms from the formula $L_a
A_{\beta,\varepsilon} = \varepsilon \sum_j L_a(j)\, z_j$ one obtains
these Ward identities in the more explicit form
\begin{eqnarray*}
    &&\left\langle L_1 F \right\rangle = \varepsilon \sum\nolimits_j
    \left\langle (\sinh t_j - {\textstyle{\frac{1}{2}}} \mathrm{e}^{t_j}
    s_j^2 - \mathrm{e}^{t_j} \bar\psi_j \psi_j) F \right\rangle \;,\\
    &&\left\langle L_2 F \right\rangle = \varepsilon \sum\nolimits_j
    \left\langle\mathrm{e}^{t_j} s_j F \right\rangle \;,
    \qquad \left\langle L_0 F \right\rangle = 0 \;.
\end{eqnarray*}
The sum rule \eqref{eq:sumrule} now follows from the identity for
$\langle L_2 F \rangle$ by taking $F = \mathrm{e}^{t_i} s_i$ and
performing the Gaussian integrals over the fields $\psi$, $\bar\psi$,
and $s$.

Another important consequence results from making the choice $F =
\mathrm{e}^{t_j} s_j\,$. Since $L_0 F = - \sinh t_j + \mathrm{e}^{
t_j}( \frac{1}{2} s_j^2 + \bar\psi_j \psi_j)$, it follows from
$\langle L_0 F \rangle = 0$ that
\begin{equation}\label{eq:B.3}
    \langle \mathrm{e}^{t_j} \rangle = \langle \cosh t_j +
    \sinh t_j \rangle = \langle \cosh t_j + \mathrm{e}^{t_j}
    ({\textstyle{\frac{1}{2}}} s_j^2 + \bar\psi_j \psi_j)
    \rangle = \langle z_j \rangle = 1 \;.
\end{equation}
The last step, $\langle z_j \rangle = 1$, is by Proposition
\ref{prop:SUSY} of Appendix \ref{app:Ward}.

\section{SUSY Ward identities}\label{app:Ward}\resetequ

The action function of our $\mathrm{H}^{2|2}$ model has a global
symmetry w.r.t.\ the Lie superalgebra $\mathfrak{g} := \mathfrak
{osp}_{2|2}$ (for any $\varepsilon \geq 0$). As a result, there exist
supersymmetric Ward identities for suitable ($\mathfrak{osp}_{2|2}$
invariant) observables. Although such identities are standard
material from the theory of localization of supersymmetric integrals
\cite{schwarz}, we nonetheless give their derivation for completeness
here, as the said identities play a central role in our analysis. The
essence of the argument can already be understood at the very special
example of a lattice $\Lambda$ consisting of just a single site. For
pedagogical reasons we first handle this simple situation and then,
in a second step, give the generalization to arbitrary lattices.

The treatment will be most transparent if we do all calculations
using the coordinates $x, y, \xi, \eta$ described at the beginning of
Section \ref{sect:defs}. As stated there, for our purposes we may
view $\mathfrak{osp}_ {2|2}$ as the space of first-order differential
operators $D$ with coefficients that are linear in the variables $x,
y, \xi, \eta$ and the property
\begin{displaymath}
    D H = 0
\end{displaymath}
of annihilating the quadratic polynomial
\begin{displaymath}
    H = x^2 + y^2 + 2 \xi \eta \;.
\end{displaymath}
Let $Q$ be the distinguished first-order differential operator
defined by
\begin{equation}\label{eq:def-Q}
    Q = x \partial_\eta - y \partial_\xi
    + \xi \partial_x + \eta \partial_y \;.
\end{equation}
Clearly $Q$ is odd, converting even coordinate generators $x,y$ into
odd generators $\xi, \eta$ and vice versa. $Q$ is also seen to
annihilate $H$, and thus represents an element of $\mathfrak{osp}
_{2|2}\,$. Notice that $Q$ squares to
\begin{displaymath}
    Q^2 = x \partial_y - y \partial_x
    + \xi \partial_\eta - \eta \partial_\xi \;,
\end{displaymath}
which is a generator from the Lie algebra part $\mathfrak{o}_2 \oplus
\mathfrak{sp}_2$ of $\mathfrak{osp}_{2|2}\,$.

Now recall from Section \ref{sect:defs} that our Berezin
superintegration form is
\begin{displaymath}
    D\mu = (2\pi)^{-1} dx dy \, \partial_\xi \partial_\eta
    \circ (1 + H)^{-1/2} \;.
\end{displaymath}
\begin{lemma}\label{lem:inv-int}
The Berezin superintegration form $D\mu$ is $Q$-invariant, i.e.,
\begin{displaymath}
    \int_{\mathbb{R}^2} D\mu\; Q f = 0
\end{displaymath}
for any bounded smooth superfunction $f = f(x,y,\xi,\eta)$.
\end{lemma}
\paragraph{Proof.}
Since $Q$ is a first-order differential operator, we have from $Q H =
0$ that $Q (1 + H)^{-1/2} = 0$. Therefore, $D\mu\, Q f = D\mu\,
(1+H)^{1/2} Q (1+H)^{-1/2} f$ and
\begin{displaymath}
    \int_{\mathbb{R}^2} D\mu\, Q f = (2\pi)^{-1}
    \int_{\mathbb{R}^2} dxdy\, \partial_\xi \partial_\eta
    \left( x \partial_\eta - y \partial_\xi + \xi \partial_x
    + \eta\partial_y \right) (1+H)^{-1/2} f \;.
\end{displaymath}
The desired result now follows because $\partial_\xi^2 =
\partial_\eta^2 = 0$ and the integral over $\mathbb{R}^2$ of the
total derivatives $\partial_x (1+H^2)^{-1/2} f$ and $\partial_y
(1+H^2)^{-1/2}f$ vanishes. \qed\smallskip

An important property of the differential operator $Q$ is that the
joint zero locus of its coefficients is the origin $x = y = 0$ and
$\xi = \eta = 0$. Denoting the origin by $o$ we write $f(x=0 , y=0 ,
\xi=0 , \eta=0) \equiv f(o)$.
\begin{lemma}\label{lem:SUSYloc}
Let $f = f(x,y,\xi,\eta)$ be a smooth superfunction which satisfies
the invariance condition $Q f = 0$ and decreases sufficiently fast at
infinity in order for the integral $\int_{\mathbb{R}^2} D\mu\, f$ to
exist. Then
\begin{displaymath}
    \int_{\mathbb{R}^2} D\mu\, f = f(o) \;.
\end{displaymath}
\end{lemma}
\paragraph{Proof.}
The idea is to `deform' the integrand $f$ (without changing the
integral) by a factor that localizes the integral at $o$. We will do
this deformation by multiplication with $\mathrm{e}^{-\tau H}$ for
some positive real parameter $\tau$. Thus we are going to show that
\begin{equation}\label{eq:deform}
    \int D\mu\, f = \int D\mu \; \mathrm{e}^{-\tau H} f \;,
\end{equation}
independent of $\tau \geq 0\,$. The desired result will then follow
by taking $\tau \to +\infty\,$.

We begin by observing that the localizing function $H$ is $Q$-exact:
it can be written as $H = Q \lambda$ with
\begin{displaymath}
    \lambda := x\, \eta - y\, \xi
\end{displaymath}
an odd superfunction. Next, using the relation $Q H = Q^2 \lambda =
0$ we do the following calculation:
\begin{displaymath}
    \mathrm{e}^{-\tau H} = 1 + \big( \mathrm{e}^{-\tau (Q \lambda)}
    - 1 \big) = 1 + Q \left( \lambda\, \frac{\mathrm{e}^{-\tau
    (Q \lambda)} - 1}{Q\lambda} \right) \;.
\end{displaymath}
Here the term in parentheses stands for
\begin{displaymath}
    \frac{\mathrm{e}^{-\tau (Q \lambda)} - 1}{Q\lambda} :=
    \sum_{n=0}^\infty \frac{(-\tau)^{n+1}}{(n+1)!}\,(Q \lambda)^n \;.
\end{displaymath}
Inserting this decomposition into the integral we obtain
\begin{displaymath}
    \int D\mu \, \mathrm{e}^{-\tau H} f = \int D\mu \, f +
    \int D\mu\, f \, Q \left( \lambda\, \frac{\mathrm{e}^{-\tau
    (Q \lambda)} - 1}{Q\lambda} \right) \;.
\end{displaymath}
Since our integrand $f$ is $Q$-invariant by assumption ($Q f = 0$),
the second integral can also be written as
\begin{displaymath}
    \int D\mu\, f \, Q \left( \lambda\, \frac{\mathrm{e}^{-\tau
    (Q \lambda)} - 1}{Q\lambda} \right) = \int D\mu \, Q
    \left( f \, \lambda\, \frac{\mathrm{e}^{-\tau (Q \lambda)} - 1}
    {Q\lambda} \right) = 0\;,
\end{displaymath}
which vanishes by Lemma \ref{lem:inv-int}. This already proves
(\ref{eq:deform}).

To complete the proof, we consider the effect of a scale
transformation $\phi_\tau^\ast$: $x \mapsto x / \sqrt{\tau}$, $y
\mapsto y / \sqrt{\tau}$, $\xi \mapsto \xi / \sqrt{\tau}$, $\eta
\mapsto \eta / \sqrt{\tau}$. Note that $\phi_\tau^\ast H = H / \tau$
and the Berezin superintegration form $dx dy \, \partial_\xi
\partial_\eta  = D\mu \circ (1 + H)^{1/2}$ is invariant by
$\phi_\tau^\ast\,$. The statement of the Lemma now results from
taking the limit
\begin{eqnarray*}
    &&\int_{\mathbb{R}^2} D\mu\, f = \lim_{\tau\to\infty}
    \int D\mu\; \mathrm{e}^{-\tau H} f = \lim_{\tau\to\infty}
    \int \phi_\tau^\ast \left(D\mu\; \mathrm{e}^{-\tau H} f
    \right)\\ &&= \lim_{\tau\to\infty} \int D\mu\,(1 + H)^{1/2}
    (1+H/\tau)^{-1/2} \mathrm{e}^{-H} \phi_\tau^\ast f = f(o)\;,
\end{eqnarray*}
where the last step is done by verifying the normalization integral
\begin{displaymath}
    \int_{\mathbb{R}^2} D\mu\, (1 + H)^{1/2} \mathrm{e}^{-H} =
    (2\pi)^{-1} \int_{\mathbb{R}^2} dx dy\, \partial_\xi
    \partial_\eta\, \mathrm{e}^{- x^2 - y^2 - 2 \xi\eta} = 1 \;,
\end{displaymath}
and observing that $\lim_{\tau \to \infty} \phi_\tau^\ast f$ is the
constant function of value $f(o)$. \qed

We finally turn to the setting of an arbitrary lattice $\Lambda$. We
have a first-order differential operator $Q_j$ for every site $j \in
\Lambda$ and we now take the symmetry generator $Q$ to be the sum of
all of these:
\begin{displaymath}
    Q = \sum_{j \in \Lambda} Q_j = \sum_{j \in \Lambda}
    \left(x_j \partial_{\eta_j} - y_j \partial_{\xi_j} +
    \xi_j \partial_{x_j} + \eta_j \partial_{y_j} \right)\;.
\end{displaymath}
By the same argument as before, one sees that $D\mu_\Lambda$ is
$Q_j$-invariant for all $j$ and hence $Q$-invariant. There still
exists $H = \sum_{j \in \Lambda} (x_j^2 + y_j^2 + 2 \xi_j \eta_j)$
and $\lambda = \sum_{j \in \Lambda} (x_j\, \eta_j - y_j\, \xi_j)$
with $Q\lambda = H$. Hence we can still localize the integral $\int
D\mu_\Lambda\, F$ for any $Q$-invariant function $F$ by deforming
with $\mathrm{e}^{-\tau H}$ and sending $\tau \to \infty$. Thus we
arrive at the following result which, though valid for any choice of
coordinate system, will be stated in terms of the horospherical
coordinates $t_j, s_j, \bar\psi_j, \psi_j$ used in the body of the
paper.
\begin{proposition}\label{prop:SUSY}
For any $Q$-invariant, smooth and integrable function $F$ of the
lattice variables $t_j , s_j , \bar\psi_j, \psi_j$ the integral of
$F$ localizes at the zero-field configuration $t_j = s_j = \bar\psi_j
= \psi_j = 0$ (for all $j \in \Lambda$):
\begin{displaymath}
    \int_{(\mathbb{R}^2)^{|\Lambda|}} D\mu_\Lambda\, F = F(o)\;.
\end{displaymath}
In particular, for the partition function \eqref{eq:partfunc} we have
\begin{displaymath}
    Z(\beta,\varepsilon) = \int_{(\mathbb{R}^2)^{|\Lambda|}}
    D\mu_\Lambda \, \mathrm{e}^{- A_{\beta,\varepsilon}} = 1 \;.
\end{displaymath}
\end{proposition}

\end{document}